%% file: SUS-20-001_temp.tex
\pdfoutput=1
\documentclass[11pt,twoside,a4paper,cmspaper,final,collab]{cms-tdr}
\def\svnVersion{36e6610-D}\def\svnDate{2021/02/26}\def\cmsCernNoTag{CERN-EP-2020-231}\def\cmsCernDate{\today}\def\cmsMessage{"Published in the Journal of High Energy Physics as \href{http://dx.doi.org/10.1007/JHEP04(2021)123}{\doi{10.1007/JHEP04(2021)123}.}"}
\begin{document}\cmsNoteHeader{SUS-20-001}

\newlength\cmsTabSkip\setlength{\cmsTabSkip}{1ex}   
\providecommand{\cmsTable}[1]{\resizebox{\textwidth}{!}{#1}}
\providecommand{\cmsTableAlt}[1]{\resizebox{\textwidth}{!}{#1}}

\newcommand{\lint}{137\fbinv}
\newcommand{\gluino}{\PSg}
\newcommand{\firstchi}{\PSGczDo}
\newcommand{\gravitino}{\PXXSG}
\newcommand{\firstcharg}{\PSGcpmDo}
\newcommand{\secondchi}{\PSGczDt}
\newcommand{\squark}{\PSq}
\newcommand{\sbottom}{\PSQb}
\newcommand{\slep}{{\HepSusyParticle{\Pell}{}{}}\Xspace}
\newcommand{\mll}{\ensuremath{m_{\Pell\Pell}}\xspace}
\newcommand{\EM}{\Pepm{}\PGmmp}
\newcommand{\ttv}{\PQt{}\PAQt{}\PV}
\newcommand{\ttz}{\PQt{}\PAQt{}\PZ}
\newcommand{\znu}{\PZ{}+\PGn}
\newcommand{\vvv}{\PV{}\PV{}\PV}
\newcommand{\vz}{\PV{}\PZ}
\newcommand{\mttwol}{\ensuremath{M_{\mathrm{T2}}(\Pell\Pell)}\xspace}
\newcommand{\mttwo}{\ensuremath{M_{\mathrm{T2}}}\xspace}
\newcommand{\mttwolb}{\ensuremath{M_{\mathrm{T2}}(\Pell\PQb\Pell\PQb)}\xspace}
\newcommand{\mt}{\ensuremath{M_\mathrm{T}}\xspace}
\newcommand{\mjj}{\ensuremath{m_{\mathrm{jj}}}\xspace}
\newcommand{\mlb}{\ensuremath{\sum m_{\Pell\PQb}}\xspace}
\newcommand{\mbb}{\ensuremath{m_{\PQb\PQb}}\xspace}
\newcommand{\nb}{\ensuremath{n_{\PQb}}\xspace}
\newcommand{\njets}{\ensuremath{n_{\mathrm{j}}}\xspace}
\newcommand{\dyjets}{DY+jets\xspace}
\newcommand{\gammajets}{\PGg{}+jets\xspace}
\newcommand{\rmue}{\ensuremath{r_{\PGm/\Pe}}\xspace}
\newcommand{\RT}{\ensuremath{R_{\mathrm{T}}}\xspace}
\newcommand{\Rsfof}{\ensuremath{R_{\mathrm{SF/DF}}}\xspace}
\newcommand{\gjets}{\gammajets}
\newcommand{\rinout}{\ensuremath{r_{\text{in/out}}}\xspace}
\newcommand{\WZ}{\PW{}\PZ}
\newcommand{\VZ}{\PV{}\PZ}
\newcommand{\ZZ}{\PZ{}\PZ}
\newcommand{\WW}{\PW{}\PW}
\newcommand{\tW}{\PQt{}\PW}
\newcommand{\ZplusX}{\ensuremath{\PZ/\PGg^*{+}\PX}\xspace}
\newcommand{\absdphi}{\ensuremath{\abs{\Delta \phi^{\Pell\Pell}}}}
\newcommand{\widthCB}{\ensuremath{\sigma_{\mathrm{GB}}}}

\cmsNoteHeader{SUS-20-001}
\title{Search for supersymmetry in final states with two oppositely charged same-flavor leptons and missing transverse momentum in proton-proton collisions at \texorpdfstring{$\sqrt{s}=13\TeV$}{sqrt(s) = 13 TeV}}

\date{\today}

\abstract{
A search for phenomena beyond the standard model in final states with two oppositely charged same-flavor leptons and missing transverse momentum is presented. The search uses a data sample of proton-proton collisions at $\sqrt{s}=13\TeV$, corresponding to an integrated luminosity of $137\fbinv$, collected by the CMS experiment at the LHC. Three potential signatures of physics beyond the standard model are explored: an excess of events with a lepton pair, whose invariant mass is consistent with
the $\PZ$ boson mass; a kinematic edge in the invariant mass distribution of the lepton pair; and the nonresonant production of two leptons. The observed event yields are consistent with those expected from  standard model backgrounds.  The results of the first search  allow the exclusion of gluino masses up to $1870\GeV$, as well as chargino (neutralino) masses up to $750\,(800)\GeV$, while those of the searches for the other two signatures allow the exclusion of light-flavor (bottom) squark masses up to $1800\,(1600)\GeV$ and slepton masses up to $700\GeV$, respectively, at $95\%$ confidence level within certain supersymmetry scenarios.
}

\hypersetup{
pdfauthor={CMS Collaboration},
pdftitle={Search for supersymmetry in final states with two oppositely charged same-flavor leptons and missing transverse momentum in proton-proton collisions at sqrt(s) = 13 TeV},
pdfsubject={CMS},
pdfkeywords={CMS, physics, supersymmetry}}

\maketitle

\section{Introduction}
\label{sec:introduction}

During the last decades, the standard model (SM) of particle physics has been proven to successfully and accurately describe most particle phenomena. 
Despite its success, the SM does not account for experimental observations
such as the existence of dark matter~\cite{Bertone:2004pz}.
The theory of supersymmetry (SUSY)~\cite{Ramond:1971gb,Golfand:1971iw,Neveu:1971rx,Volkov:1972jx,Wess:1973kz,Wess:1974tw,Fayet:1974pd,Nilles:1983ge}
extends the SM through an additional symmetry that relates fermions and bosons: 
for each fermion (boson) of the SM, SUSY predicts the existence of a bosonic (fermionic) partner.
The SUSY partners of the SM particles can contribute to the stabilization of the electroweak loop corrections to the Higgs boson (\PH) mass and
allows the unification of the electroweak (EW) and strong interactions~\cite{unification2}.
Moreover, 
if $R$-parity~\cite{Farrar:1978xj} is conserved, the lightest SUSY particle (LSP) is predicted to be stable, likely neutral, and
possibly massive, 
representing thereby a suitable candidate for dark matter.

We present a search for  physics beyond the SM (BSM) in events with two oppositely charged (or opposite-sign, OS), same-flavor (SF) leptons
(denoted $\ell$, representing either electrons or muons), referred to as OSSF leptons, and an imbalance of transverse momentum, \ptmiss. The data are obtained from proton-proton ($\Pp{}\Pp$) collisions at a center-of-mass energy of $\sqrt{s}=13\TeV$ and correspond to an integrated luminosity
of \lint collected with the CMS detector at the CERN LHC in 2016--2018.

The search results are interpreted in the context of $R$-parity conserving SUSY models
that predict pairs of OSSF leptons in the  final state.
This signature is expected in a variety of SUSY models where the leptons
emerge either from on- or off-shell \PZ boson decays, or from the decay of the SUSY partners of SM leptons (sleptons, \slep). 
Leptons from the decay of an on-shell \PZ boson can produce an excess of events with a dilepton invariant
mass, \mll, close to the \PZ boson mass. 
In off-shell \PZ boson decays, the excess can present a characteristic edge-like distribution in the \mll spectrum~\cite{Hinchliffe:1996iu}. 

\begin{figure}[!ht]
\centering
\includegraphics[width=0.3\textwidth, height=3cm]{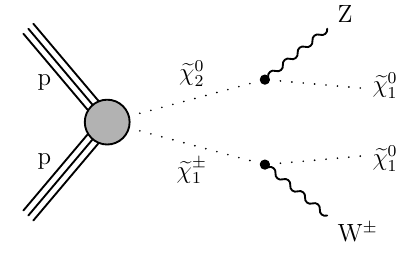}
\includegraphics[width=0.3\textwidth, height=3cm]{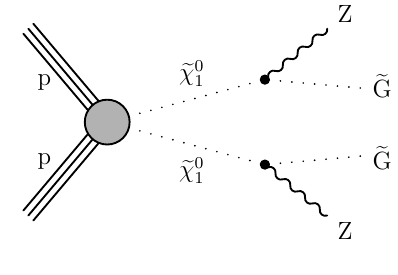}

\includegraphics[width=0.3\textwidth, height=3cm]{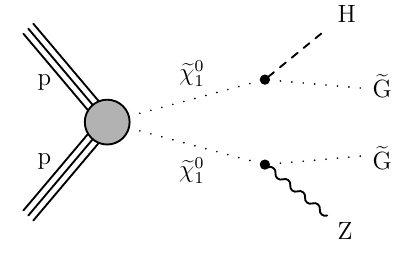}
\includegraphics[width=0.3\textwidth, height=3cm]{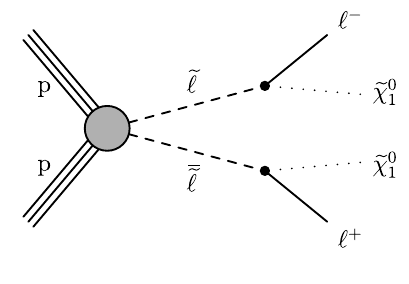}
\caption{Diagrams for models of neutralino/chargino production (upper left), GMSB neutralino pair production with  $\PZ\PZ$ (upper right) and  $\PZ\PH$ bosons (lower left) in the final state, and direct slepton pair production (lower right). In the first GMSB neutralino pair production model, the \firstchi is assumed to decay exclusively into a \PZ boson, while in the latter, the $\PZ\PH$ final state is accompanied by the $\PZ\PZ$ final state with 50\% branching fractions of the \firstchi decaying into an \PH or a \PZ boson. Only $\PZ\PH$ and $\PZ\PZ$ final states are taken into account in the analysis, since the contribution of the $\PH\PH$ topology to our signal regions is expected to be negligible.  Such models predict the SUSY particles to be produced via EW interactions, with limited if any production of accompanying quarks in the final state.  }
\label{sig:feynman_ewk}
\end{figure}

\begin{figure}[!ht]
\centering
\includegraphics[width=0.30\textwidth]{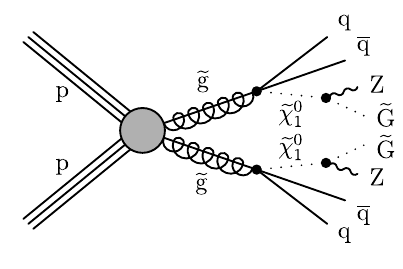}
\includegraphics[width=0.30\textwidth]{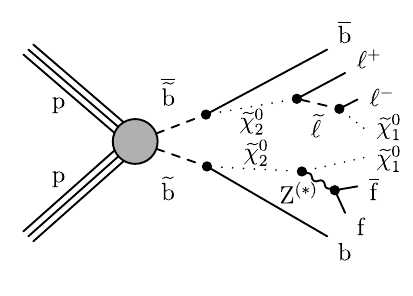}
\includegraphics[width=0.30\textwidth]{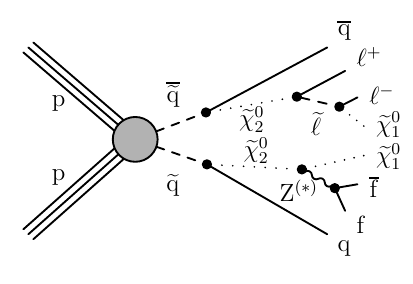}
\caption{
  Diagram for GMSB gluino (\gluino) pair production (left), where each \gluino decays into a pair
  of quarks and a neutralino. The neutralino then decays to a \PZ boson and an LSP.
  Diagrams for sbottom \sbottom (center) and squark \squark (right) pair production are also shown.
  Such models feature a mass edge from the decay of a \secondchi via an intermediate slepton,
  \slep. In the central diagram, a pair of \PQb quarks is present in the final state. In these
  models we assume a fixed \firstchi mass of 100\GeV, while the mass of the slepton is taken to be
  equidistant from the masses of the two neutralinos. Only the lightest \sbottom mass eigenstate,
  $\sbottom_1$, is assumed to be involved in the models considered.
  All these models assume strong production of SUSY particles and predict an abundance of quarks
  in the final state.}
\label{sig:feynman_strong}
\end{figure}

The search is designed to cover a range of simplified model spectra
(SMS)~\cite{ArkaniHamed:2007fw,Alwall:2008ag,Alwall:2008ag,Alwall:2008va,Alves:2011wf} 
that are classified according to the underlying SUSY model, 
the production mechanism (EW or strong production), 
and the abundance of quarks in the final state.
These models assume the production and subsequent decay of SUSY
particles in specific modes.
Some of these models are inspired by gauge-mediated SUSY breaking (GMSB) with the gravitino (\gravitino) as the LSP,
while in the others the lightest neutralino (\firstchi) is the LSP.
Diagrams for EW and strong production are shown in Figs.~\ref{sig:feynman_ewk}
and~\ref{sig:feynman_strong}.
The SMS models assume that all SUSY particles other than those directly involved in the specified process
are decoupled, \ie,  too heavy to be produced or affect the decays of the particles of interest.

Particles resulting from the decay of an object produced with a large Lorentz boost are present in models with a large mass splitting between SUSY particles and their decay products. Such particles are expected to be emitted in collinear configurations in the laboratory frame.
As shown in the upcoming sections, this feature is taken into account in the object and event selections 
to enhance the sensitivity of the search to such signatures.

Searches in this final state have been performed by the ATLAS~\cite{Aad:2019vnb,ATLAS:edge,ATLASewk8tev,ATLASOS13tev} and CMS~\cite{SUS-16-034,Sirunyan:2018nwe,CMS:edge,CMS:Zedge2015,OSpaperCMS7TeV,OSpaperCMS2011,2012ewk,2012ewkhiggs} experiments using data collected at $\sqrt{s}=8$ and 13\TeV. 
None of these searches reported evidence for BSM physics. Their results were used to constrain a range of (simplified) SUSY models.

Compared to previous work performed by the CMS experiment~\cite{SUS-16-034,Sirunyan:2018nwe} the search described in this paper is expanded by the addition of signal regions (SRs) targeting supersymmetry models with higher sparticle masses, and by improvements in the background estimations. This, together with the increase on luminosity, enhances the sensitivity to the models studied.

This paper is organized as follows. Section~\ref{sec:cmsdetector} provides a brief description of the CMS detector, while Section~\ref{sec:samplesObjects} describes the datasets, triggers and object reconstruction in CMS. Section~\ref{subsec:signalregions} describes the event selection criteria and the SRs used in the search, while the estimation of the SM background contribution is described in Section~\ref{sec:backgrounds}. Section~\ref{sec:kinfit} describes the fit to the  \mll distribution, used to extract a possible edge-like signal. The results of the search are described in Section~\ref{sec:results}, and are interpreted in terms of constraints on the cross sections of the SMS models, as described in Section~\ref{sec:interpretation}. Finally, a summary of the analysis is given in Section~\ref{sec:summary}.

\section{The CMS detector}
\label{sec:cmsdetector}

The central feature of the CMS apparatus is a superconducting solenoid of 6\unit{m}
internal diameter, providing a magnetic field of 3.8\unit{T}. Within the solenoid
volume are a silicon pixel and strip tracker, a lead tungstate crystal electromagnetic
calorimeter (ECAL), and a brass and scintillator hadron calorimeter (HCAL), each composed
of a barrel and two endcap sections. 
The tracker system measures charged particles within the pseudorapidity range $\abs{\eta}<2.5$.
The ECAL is a fine-grained hermetic calorimeter with quasi-projective geometry,
and is segmented into the barrel region of $\abs{\eta} < 1.48$ and in two endcaps that extend up to $\abs{\eta} < 3.0$.
The HCAL barrel and endcaps similarly cover the region $\abs{\eta} < 3.0$.
Forward calorimeters extend the coverage up to $\abs{\eta} < 5.0$.
Muons are measured and identified in the range $\abs{\eta} < 2.4$
by gas-ionization detectors embedded in the steel flux-return yoke outside the solenoid.
 A two-tier trigger system selects events of interest for physics analysis.
The first level of the CMS trigger system, composed of custom hardware processors, uses information from the
calorimeters and muon detectors to select the most interesting events in a fixed time interval of less than 4\mus. The high-level
 trigger processor farm further reduces the event rate from around 100\unit{kHz} to about 1\unit{kHz}, before data storage.
A more detailed description of the CMS detector and trigger system, together with a definition of the coordinate system used and the relevant
kinematic variables, can be found in Refs.~\cite{Chatrchyan:2008zzk,CMStrigger}.

\section{Data, triggers, and object reconstruction}
\label{sec:samplesObjects}

We use events containing at least two OS leptons ($\EE$, $\MM$, or $\EM$).
Only SF leptons ($\EE$ or $\MM$) are used to define SRs,
while $\EM$ events are used in control regions (CRs).
These events are preselected using dilepton triggers
that require the leptons with the highest (next-to-highest) transverse momentum \pt
to pass respective thresholds ranging from 17--23\,(8--12)\GeV, depending on the data taking period and lepton flavor.
In addition, these triggers require the leptons to pass isolation criteria.
To retain high efficiency for highly boosted topologies that contain nearly collinear lepton pairs, 
we also use a second set of dilepton triggers with higher respective \pt thresholds of 25--37\,(8--33)\GeV but
without any isolation requirement.
Trigger efficiencies are measured in events selected using triggers
based on the \ptmiss and found to be 85--95\%.
In addition, a \gammajets event sample is used as a CR to estimate the Drell--Yan (DY) background (as discussed in Section~\ref{sec:backgrounds}).
This sample is collected using a set of photon triggers with \pt thresholds ranging between 50 and 200\GeV.
A subset of these photon triggers with lower \pt thresholds are prescaled to keep the rate under control.
Events collected with prescaled triggers are reweighed accordingly.

The particle-flow algorithm~\cite{Sirunyan:2017ulk} aims to reconstruct and identify each individual particle in an event, with an optimized combination of information from the various elements of the CMS detector. The energy of photons is obtained from the ECAL measurement. The energy of electrons is determined from a combination of the electron momentum at the primary interaction vertex as determined by the tracker, the energy of the corresponding ECAL cluster, and the energy sum of all bremsstrahlung photons spatially compatible with originating from the electron track. The momentum of muons is obtained from the curvature of the corresponding track. The energy of charged hadrons is determined from a combination of their momentum measured in the tracker and the matching ECAL and HCAL energy deposits, corrected for the response function of the calorimeters to hadronic showers. Finally, the energy of neutral hadrons is obtained from the corresponding corrected ECAL and HCAL energies.
The particles reconstructed with this algorithm are referred to as PF candidates.
Events selected for further study require the presence of at least one reconstructed
vertex. Due to the presence of additional $\Pp\Pp$ interactions within the same or nearby bunch crossings (pileup) several vertices
are reconstructed. The candidate vertex with the largest value of summed object $\pt^2$ is taken to be the primary
$\Pp\Pp$ interaction vertex (PV). The physics objects considered for the construction of the candidate vertex
are the jets, clustered using the anti-\kt jet finding
algorithm~\cite{Cacciari:2008gp,Cacciari:2011ma} with the PF tracks assigned as inputs,
and the associated missing transverse momentum, taken as the negative of the vector \pt sum of those jets.

Electrons and muons are identified among the PF candidates by
exploiting specific signatures in the CMS subdetectors~\cite{MUO-16-001,Khachatryan:2015hwa}. 
Leptons reconstructed in the transition region between
the barrel and endcap of the ECAL ($1.4<\abs{\eta}<1.6$) are rejected
to reduce efficiency differences between electrons and muons.
Muons are required to pass the medium identification criteria described in Ref.~\cite{MUO-16-001},
while electrons are selected according to a multivariate discriminant based on the shower shape
and track quality variables~\cite{Khachatryan:2015hwa}.
These criteria maintain approximately 99 (90)\% efficiency for muons
(electrons) produced in the decay of \PW or \PZ bosons~\cite{MUO-16-001,Khachatryan:2016kod}.
For both lepton flavors, the impact parameter relative to the PV
is required to be $<$0.5\unit{mm} in the transverse
plane and $<$1\unit{mm} along the beam direction.
To reject lepton candidates within jets, leptons are required to be
isolated from other particles in the event. The lepton isolation variable is
defined as the scalar \pt sum of all PF candidates in a cone around
the lepton. The cone size, defined as $\DR=\sqrt{\smash[b]{(\Delta\eta)^2+(\Delta\phi)^2}}$,
where $\phi$ is the azimuthal angle in radians,
changes as a function of the lepton \pt:  $\DR=0.2$ when $\pt<50\GeV$, $\DR=10\GeV/\pt$
when $50<\pt<200\GeV$, and $\DR=0.05$ otherwise.
This choice prevents efficiency loss due to the overlap of leptons and jets
in events with high jet multiplicity.
In order to mitigate the effect of pileup, charged particles that originate only from the
primary vertex are taken into account in the calculation of the isolation variable. In addition,
residual contributions from pileup to the neutral component of the isolation
are subtracted using the method described in Ref.~\cite{Khachatryan:2015hwa}. The isolation
variable is required to be $<$10 (20)\%
of the electron (muon) \pt. 
The electron and muon selections are optimized 
to maximize the corresponding selection efficiencies, in addition to retaining similar selection efficiencies for the two flavors, in order to enhance the statistical power of some of the CRs described in Section~\ref{sec:backgrounds} that are employed to estimate SM backgrounds.

Photons are required to pass identification criteria based on the cluster energy distribution in the ECAL
and on the fraction of their energy deposited in the HCAL~\cite{CMSPhotonID}.
Photons must have $\pt>50\GeV$, and be within $\abs{\eta}<2.4$, excluding the ``transition region"
of $1.4<\abs{\eta}<1.6$ between the ECAL barrel and endcap.
Photons are required to be isolated from other PF candidates within a cone of $\DR=0.3$.
To distinguish photons from electrons, we reject photons that can be associated to a pattern of hits in the pixel detector
indicating the presence of a charged-particle track.
To reduce the contamination due to mismeasurements of the
photon energy that can create a significant \ptmiss, events with $\Delta\phi(\vec{p}^{\,\cPgg}_\mathrm{T},\ptvecmiss)<0.4$ are rejected.
The vector \ptvecmiss is defined as the negative vector \pt sum of all the PF candidates in the event.

To further identify additional leptons and isolated charged hadrons in the final state, isolated charged particle tracks
that are identified by the PF algorithm as leptons (charged hadrons) and having $\pt>5$ (10)\GeV are used.

Jets are clustered from PF candidates
using the anti-\kt\ clustering algorithm~\cite{Cacciari:2008gp} with a distance parameter of 0.4,
unless specified otherwise, implemented in the \FASTJET\ package~\cite{Cacciari:2011ma,Cacciari:2005hq}.
Jet momentum is determined as the vectorial sum of all particle momenta in the jet, and is found from simulation to be, on average, within 5 to 10\% of the true momentum over the whole \pt spectrum and detector acceptance. Pileup interactions can contribute additional tracks and calorimetric energy depositions, increasing the apparent jet momentum. To mitigate this effect, tracks identified to be originating from pileup vertices are discarded and an offset correction is applied to correct for remaining contributions. Jet energy corrections are derived from simulation studies so that the average measured energy of jets becomes identical to that of particle level jets. In situ measurements of the momentum balance in dijet, photon+jet, Z+jet, and multijet events are used to determine any residual differences between the jet energy scale in data and in simulation, and appropriate corrections are made~\cite{Khachatryan:2016kdb}. Additional selection criteria are applied to each jet to remove jets potentially dominated by instrumental effects or reconstruction failures.
Jets are required to satisfy $\abs{\eta}<2.4$ and $\pt>25$ or $35\GeV$,
where the 25\GeV threshold is considered in regions in which the presence of jets is vetoed, in
order to efficiently reject SM processes with jets, while the 35\GeV threshold is used to construct
regions aiming for topologies with jets.
Corrections to the jet energy are propagated to \ptvecmiss
using the procedure developed in Ref.~\cite{1748-0221-6-11-P11002}.
As isolated prompt leptons or photons may be included in the jet definition,
jets are removed from the event if they point within $\DR<0.4$ of any of the selected leptons or the highest \pt\ photon.
The DeepCSV algorithm~\cite{Sirunyan:2017ezt}
is used to identify jets produced by the hadronization of \PQb quarks, 
using a working point that yields an identification efficiency of about 70\% and
misidentification probabilities of 1 and 12\% for light-flavor or gluon jets and charm jets,
respectively. These efficiencies are measured in data samples enriched in \ttbar and multijet events
as a function of jet \pt and $\eta$~\cite{Sirunyan:2017ezt} and are used to correct the prediction from simulated events. 
Jets passing the \PQb-tagging criteria are required to have $\abs{\eta}<2.4$ and $\pt>25$ or 35\GeV, depending
on the SR,
as described in Section~\ref{subsec:signalregions}.

Jets reconstructed using the anti-\kt clustering algorithm with a distance parameter of 0.8 are used
to identify energetic \PW and \PZ bosons that decay to $\PQq\PAQq'$, since their decay
products are collimated into a single large radius jet. The \PV ($\PV=\PW$ or $\PZ$) boson candidates  have $\pt>200\GeV$ and soft-drop masses between 65 and 105\GeV; the soft-drop
mass is a groomed jet mass calculated using the mass drop algorithm~\cite{Butterworth:2008iy,Dasgupta:2013ihk} with the angular
exponent $\beta=0$ and a soft cutoff threshold $z_\text{cut}<0.1$. 
Additional selection criteria are imposed on the ratio of the 2- to the 1-subjettiness variable~\cite{Thaler:2010tr}, 
$\tau_{21}=\tau_{2}/\tau_{1}$,
to select jets compatible with a 2-prong structure expected in \PV boson decays~\cite{Sirunyan:2020lcu}.
These variables are calibrated in a \ttbar sample enriched in hadronically decaying $\PW$ bosons~\cite{Khachatryan:2014vla}.

Samples of simulated events are used to model signal and background processes.
The BSM signal events are generated using the 
\MGvATNLO
program~2.3.3~\cite{Alwall:2014hca} at leading order (LO) precision, with up to two additional partons in the matrix element calculation. 
Samples of DY processes and photons produced in association with jets (\gammajets)
are generated using the \MGvATNLO
event generator at LO precision,
with up to four additional partons in the matrix element. 
The \ttv and \vvv  events are produced with the same generator at next-to-LO (NLO) precision.
Other SM processes, such as \WW, $\PQq\PAQq\to\PZ\PZ$, \ttbar, and single top quark production, are generated at NLO precision 
using \POWHEG~(v1.0, or v2.0)~\cite{powheg0,powheg1,powheg2}.
A generator-level \pt-dependent next-to-NLO (NNLO)/NLO k-factor~\cite{ZZ_1,ZZ_2,ZZ_3}, ranging from 1.1 to 1.3, is applied
to simulated $\PQq\PAQq\to\PZ\PZ$ events 
to account for the missing higher-order matrix element contributions.
Finally, the $\Pg\Pg\to\PZ\PZ$ process is generated at LO using \MCFM~7.0~\cite{mcfm0,mcfm1,mcfm2}.

For modeling fragmentation and parton showering, generators described above are interfaced to \PYTHIA~\cite{pythia82}~8.205 for 2016 samples and \PYTHIA~8.230 for 2017 and 2018 samples.  
For samples generated at LO (NLO) precision, the MLM~\cite{Alwall:2007fs} (FxFx~\cite{fxfx}) prescription is used to match partons from 
the matrix element calculation to those from parton showers.
The CUETP8M1 underlying event tune~\cite{CUETP8M1bib} is used for the 2016 SM background and signal.
For 2017 and 2018, the CP5 and CP2 tunes~\cite{CP15bib} are used for the SM background and signal samples, respectively.
The NNPDF3.0LO (NNPDF3.0NLO)~\cite{Ball:2014uwa} parton distribution functions (PDFs) are used to generate the 2016 LO (NLO) samples,
while the NNPDF3.1LO (NNPDF3.1NNLO)~\cite{nnpdf3bib} PDFs are used for the 2017 and 2018 samples.

For all SM processes, the detector response is simulated through a \GEANTfour model~\cite{Geant} of the CMS detector,
while BSM samples are processed using the CMS fast simulation framework~\cite{fastsim,fastsimrun2}.
The simulation programs account for different detector conditions in the three years of data taking.
Multiple $\Pp\Pp$~interactions are superimposed on the hard
collision, and the simulated events are reweighed in a way that the number of collisions per bunch crossing accurately reflects
the observed distribution.

Cross sections at NLO and NNLO~\cite{Alwall:2014hca,powheg2,Re:2010bp,Czakon:2011xx,ggZZ,Li:2012wna}
are used to normalize the simulated background samples,
while signal cross sections are implemented at NLO using next-to-leading-logarithmic (NLL) order in \alpS~\cite{ewkxsec0,ewkxsec1,ewkxsec2,ewkxsec3,ewkxsec4,ewkxsec5,ewkxsec6,ewkxsec7} soft-gluon
for the EW processes,
or at approximately NNLO + next-to-NLL (NNLL) 
order in \alpS~\cite{bib-nlo-nll-01,bib-nnll-05,bib-nlo-nll-02,bib-nlo-nll-03,bib-nlo-nll-04,bib-nnll-06,bib-nlo-nll-05,bib-nnll-02,bib-nnll-03,bib-nnll-04,bib-nnll-07,bib-nnll} 
for the strong production.
The production cross sections for the EW GMSB model
are computed in a limit of mass-degenerate higgsino states, the lightest chargino (\firstcharg), the next-to-lightest neutralino (\secondchi), and \firstchi
with all the other SUSY particles assumed to be heavy and decoupled.

\section{Event selection}
\label{subsec:signalregions}

The SRs are designed to be sensitive to a range of BSM  models
while keeping moderate SM background rates.
Four main samples are defined starting from a baseline selection
and are tuned to maximize the sensitivity to specific SUSY processes.
Since the statistical interpretation of the results is performed separately in each sample, we do not require the samples to be disjoint.
The first (second) sample targets strong (EW)-production SUSY processes with an on-shell \PZ boson in the decay chain.
Another sample, referred to as the ``edge'' sample, targets strong SUSY production with an off-shell \PZ boson or a slepton in the decay chain.
The requirements for the fourth sample
are designed to be sensitive to the direct production of a slepton pair.
The selections used to define all samples are summarized in Table~\ref{tab:selections_signalRegions}.
In addition to the SRs, we also define a set of CRs to be used in the estimation of the main SM backgrounds. 

\begin{table}[htb]
\topcaption{\label{tab:selections_signalRegions} Summary of search category selections. In regions with the additional lepton
  veto selection, events containing additional leptons or charged isolated tracks are rejected. All the regions besides the edge search samples implement a veto to an additional lepton. The numbers in the rightmost
column represent the edges of the bins that define the signal regions. Events in the edge search sample are further categorized as \ttbar-like and non-\ttbar-like as described in Section~\ref{subsub:edge}. }
\centering
\cmsTableAlt{
\begin{tabular}{llllll}

\multicolumn{6}{c}{ \textit{Strong-production on-\PZ search sample} ($86 < \mll < 96\GeV$)  }  \\
Region & \njets & \nb & \HT [\GeVns{}]& \mttwol [\GeVns{}]  &\ptmiss bins [\GeVns{}]\\ \hline 
SRA \PQb veto & 2--3 & $=$0 & $>$500 & $>$80   & [100, 150, 230, 300, $\infty$) \\
SRB \PQb veto & 4--5 & $=$0 & $>$500 & $>$80   & [100, 150, 230, 300, $\infty$) \\
SRC \PQb veto & $>$5 & $=$0 & \NA & $>$80      & [100, 150, 250, $\infty$)  \\
SRA \PQb tag  & 2--3 & $>$0 & $>$200 & $>$100  & [100, 150, 230, 300, $\infty$) \\
SRB \PQb tag  & 4--5 & $>$0 & $>$200 & $>$100  & [100, 150, 230, 300, $\infty$) \\
SRC \PQb tag  & $>$5 & $>$0 & \NA & $>$100     & [100, 150, 250, $\infty$) \\ [\cmsTabSkip]
\multicolumn{6}{c}{ \textit{EW-production on-\PZ search sample} ($86 < \mll < 96\GeV$)}  \\ 
\multirow{2}{*}{Region} & \multirow{2}{*}{\njets ($n^{\text{boosted}}_{\PV}$)} & \multirow{2}{*}{\nb} & Dijet mass & \multirow{2}{*}{\mttwo [\GeVns{}]} & \multirow{2}{*}{\ptmiss bins [\GeVns{}]}\\
       &                                       &     &  [\GeVns{}]&                   &                       \\ \hline 
Boosted  \vz & $<$2 \ ($>$0) & $=$0 & \NA          & \NA                & [100, 200, 300, 400, 500, $\infty$) \\
Resolved \vz & $>$1          & $=$0 & $\mjj < 110$ & $\mttwol > 80$     & [100, 150, 250, 350, $\infty$)      \\
$\PH\PZ$     & $>$1          & $=$2 & $\mbb < 150$ & $\mttwolb > 200$   & [100, 150, 250, $\infty$)           \\ [\cmsTabSkip]
\multicolumn{6}{c}{\textit{Edge search sample}  ($20<\mll< 86$ or  $\mll > 96\GeV$)  }  \\ 
Region & \njets &  \nb & \mttwol [\GeVns{}]& \ptmiss  [\GeVns{}]  & \mll bins [\GeVns{}]\\ \hline 
Edge fit & $>1$ & \NA &$>$80 & $>$200 & $>$20\\
\PQb veto & $>1$ & $=$0 &$>$80 & $>$150  & [20, 60, 86]$+$[96, 150, 200, 300, 400, $\infty$)\\
\PQb tag & $>1$ & $>$0 &$>$80 & $>$150 & [20, 60, 86]$+$[96, 150, 200, 300, 400, $\infty$)\\
\multicolumn{6}{c}{\textit{Slepton search sample} ($20<\mll < 65$ or  $\mll > 120\GeV$)}  \\
Region & \njets &\nb & $\pt^{\Pell_{2}}/ \pt^{\mathrm{j}_{1}}$ & \mttwo [\GeVns{}] &\ptmiss bins [\GeVns{}]\\ \hline 
Slepton jet-less & $=$0 & $=$0 & \NA & \mttwol $>$100      & [100, 150, 225, 300, $\infty$) \\
Slepton with jets & $>$0 & $=$0 & $>$1.2 & \mttwol $>$100  & [100, 150, 225, 300, $\infty$) \\ 
\end{tabular}}
\end{table}

The baseline selection requires the presence of two OS leptons within $\abs{\eta}<2.4$
and with $\pt>25$ (20)\GeV for the highest (next-to-highest) \pt lepton. 
Each event must contain lepton flavors consistent with the corresponding requirement at the trigger level; \eg,
if an event is preselected using a dilepton $\EE$ trigger, both leptons are required to be electrons.
To avoid differences in reconstruction and isolation efficiencies between electrons and muons in boosted topologies,
the two highest \pt leptons are required to be separated by a distance $\DR>0.1$.
The \mll of the dilepton system, its transverse momentum  $\pt^{\Pell\Pell}$, and \ptmiss are required to be greater than 20, 50 and 50\GeV, respectively.
In the SRs, the two highest \pt leptons are also required to have the same flavor, $\EE$ or $\MM$, 
while for a number of CRs we require the presence of different-flavor (DF) leptons, $\EM$.

To suppress backgrounds where instrumental \ptmiss arises from mismeasurements of jet energies, 
the two highest \pt jets in the event are required to have a separation in $\phi$ from \ptvecmiss of at least 0.4, or $\Delta\phi(\vec{p}^{\,\mathrm{j}_{1,2}}_\mathrm{T},\ptvecmiss)>0.4$. 
In regions with only one jet, this criterion is only applied to the single jet. 
If the aforementioned jet is a $\PV$ boson candidate, the selection is modified to $\Delta\phi>0.8$.

\subsection{The on-\texorpdfstring{\PZ}{Z} search sample}

Events with a \PZ boson candidate define the ``on-\PZ'' SRs and must have an invariant mass of $86<\mll<96\GeV$. 
Events containing additional leptons or isolated tracks, as described in Section~\ref{sec:samplesObjects}, are rejected.

\subsubsection{Strong-production on-\texorpdfstring{\PZ}{Z} search samples}

Six disjoint event categories are defined that are expected to be sensitive
to strong production of SUSY particles.
These are defined
on the basis of the number of jets (SRA, SRB and SRC) reconstructed with a distance parameter
of 0.4 having $\pt\geq35\GeV$ (henceforth called $\njets$) and the presence of \PQb-tagged jets (\PQb veto and \PQb tag).
This selection is made targeting the gluino (\gluino) pair production mode considered in Section~\ref{sec:introduction}, in cases where one of the \PZ boson decays leptonically and the remaining, hadronically. 
Further requirements are made on the \mttwo variable defined below,
as well as \HT, the scalar sum of jet \pt.
Each category is divided into multiple bins of \ptmiss, as
indicated in Table~\ref{tab:selections_signalRegions}.

The \mttwo variable~\cite{MT2variable,MT2variable2} is used
to reduce the \ttbar background contribution.
It is constructed from \ptvecmiss and two visible objects, as:
\begin{linenomath}
\begin{equation}
\mttwo = \min_{\ptvecmiss{}^{(1)} + \ptvecmiss{}^{(2)} = \ptvecmiss}
  \left[ \max \left( \mt^{(1)} , \mt^{(2)} \right) \right],
\label{eq:MT2}
\end{equation}
\end{linenomath}
where $\ptvecmiss{}^{(i)}$ ($i=1,2$) are two vectors in the transverse plane that represent
an hypothesis for the invisible objects and whose sum is equal to \ptvecmiss. The $\mt^{(i)}$ are
the transverse masses obtained by pairing the $\ptvecmiss{}^{(i)}$ with either of the two visible objects.  
When evaluated using the two selected leptons as the visible objects, 
the resulting quantity is referred to as \mttwol and exhibits
an endpoint at the \PW boson mass in \ttbar events. 
A requirement of $\mttwol>100$ (80)\GeV is applied in the \PQb-tagged jet (veto)  SRs 
to suppress such background contributions.

\subsubsection{Electroweak-production on-\texorpdfstring{\PZ}{Z} search samples}

The first EW on-\PZ event category (referred to as ``\vz'' category)
targets final states with a diboson pair ($\PZ\PZ$ or $\PZ\PW$),
with one leptonically decaying \PZ boson, and with the second boson decaying
into jets.
Depending on its momentum, the decay products of the
decaying boson can either be collimated and reconstructed within a large radius jet,
or resolved into
two jets.  For this reason, we
define two subcategories, ``boosted'' and ``resolved'' that are subdivided
into several bins of \ptmiss.

For the resolved subcategory we require the presence of at least two jets,
and require the two that are closest in $\phi$ to have an invariant mass $\mjj < 110\GeV$,
consistently with being a \PV boson decaying into jets.  To reduce the
\ttbar background contribution, we reject events
that have \PQb-tagged jets with $\pt>25\GeV$ or 
$\mttwol<80\GeV$.

In the boosted subcategory we require the presence of a
large-radius jet with $\pt>200\GeV$,
consistent with a hadronically decaying $\PV$ boson
candidate ($n^{\text{boosted}}_{\PV}>0$).
In order to ensure that the boosted and resolved categories are disjoint,
events with $\njets>1$ are not accepted.

The last EW-production on-\PZ category, referred to as ``$\PH\PZ$'',
is designed to be sensitive to events with an
$\PH\to\bbbar$ decay.  
Events in this category must have exactly two \PQb-tagged jets with $\pt>25\GeV$ and an invariant mass $\mbb<150\GeV$. 
To reduce the \ttbar background contribution, the \mttwo variable is calculated
using combinations consisting of one lepton and one \PQb-tagged jet as visible objects. 
Each lepton is paired with a \PQb-tagged jet, and \mttwo is evaluated for
all possible $\ell\PQb$-$\ell\PQb$ combinations. The
smallest value of \mttwo is denoted by \mttwolb.
We require $\mttwolb > 200\GeV$, since
in \ttbar events this variable has an endpoint at the top mass.
The events are finally subdivided in bins of \ptmiss.

\subsection{The off-\texorpdfstring{\PZ}{Z} search samples}

Additional samples (``edge'' and ``slepton'')
are defined targeting models without on-shell \PZ bosons in the final state. 
The edge SRs are designed for signals with several jets in the
final state and 
with a kinematic edge in the dilepton invariant mass distribution.
The slepton SRs do not require significant jet activity in the final state.

\subsubsection{Edge search sample}
\label{subsub:edge}
The edge sample is constructed with events with at least two jets, $\ptmiss>150$ or 200\GeV, and ${\mttwol>80\GeV}$ to reject DY and \ttbar events.
Two approaches are used to search for a kinematic edge in the \mll spectrum.
The first one is based on 
a fit to the \mll distribution
in events with $\ptmiss>200\GeV$ as described in 
Section~\ref{sec:kinfit}.
In the second approach, we count 
the number of events with $\ptmiss>150\GeV$ distributed across 28 disjoint regions
as described below. A looser selection on \ptmiss is applied, since with this categorization
we can define regions with improved signal purity.
First, we define seven bins in \mll,
excluding the region $86<\mll<96\GeV$, 
to be able to probe different positions of a possible kinematic edge.
For each \mll bin, events are further categorized according to the \PQb-tagged jet multiplicity,
counting \PQb-tagged jets with $\pt>25\GeV$.
Events are also categorized as \ttbar-like or non-\ttbar-like based on a
likelihood discriminant 
that exploits different kinematic properties of \ttbar events relative to a range of possible BSM contributions.
We construct this discriminant as a product of probability density functions in the observables
\ptmiss, $\pt^{\Pell\Pell}$, the $\Delta\phi$ between the two leptons \absdphi, and \mlb.

The \mlb variable is defined as the sum of the invariant masses of  
two lepton-jet pairs.
Priority is given to pairs consisting of a lepton and a \PQb-tagged jet. However, if there are no \PQb-tagged jets in the event, we
use jets
without \PQb tags.
The first lepton-jet pair is selected as the one with the minimum invariant mass.
The second pair is obtained by repeating the same procedure, after the exclusion of the already selected lepton and jet.

The likelihood is constructed from probability density functions 
for each observable obtained from DF CR enriched in \ttbar events.
We use
a sum of two exponential distributions for \ptmiss,
a third-order polynomial for \absdphi,
and Crystal Ball (CB)~\cite{Crystal} functions for both $\pt^{\Pell\Pell}$ and \mlb.
These distributions are found to model well those observed.
The negative logarithm of the likelihood is then taken as the discriminator value
used to categorize the event as either \ttbar-like or non-\ttbar-like.

\subsubsection{Slepton search sample}

The slepton SRs seek BSM signatures with two leptons, with
\ptmiss ($>100\GeV$),
no \PQb-tagged jets,
and moderate jet activity.
The threshold on the highest \pt lepton is
raised from the baseline requirement of 25 to 50\GeV, in order to further suppress the \dyjets contribution.
In addition, \mll is required to be $<65$ or $>$120\GeV
and \mttwol must be $>$100\GeV.  
Events are categorized on the basis of the jet multiplicity ($\njets=0$ or $\njets > 0$),
but events with one jet or more are kept only if $\pt^{\Pell_{2}}/\pt^{\mathrm{j}_{1}}>1.2$. 
The $\njets > 0$ category serves to recover possible BSM events characterized by moderate initial-state radiation (ISR).
Events are then further split into bins of \ptmiss, as shown in Table~\ref{tab:selections_signalRegions}.

\section{Standard model background}
\label{sec:backgrounds}

Three independent sources of SM backgrounds contribute to the SRs. 
The first consists of flavor-symmetric backgrounds 
from SM processes where SF and DF lepton pairs are produced at the same rate.
The dominant process contributing to such a category is \ttbar production. 
Additional contributions arise from \WW, $\PZ/\PGg^*\to\PGtp\PGtm$ and \tW production 
as well as events with leptons from hadron decays.
Flavor-symmetric backgrounds are estimated by constructing DF control samples in data.

The second source of backgrounds results from
\dyjets events with significantly mismeasured \ptmiss (referred to as instrumental \ptmiss in what follows).
This background is estimated from photon data samples in combination with CRs enriched in \dyjets events.

The third type of SM backgrounds consists of processes yielding final states with an SF lepton pair produced
in the decay of a \PZ boson or a virtual photon accompanied by neutrinos (\PGn) produced in the decay of a \PW or \PZ bosons. 
The main process contributing here is \VZ production.
Rarer processes, such as \ttz production, also contribute to certain SRs. 
These backgrounds are referred to as \znu backgrounds and
are estimated from simulation. The prediction is validated in dedicated data control regions.

\subsection{Flavor-symmetric backgrounds}
\label{sub:fsbkg}

As already mentioned,
the estimation of flavor-symmetric backgrounds exploits the fact
that in such processes, the DF and SF events are produced at the same rate. 
The CRs are defined in data with the same selections as the corresponding SRs, 
but requiring the presence of a DF lepton pair instead of an SF pair.
The background contribution in the SR is then predicted by means of
a transfer factor, denoted by \Rsfof,
that accounts for the differences in reconstruction,
identification and trigger efficiencies between DF and SF events.
These are caused by the residual differences in the 
efficiencies between electrons and muons.
The transfer factor consists of the product of two correction factors, determined from CR data.

The first correction factor, \rmue, is the ratio of muon and electron reconstruction and identification efficiencies
measured in a region
enriched in DY events, requiring two SF leptons, at least two jets, 
$\ptmiss<50\GeV$, and $60<\mll<120\GeV$. 
Assuming that the efficiency for each of the two leptons in the event is independent of the other lepton, 
\rmue can be defined as $\rmue=\sqrt{\smash[b]{N_{\PGmp\PGmm}/N_{\Pep\Pem}}}$, where $N_{\PGmp\PGmm(\Pep\Pem)}$ is the number of $\PGmp\PGmm$ $(\Pep\Pem)$ events.
The \rmue factor is
parametrized as a function of the lepton \pt and $\eta$ by
the following empirical form:
\begin{linenomath}
\begin{equation}
\begin{aligned}
\rmue(\pt,\eta)  &= \rmue^0 \, f\left(\pt\right) \, g\left(\eta\right),  \\
\end{aligned}
\end{equation}
\end{linenomath}
where
\begin{linenomath}
\begin{equation}
\begin{aligned}
    f(\pt) &= ( a_1 + b_1/\pt ), \\ 
\end{aligned}
\end{equation}
\end{linenomath}
and
\begin{linenomath}
\begin{equation}
\begin{aligned}
    g(\eta) &=   a_2 + \begin{cases} 0 & \abs{\eta} < 1.6 \\ c_1 \, (\eta - 1.6)^2 & \eta > 1.6 \\ c_2 \, (\eta + 1.6)^2 & \eta < -1.6 \end{cases}.
\end{aligned}
\end{equation}
\end{linenomath}
The constants $a_1$, $a_2$, $b_1$, $c_1$, $c_2$, and $\rmue^0$
are extracted in a fit to the \rmue computed in data in bins 
of the $\eta$ and \pt  of the positive lepton in the DY-enriched sample.
The fit is performed iteratively, in which the \pt and $\eta$ dependencies, and the
normalizations, are extracted in separate steps. 
These values,  shown in Table~\ref{tab:rmue_values},  are obtained separately for each data taking year
and found to be statistically  consistent with those predicted from simulation. 
A greater dependency on $\eta$ is observed in the $\rmue$ factor in data collected in  2016 and 2017 that is caused by
a loss in the transparency of the ECAL endcap crystals, which affected trigger performance and was corrected in the 2018 data.
The transparency loss and its effects are stronger in data collected in 2017.
We assign systematic uncertainties of 5\% to the measured \rmue value
and an additional 5\% for each of its \pt and $\eta$ parametrizations 
that cover possible residual kinematic dependence.

\begin{table}[hbtp!]
 \renewcommand{\arraystretch}{1.3}
 \setlength{\belowcaptionskip}{6pt}
 \centering
 \topcaption{Summary of the \rmue parameters obtained by fitting the lepton \pt and $\eta$, in a DY-enriched control region, for different data taking years. 
 Only statistical uncertainties are shown.}
   \label{tab:rmue_values}
   \cmsTable{
  \begin{tabular}{lcccccc }
\hline 
Year   & $\rmue^0$ & $a_1$ & $b_1$ & $a_2$ & $c_1$ & $c_2$ \\    \hline
2016      &  $1.277\pm0.001$ &  $1.493\pm0.008$ &  $6.135\pm0.364$ &  $0.600\pm0.001$ &  $0.356\pm0.022$  &  $0.476\pm0.024$    \\
2017      &  $1.226\pm0.001$ &  $1.356\pm0.008$ &  $6.665\pm0.325$ &  $0.647\pm0.002$ &  $0.462\pm0.024$  &  $0.690\pm0.027$    \\
2018      &  $1.234\pm0.001$ &  $1.437\pm0.006$ &  $3.870\pm0.266$ &  $0.653\pm0.001$ &  $0.097\pm0.015$  &  $0.099\pm0.015$    \\ \hline 
  \end{tabular}
  }
\end{table}

Neglecting differences in trigger efficiencies, 
\rmue can be used to estimate the number of SF ($\EE$ and $\MM$) events from the observed number of DF events ($N_{\mathrm{DF}}$) 
in the DF CR as follows:
$N^{\text{est.}}_{\EE}=(1/2)(r_{\Pgm\mathrm{/}\Pe}(p_{\mathrm{T},\PGm},\eta_{\PGm} )^{-1}) N_{\mathrm{DF}}$ and
$N^{\text{est.}}_{\MM}=(1/2)r_{\Pgm\mathrm{/}\Pe}(p_{\mathrm{T},\Pe},\eta_{\Pe}) N_{\mathrm{DF}}$, leading to an estimated SF yield of
$N^{\text{est.}}_{\mathrm{SF}}=(1/2)(r_{\Pgm\mathrm{/}\Pe}(p_{\mathrm{T},\Pe},\eta_{\Pe})+r_{\Pgm\mathrm{/}\Pe}(p_{\mathrm{T},\PGm},\eta_{\PGm})^{-1})N_{\mathrm{DF}}$. 

Another correction factor, \RT, 
defined as $\sqrt{\smash[b]{\epsilon^{\mathrm{T}}_{\PGm\PGm}\epsilon^{\mathrm{T}}_{\Pe\Pe}}}/\epsilon^{\mathrm{T}}_{\Pe\PGm}$, where $\epsilon^{\mathrm{T}}_{\PGm\PGm}$, $\epsilon^{\mathrm{T}}_{\Pe\Pe}$ and $\epsilon^{\mathrm{T}}_{\Pe\PGm}$ are the trigger efficiency of the di-muon, di-electron and muon-electron triggers respectively,
is then used to account for differences between SF and DF dilepton trigger efficiencies $\epsilon^{\mathrm{T}}_{\ell\ell'}$.
These efficiencies are measured in data and found to be 85--95\%, depending on the lepton flavor and the data taking period.
The resulting \RT coefficient is measured to be 1.03--1.05, with an uncertainty of 4--5\%.

The transfer factor \Rsfof, used to predict SF events from DF ones, is finally defined as:
\begin{linenomath}
\begin{equation}
\label{eq:rsfof}
\Rsfof = (1/2)(r_{\Pgm\mathrm{/}\Pe}(\PGm)+r_{\Pgm{/}\Pe}(\Pe)^{-1}) \RT. 
\end{equation}
\end{linenomath}

\begin{figure}[htb!]
\centering
\includegraphics[width=0.49\textwidth]{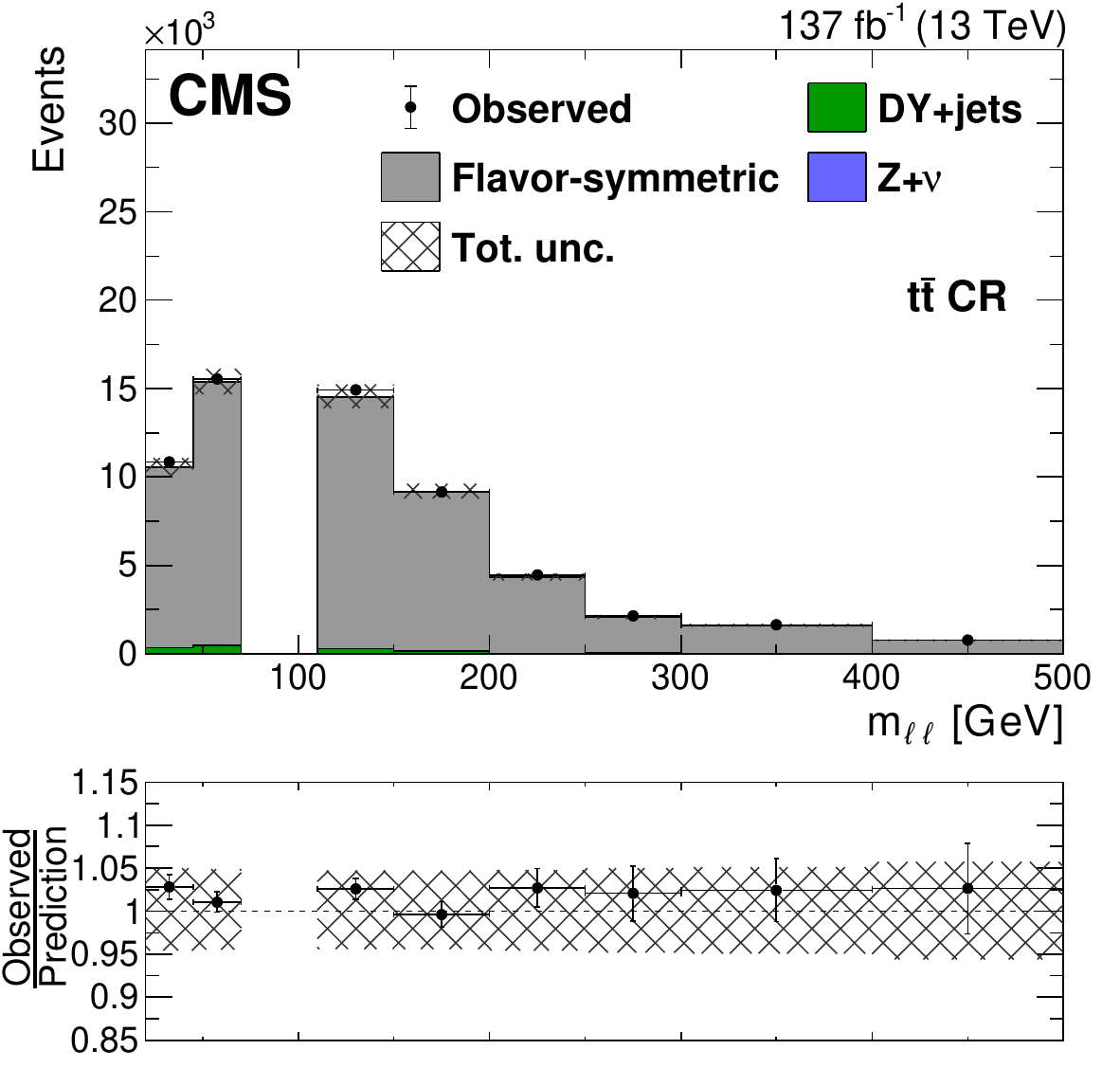}
\includegraphics[width=0.49\textwidth]{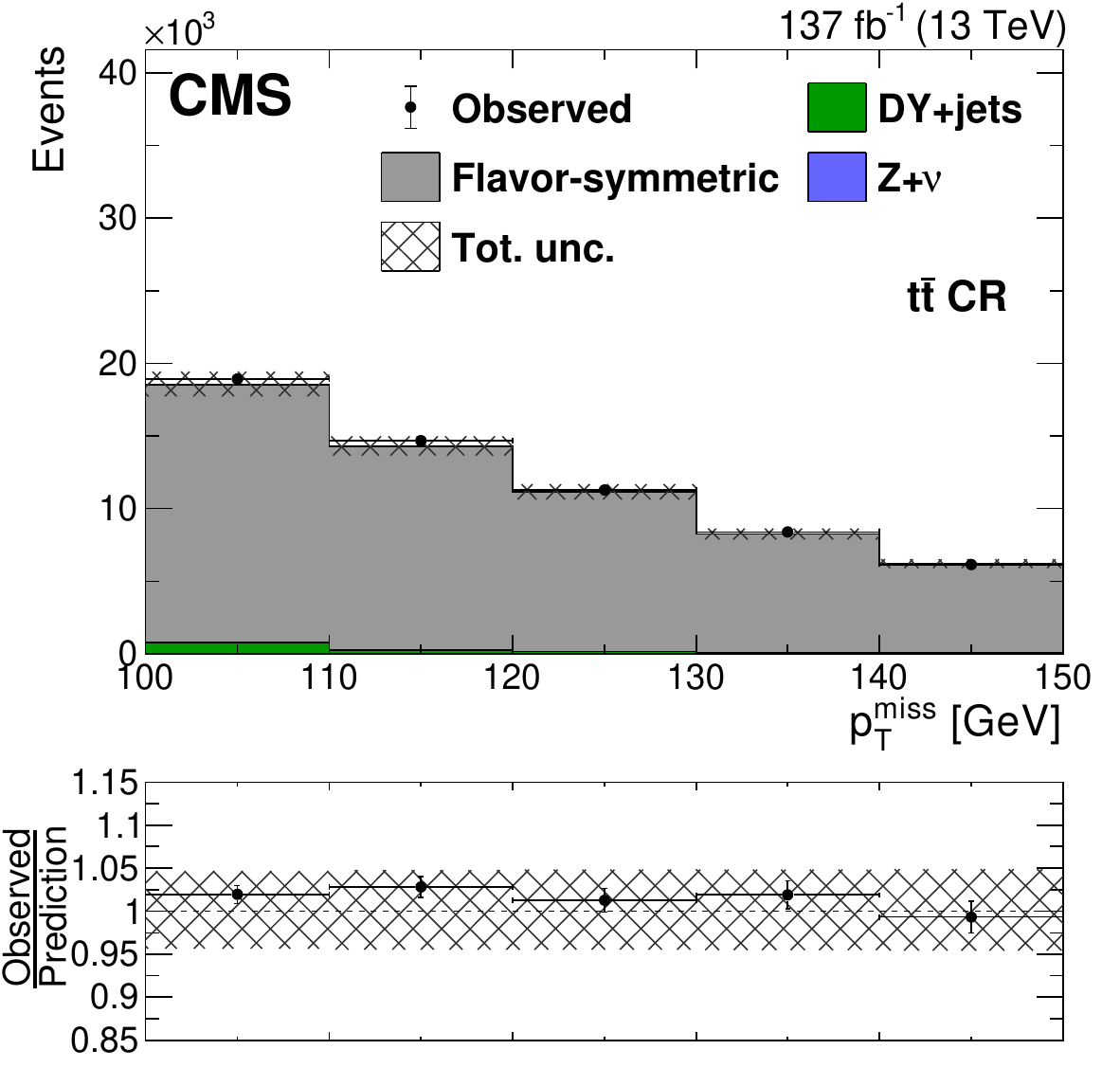}

\includegraphics[width=0.49\textwidth]{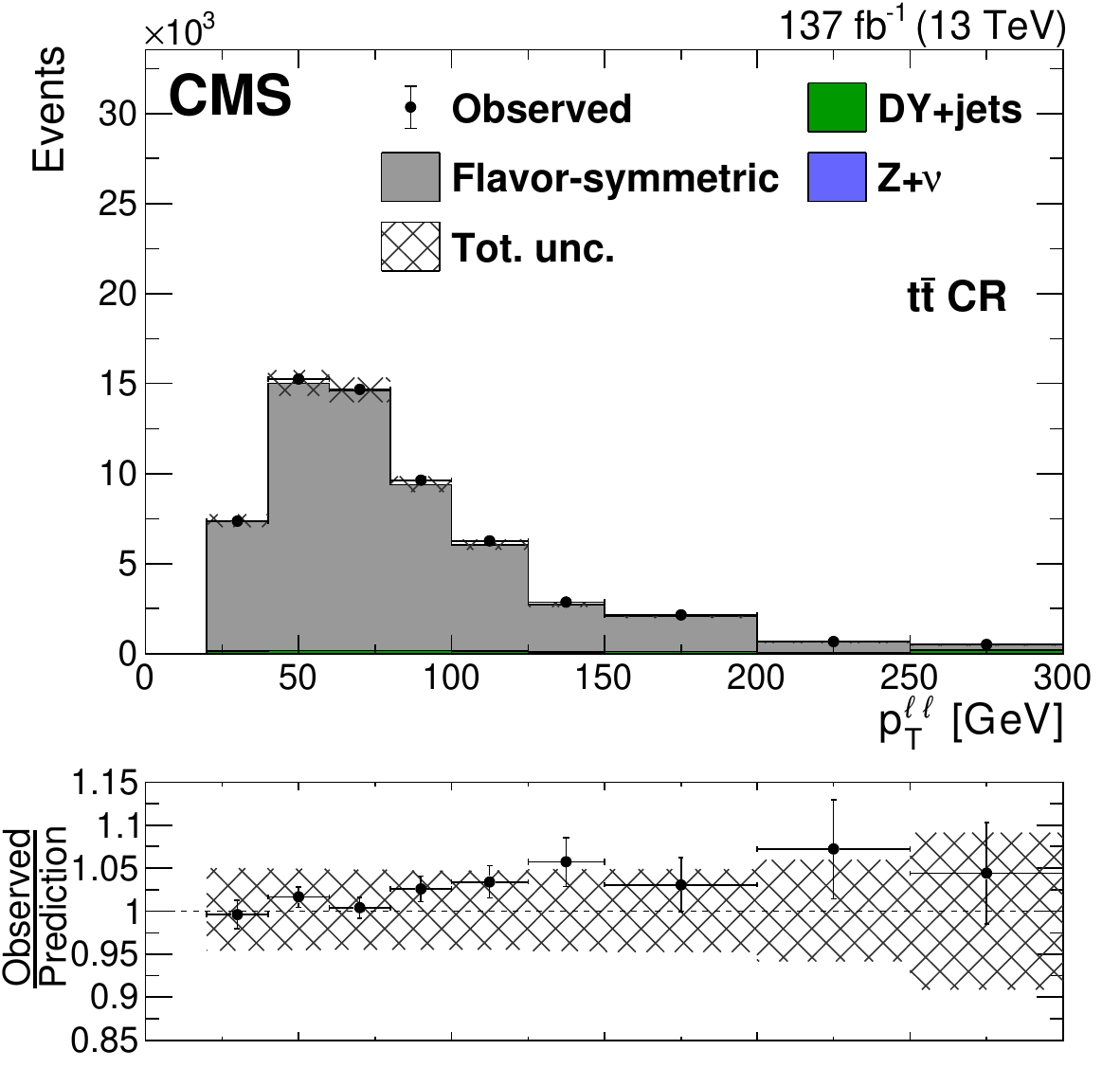}
\caption{ Distributions in \mll (upper left),  \ptmiss (upper right), and  $\pt^{\ell\ell}$ (lower) in a \ttbar-enriched CR in data. 
    The  flavor-symmetric background prediction obtained from data, as discussed in the main text, (gray solid histogram) is compared to data (black markers). Other backgrounds are estimated directly from simulation (green and blue solid histograms). 
    The uncertainty band includes both the statistical and systematic contributions to the prediction.
    The last bin includes overflow events. 
}
  \label{fig:rsfof_validation}
\end{figure}

The background estimation method is validated in data in a CR enriched
in flavor-symmetric \ttbar events.
This region is defined by requiring an SF lepton pair, exactly two jets, and $100<\ptmiss<150\GeV$. 
Events with $70<\mll<110\GeV$ are rejected to reduce the contribution from DY events. 
Figure~\ref{fig:rsfof_validation} compares the prediction
from a DF selection in SF data in this
region, as a function of different kinematic variables.
An agreement within the uncertainties is observed,
thus validating the background estimation method.  

The statistical uncertainty arising from the limited size of the DF control sample represents the dominant contribution to the total uncertainty in the flavor-symmetric background prediction.
For the estimation of this background in the on-\PZ SRs, where $86<\mll<96\GeV$, 
the \mll requirement in the DF control sample is relaxed to $\mll>20\GeV$, and an additional multiplicative factor, 
$\kappa = N_{\mathrm{DF}}(86 < \mll < 96 \GeV)/N_{\mathrm{DF}}(\mll > 20 \GeV)$, is used to account for the different \mll selection in CRs and SRs. 
This factor  is determined from dedicated DF CRs in data, 
defined by relaxing or merging a subset of selection requirements described in Section~\ref{subsec:signalregions}. The regions of interest (SRA, SRB
and SRC strong-production SRs,  and the $\PH\PZ$ and resolved $\vz$  SRs) are defined in Table~\ref{tab:selections_signalRegions}. The boosted $\vz$ SR is also considered, relaxing the veto of additional jets. In these regions, $\kappa$ is measured to be in the range 0.045--0.067.
We also determine $\kappa$ as a function of several kinematic variables
to assess the possible dependencies. Based on these measurements, we assign a
systematic uncertainty of 20\% to the value of $\kappa$ to cover such effects.

\subsection{Drell--Yan+jets backgrounds}
\label{sub:dybkg}

The contribution from \dyjets events to the SRs mainly arises from mismeasurements of momenta of reconstructed
objects affecting \ptvecmiss. 
In regions where jets in the final state are required,
instrumental \ptmiss arises mainly from jet energy mismeasurement,
and the \ptmiss ``templates'' method~\cite{OSpaperCMS2011,OSpaperCMS7TeV,CMS:edge,CMS:Zedge2015} is used to estimate the resulting background contribution. 
In the slepton SRs, since only jets with low \pt are present, we use a different method 
exploiting a CR enriched in \dyjets events.

The \ptmiss ``templates'' method relies on the fact that instrumental \ptmiss in \dyjets events is caused by
limited detector resolution in measuring the \pt of the jets recoiling against the leptonically decaying \PZ boson. 
Since the \pt resolution of
leptons and photons is much better than that of jets, 
the \ptmiss distribution in \dyjets events
can be estimated directly from \gammajets data.

The \gammajets events are selected with
jet requirements identical as those used in defining the SRs in
Section~\ref{subsec:signalregions}.
We assume that the \gammajets events
are not affected by potential contamination from any of the BSM physics considered in this search.

The \mttwo variable used to select events in several SRs requires the presence of two visible
objects and therefore cannot be defined in the \gammajets sample. 
Instead, its behavior is emulated by mimicking the decay of the photon into
two leptons.
The decay is modeled assuming that the leptons arise from a particle that
has the mass of a \PZ boson and the momentum of the selected photon, with
the angular distributions in the decay as 
expected at LO in perturbation theory. The simulated leptons are used to calculate the \mttwol variable in the \gammajets data sample. 

Events with genuine \ptmiss may, in fact, be present in the \gammajets sample,
originating from EW processes such as \PW{}\PGg  + jets production,
where the \PW boson decays to $\Pell\PGn$. However, such contributions can be suppressed by rejecting events that contain additional leptons.
The residual EW contamination in the \gammajets sample, which is
larger at large \ptmiss, is subtracted using simulation.

The photon \pt distribution in \gammajets events is expected to differ
from that 
of the \PZ boson in \dyjets, 
mainly because of the different boson masses.
Thus, simulation is used in each SR to obtain a set of weights that match the
photon \pt distribution to the expected \PZ boson \pt distribution.
These weights are then used to reweigh the \ptmiss templates in \gjets data in the SRs. After this correction, the corrected \ptmiss template in each SR is
normalized based on
the observed yield in dilepton data in the range $50<\ptmiss<100\GeV$, 
where \dyjets events dominate the data sample.
We note that to account for potential contamination from BSM physics 
the $50<\ptmiss<100\GeV$ bin in each SR is included in the signal extraction fit described in Section~\ref{sec:interpretation}.

Several sources of uncertainty are considered for the \dyjets background prediction: the 
statistical uncertainty arising from the limited size of the \gammajets sample in each \ptmiss bin,
the systematic uncertainty in the EW subtraction, and the 
statistical uncertainty in the template normalization arising from the dilepton data yield in the range $50<\ptmiss<100\GeV$.
An additional systematic uncertainty is assessed through
a closure test of the method in simulation,
where the \ptmiss distribution in simulated \dyjets events
is compared to the distribution obtained by applying the
background prediction method to a \gjets simulated sample.
In each \ptmiss bin, we assign an uncertainty
equal to the largest of the differences between the predicted and simulated yields,
and the statistical uncertainty reflecting the size of the samples.
The resulting uncertainty ranges between 20 and 100\% across the search bins 
with the largest values obtained in bins affected by the limited number of simulated events.

\begin{figure}
  \centering
  \includegraphics[width=0.4\linewidth]{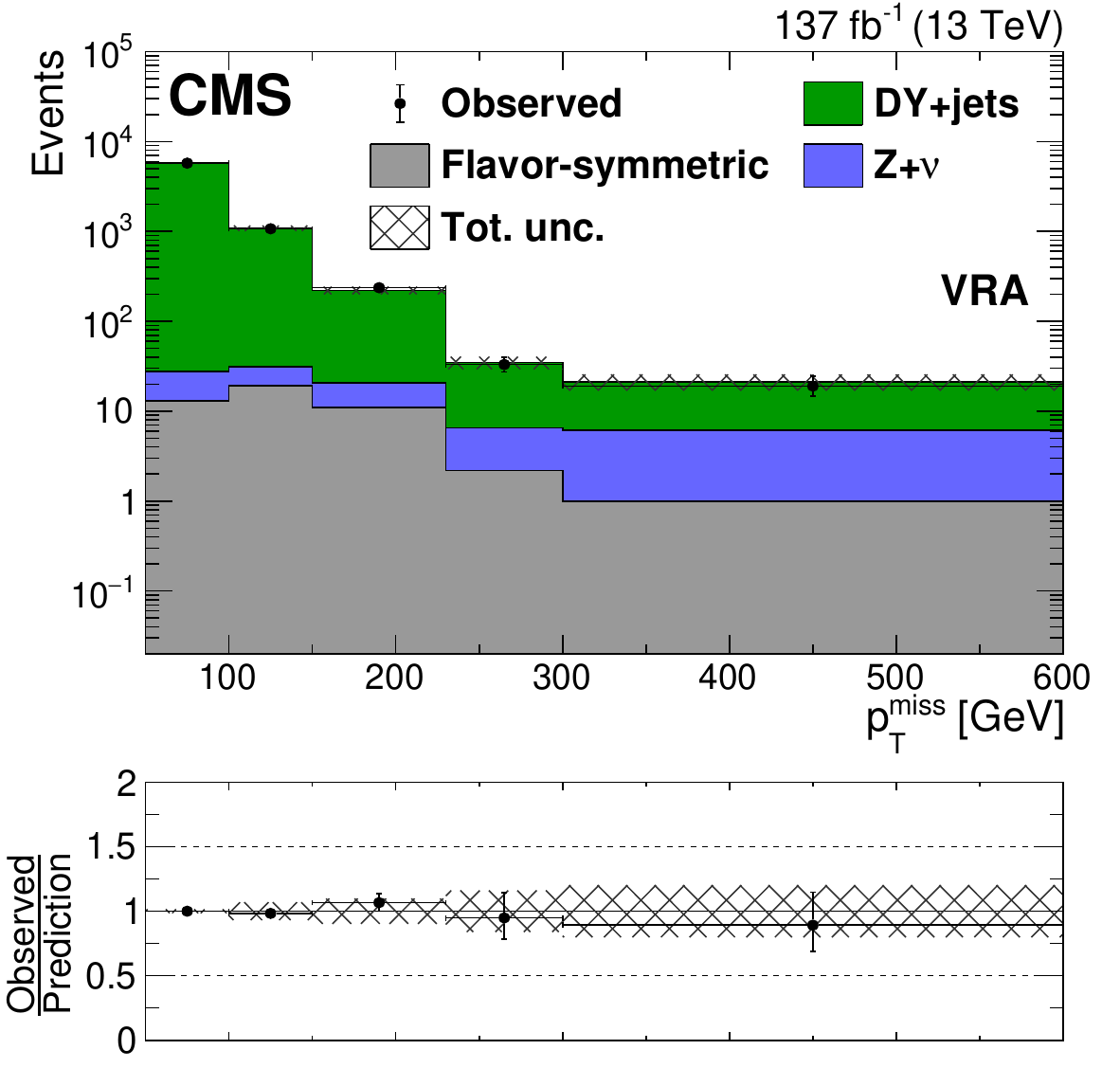}
  \includegraphics[width=0.4\linewidth]{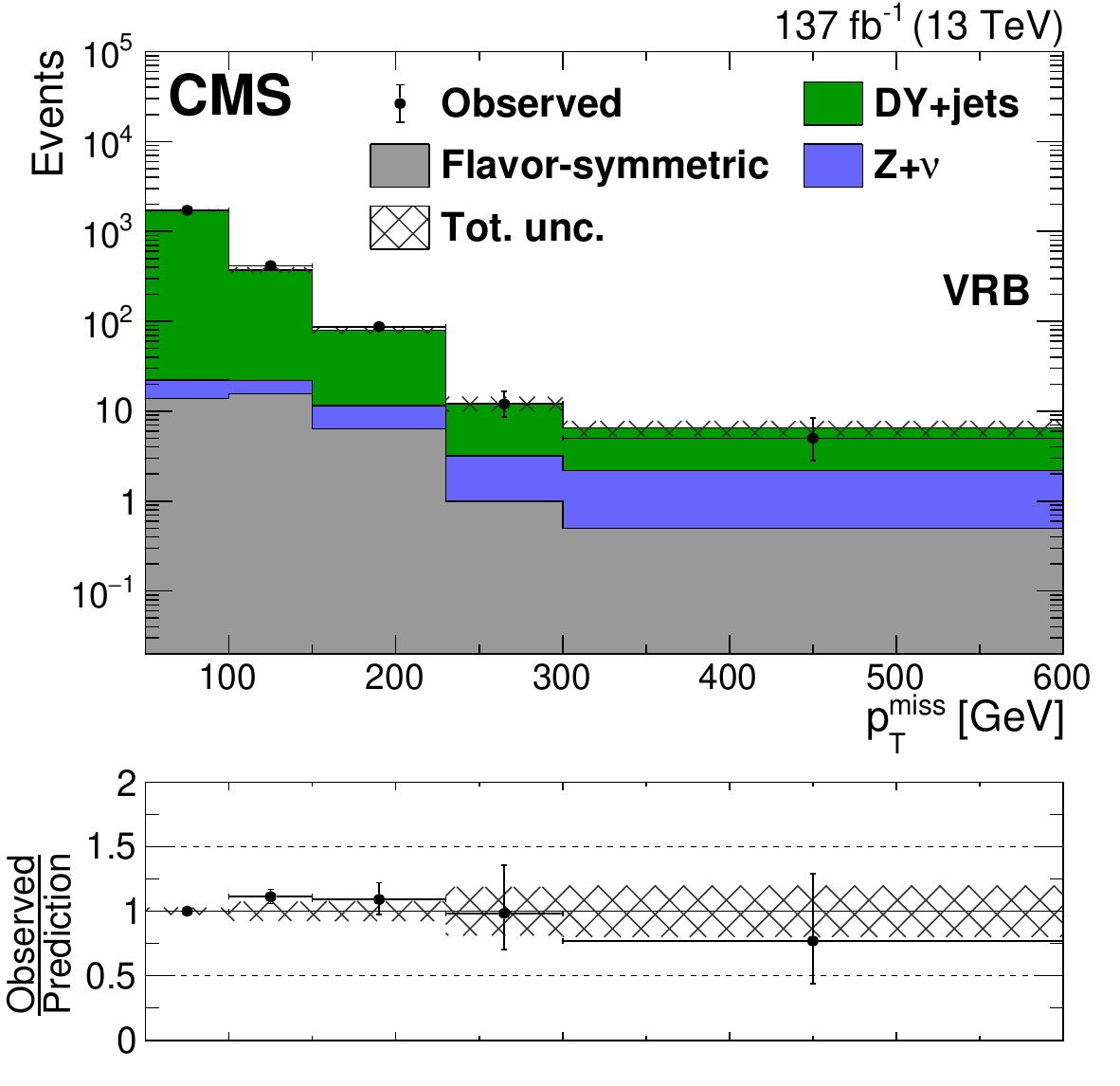}
                           
  \includegraphics[width=0.4\linewidth]{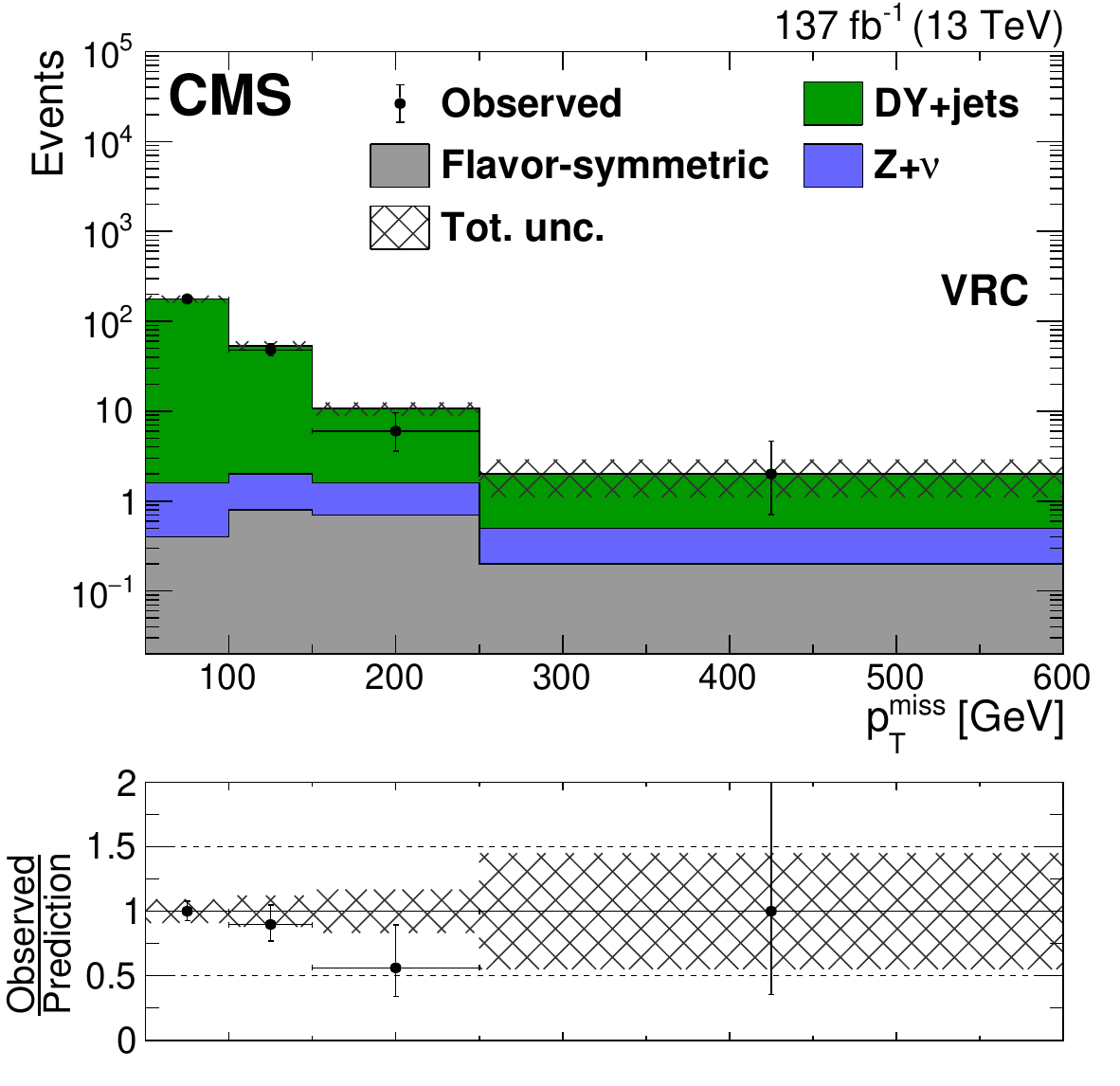}
  \includegraphics[width=0.4\linewidth]{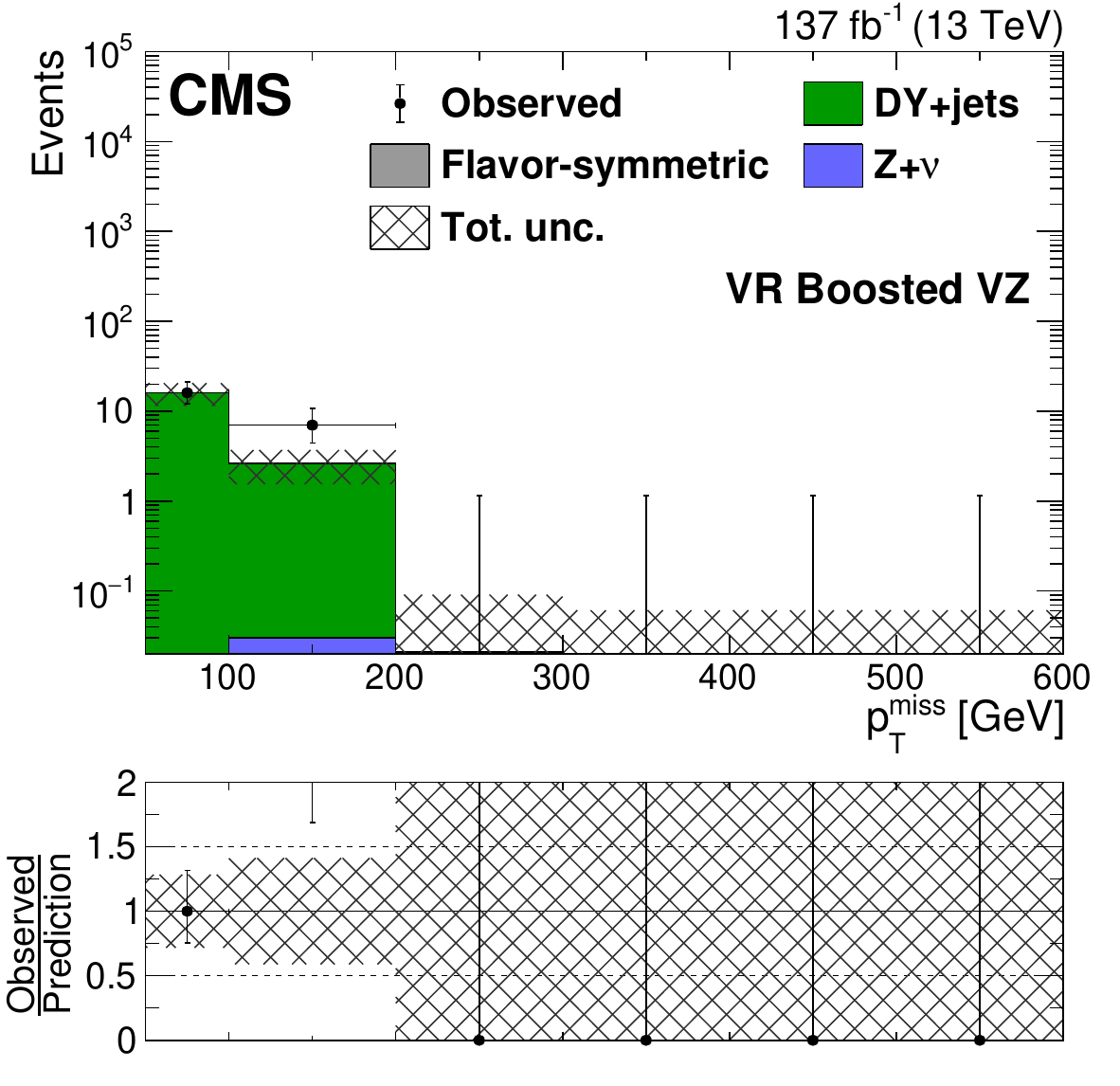}
                           
  \includegraphics[width=0.4\linewidth]{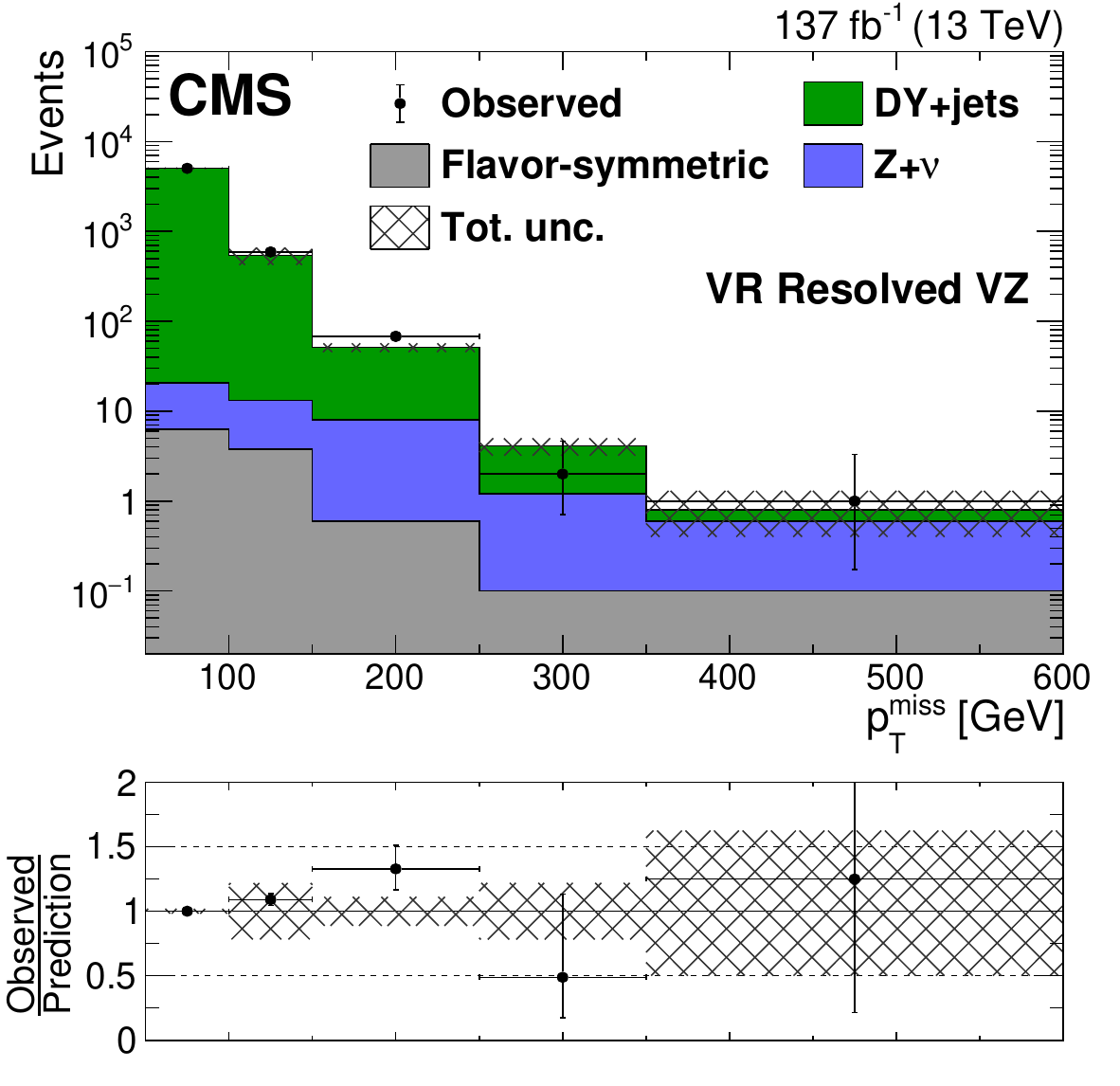}
  \includegraphics[width=0.4\linewidth]{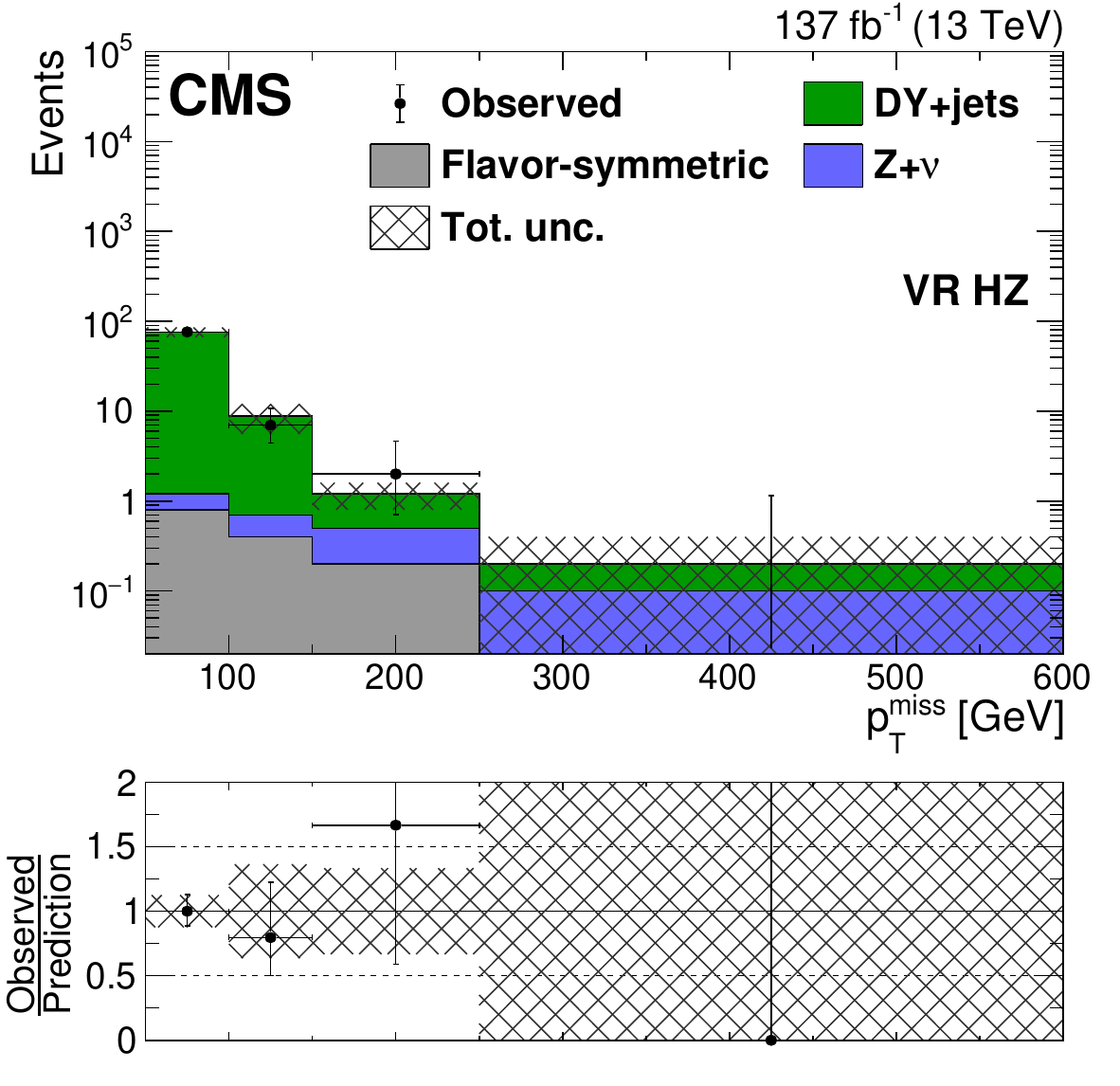}
  \caption{\label{fig:templates_validation} The \ptmiss distribution observed in data (black markers) is compared to the
    background prediction (solid histograms) in the on-\PZ VRs. 
    Comparison in the
    strong on-\PZ VRs associated to (upper left) SRA, (upper right) SRB , and (middle left) SRC. 
    Comparison in the
    EW on-\PZ VRs: (middle right) boosted \vz, (lower left) resolved \vz, and (lower right) $\PH\PZ$.
    The uncertainty band includes both the systematic and statistical components of the uncertainty in the prediction. The last bin includes overflow events. 
  }
\end{figure}

The validity of the method is further tested in data CRs enriched in events containing instrumental \ptmiss.
These samples are defined by inverting the $\Delta\phi(\vec{p}^{\,\mathrm{j}_{1,2}}_\mathrm{T},\ptvecmiss)$ selection
(or, in the boosted \vz region, $\Delta\phi(\text{V boson candidate},\ptvecmiss)$).
In addition, the \PQb-tagged jet multiplicity categorization is removed from the on-\PZ strong-production regions 
yielding a total of six validation regions (VRs) with the same \ptmiss binning as used in the corresponding SRs.
The observed \ptmiss distribution  is compared to the prediction in the VRs in Fig.~\ref{fig:templates_validation}
showing agreement within the uncertainties.

The method described above is also used to predict the \dyjets background in the edge SRs, 
where events with $86<\mll<96\GeV$ are rejected, and therefore the contribution from \dyjets events is expected to be small.
In this case, the prediction is obtained from a CR with inverted \mll selection, 
by means of a transfer factor \rinout defined as the ratio of the \dyjets yield in a given \mll bin over the yield in the range $86<\mll<96\GeV$.
The \rinout ratio is measured in a data control sample enriched in \dyjets events, obtained by requiring at least two jets, with $\ptmiss<50\GeV$ and $\mttwol>80\GeV$, 
after subtracting the flavor-symmetric contribution estimated as described in Section~\ref{sub:fsbkg}. 
The \rinout value is measured to be in the range
0.003--0.06, depending on the \mll bin. 
We assign a systematic uncertainty in \rinout
to cover its possible dependence on \ptmiss and \njets, 
of 50\,(100)\% in \mll bins below (above) 150\GeV.

In the slepton SRs, the \dyjets background is estimated in each \ptmiss bin 
using a CR in data enriched with \dyjets events, 
obtained by applying the same selection criteria as used in the SRs, but inverting the selection on \mll ($65<\mll<120\GeV$).
The prediction is then obtained by means of a transfer factor, \rinout, which is measured in data,
after relaxing the \ptmiss and \njets selections applied in the SRs.
The \rinout value is measured to be 0.07, with a 50\%
uncertainty obtained from
a closure test performed using simulated \dyjets events.
To account for possible contamination from BSM physics in the $65<\mll<120\GeV$ region,
that region is included in the signal extraction fit described in Section~\ref{sec:interpretation}.

The systematic uncertainties associated with the flavor-symmetric  and  \dyjets background
estimation are summarized in Table~\ref{tab:data_driven_sys}.

\begin{table}
  \renewcommand*{\arraystretch}{1.1}
  \topcaption{Summary of the uncertainties in background estimations performed on data.}
  \centering
  \label{tab:data_driven_sys}

  \begin{tabular}{ll}
  \hline
    Source & Size \\ \hline
    \multicolumn{2}{c}{\textit{Flavor-symmetric backgrounds}} \\
    \rmue residual dependencies & 5\% flat   \\
    & 5\% \pt-dependent \\
    & 5\% $\eta$-dependent \\ 
    \RT uncertainty & 4--5\% \\
    Statistical uncertainty in DF sideband & \checkmark \\
    $\kappa$ uncertainty (on-\PZ SRs only) & 20\%\\ [\cmsTabSkip]
    \multicolumn{2}{c}{\textit{\ptmiss templates}} \\
    Closure in simulations & 20--100\%\\
    Statistical uncertainty in \gjets sample & \checkmark \\
    Statistical uncertainty in normalization bin & \checkmark \\
    EW subtraction &  30\% of EW yield  \\
    & in \gjets sample \\ 
    \rinout (edge SRs only) & 50--100\% \\ [\cmsTabSkip]
    \multicolumn{2}{c}{\textit{\dyjets in slepton SRs}} \\ 
    \rinout (slepton SRs only) & 50\% \\\hline
  \end{tabular}
\end{table}

\subsection{Backgrounds containing \texorpdfstring{\PZ}{Z} bosons and genuine \texorpdfstring{\ptmiss}{missing transverse momentum}}
\label{sub:zmetbkg}

Backgrounds from events with $\PZ/\PGg^*$ bosons and genuine \ptmiss such
as \WZ, \ZZ, and \ttz can be important in SRs of
large \ptmiss, and are estimated directly from simulation.
Dedicated data CRs of trileptons and two pairs of OSSF
leptons are used to determine the overall normalization
and to check the modeling of such events in simulation.
Systematic uncertainties as large as 50\% are assessed for each process
to cover differences between data and simulation.
In predicting the \ZZ yield we also assign an additional
uncertainty given by 
the difference between the nominal NLO simulation and the
NNLO prediction achieved applying the k-factor as
described in Section~\ref{sec:samplesObjects}.
Finally, we include statistical
uncertainties associated with the limited size of the simulated
event samples, and systematic uncertainties arising from
the modeling of pileup, 
lepton reconstruction and isolation efficiencies,
\PQb tagging efficiency, 
and jet energy scale (JES),
as well as the choice of the renormalization ($\mu_{\mathrm{R}}$)
and factorization ($\mu_{\mathrm{F}}$) scales used in the event generation.
The uncertainties are summarized in Table~\ref{tab:znusysts},
together with their typical size in the SRs.

For each data sample corresponding to different periods of data taking, 
uncertainties in the trigger, \PQb tagging and lepton efficiencies are treated as correlated across the SRs. 
Uncertainties in the ISR modeling, fast simulation \ptmiss distributions,
JES, and trigger, \PQb tagging, and lepton efficiencies 
are treated as correlated also across the data samples.
Uncertainties in the integrated luminosity have a correlated and uncorrelated components.
The remaining uncertainties are taken as uncorrelated.

\begin{table}[htb]
\renewcommand*{\arraystretch}{1.1}
\centering
\topcaption{\label{tab:znusysts}
Summary of systematic uncertainties in the predicted \znu background yields, 
together with their typical sizes across the SRs.}

\begin{tabular}{l c}
\hline
Source of uncertainty                  & Uncertainty (\%)   \\ \hline 
Integrated luminosity                  & 1.8                \\
Limited size of simulated samples      & 1--15              \\
Simulation modeling in data CRs     & 30--50             \\
Trigger efficiency                     & 3                  \\
NNLO/NLO $\kappa$-factor (for \ZZ)          & 10--30             \\
Lepton efficiency                      & 5                  \\
\PQb tagging efficiency                    & 0--5               \\
JES                       & 0--5               \\
Pileup modeling                        & 1--2               \\
$\mu_{\mathrm{R}}$ and $\mu_{\mathrm{F}}$ dependence & 1--3  \\ \hline
\end{tabular}
\end{table}

\section{Edge fit to the dilepton invariant mass distribution}
\label{sec:kinfit}

We perform a simultaneous unbinned maximum likelihood fit as a function of \mll in $\EE$, $\MM$, and $\EM$ data 
to search for a kinematic edge. 
The fit is performed in the ``edge fit'' SR defined in Section~\ref{subsec:signalregions}. 
The functional forms used to model the signal and the two main SM background components (flavor-symmetric background 
and backgrounds arising from other SM processes containing a \PZ boson) 
are described below.

The flavor-symmetric background component is modeled
using the CB function:
\begin{linenomath}
\begin{equation}
\begin{aligned}
\mathcal{P}_{\mathrm{CB}}(\mll) = \begin{cases}
\exp\left[-\frac{(\mll-\mu_{\mathrm{CB}})^2}{2\widthCB^2}\right] &\text{ if } \frac{\mll-\mu_{\mathrm{CB}}}{\widthCB}\leq\alpha \\
A (B+\frac{\mll-\mu_{\mathrm{CB}}}{\widthCB})^{-n} &\text{ if } \frac{\mll-\mu_{\mathrm{CB}}}{\widthCB}>\alpha \\
\end{cases},
\end{aligned}
\end{equation}
\end{linenomath}
where
\begin{linenomath}
\begin{equation}
A = \left(\frac{n}{\abs{\alpha}}\right)^{n} \exp\left(-\frac{\abs{\alpha}^2}{2}\right) \quad \text{and}\quad B = \frac{n}{\abs{\alpha}}-\abs{\alpha}.
\end{equation}
\end{linenomath}
This model has five free parameters: the overall normalization, the mean $\mu_{\mathrm{CB}}$ and the full width $\widthCB$ at half maximum of the Gaussian core component, 
the transition point $\alpha$ between the Gaussian core and the power-law tail, and the power-law parameter $n$.

Backgrounds containing a leptonically decaying \PZ boson ($\ZplusX$) are modeled through a sum of an exponential function, which describes the rise at small mass, and a Breit--Wigner
function with the mean and the width set to the nominal \PZ boson values~\cite{PDG2020}
convolved with a double-sided CB function, $\mathcal{P}_{\mathrm{DSCB}}(\mll)$ to
account for the experimental resolution:
\begin{linenomath}
\begin{equation}
\begin{aligned}
\mathcal{P}_{\mathrm{DSCB}}(\mll) \propto \begin{cases} A_{1} (B_{1}-\frac{\mll-\mu_{\mathrm{DSCB}}}{\Gamma_{\mathrm{DSCB}}})^{-n_{1}} &\text{ if } \frac{\mll-\mu_{\mathrm{DSCB}}}{\Gamma_{\mathrm{DSCB}}}\leq-\alpha_{1} \\
\exp\left[-\frac{(\mll-\mu_{\mathrm{DSCB}})^2}{2\Gamma_{\mathrm{DSCB}}^2}\right] &\text{ if } -\alpha_{1}<\frac{\mll-\mu_{\mathrm{DSCB}}}{\Gamma_{\mathrm{DSCB}}}\leq\alpha_{2} \\
A_{2} (B_{2}+\frac{\mll-\mu_{\mathrm{DSCB}}}{\Gamma_{\mathrm{DSCB}}})^{-n_{2}} &\text{ if } \frac{\mll-\mu_{\mathrm{DSCB}}}{\Gamma_{\mathrm{DSCB}}}>\alpha_{2} \\
\end{cases},
\end{aligned}
\end{equation}
\end{linenomath}
where $\mu_{\mathrm{DSCB}}$ and $\Gamma_{\mathrm{DSCB}}$ are the mean and width, respectively, of the CB function,
and $\alpha_{1}$ and $\alpha_{2}$ are the transition points.
The model for the \ZplusX background line shape is thus:
\begin{linenomath}
\begin{equation}
\mathcal{P}_{\ZplusX} (\mll) = (1-C) \int \mathcal{P}_{\mathrm{DSCB}}(\mll)\mathcal{P}_{\mathrm{BW}}(\mll-m') \rd{}m' + C\mathcal{P}_{\mathrm{exp}} (\mll),
\end{equation}
\end{linenomath}
where $\mathcal{P}_{\mathrm{BW}}$  and $\mathcal{P}_{\mathrm{exp}}$ are the Breit--Wigner and exponential functions, respectively. The complete \dyjets
background model has therefore nine free parameters each for the $\EE$ and $\MM$ final states.

The signal component is described by a triangular form, inspired by the slepton edge models~\cite{Hinchliffe:1996iu}, convolved with a Gaussian function to account for the experimental resolution:
\begin{linenomath}
\begin{equation}
 {\mathcal{P}}_{\mathrm{S}}(\mll) \propto \frac{1}{\sqrt{2\pi}\Gamma_{\Pell\Pell}} \int_{0}^{\mll^{\text{edge}}} y \exp\left[ -\frac{(\mll-y)^2}{2\Gamma_{\Pell\Pell}^{2}}\right]\, \rd{}y.
\end{equation}
\end{linenomath}
The signal model has two free parameters: the fitted signal yield and the position of the kinematic endpoint, $\mll^{\text{edge}}$, as the experimental resolution $\Gamma_{\Pell\Pell}$ for each leptonic final state is obtained from the CB function of the \dyjets model.

In an initial step, a fit to data is performed in a \dyjets-enriched CR with at least two jets, $\mttwol>80\GeV$, and $\ptmiss<50\GeV$,
separately for $\EE$ and $\MM$ events, to determine the parameters for backgrounds containing a \PZ boson.
The final fit is then performed simultaneously to the invariant mass distributions in the $\EE$, $\MM$, and $\EM$ data samples. 
The model for the flavor-symmetric background is varied consistently in the SF and DF samples.
The relative normalization of SF and DF events is given by the \Rsfof factor, which is treated
as a nuisance parameter,
constrained by a Gaussian prior with the mean value and standard deviation (s.d.), as determined in
Section~\ref{sub:fsbkg}. 
In total, the final fit has ten parameters: a normalization parameter for each of the three
fit components,
four parameters for the distribution of the flavor-symmetric background, \Rsfof, the relative fraction of dielectron and dimuon events in the flavor-symmetric prediction, 
and the position of the signal edge. Out of these, only \Rsfof is constrained, while the others are treated as free parameters of the fit.

\section{Results}
\label{sec:results}

The observed yields in each SR are compared to the SM predictions for the on-\PZ,
edge, and slepton SRs. 
In the search for an edge, a fit is also performed to the \mll distribution in data to find a kinematic edge in the \mll spectrum as discussed in Section \ref{sec:kinfit}. 

\subsection{Results for the \texorpdfstring{on-\PZ}{on-Z} samples}
\label{sub:onZResults}

\begin{figure}[htbp!]
\centering
\includegraphics[width=0.42\linewidth]{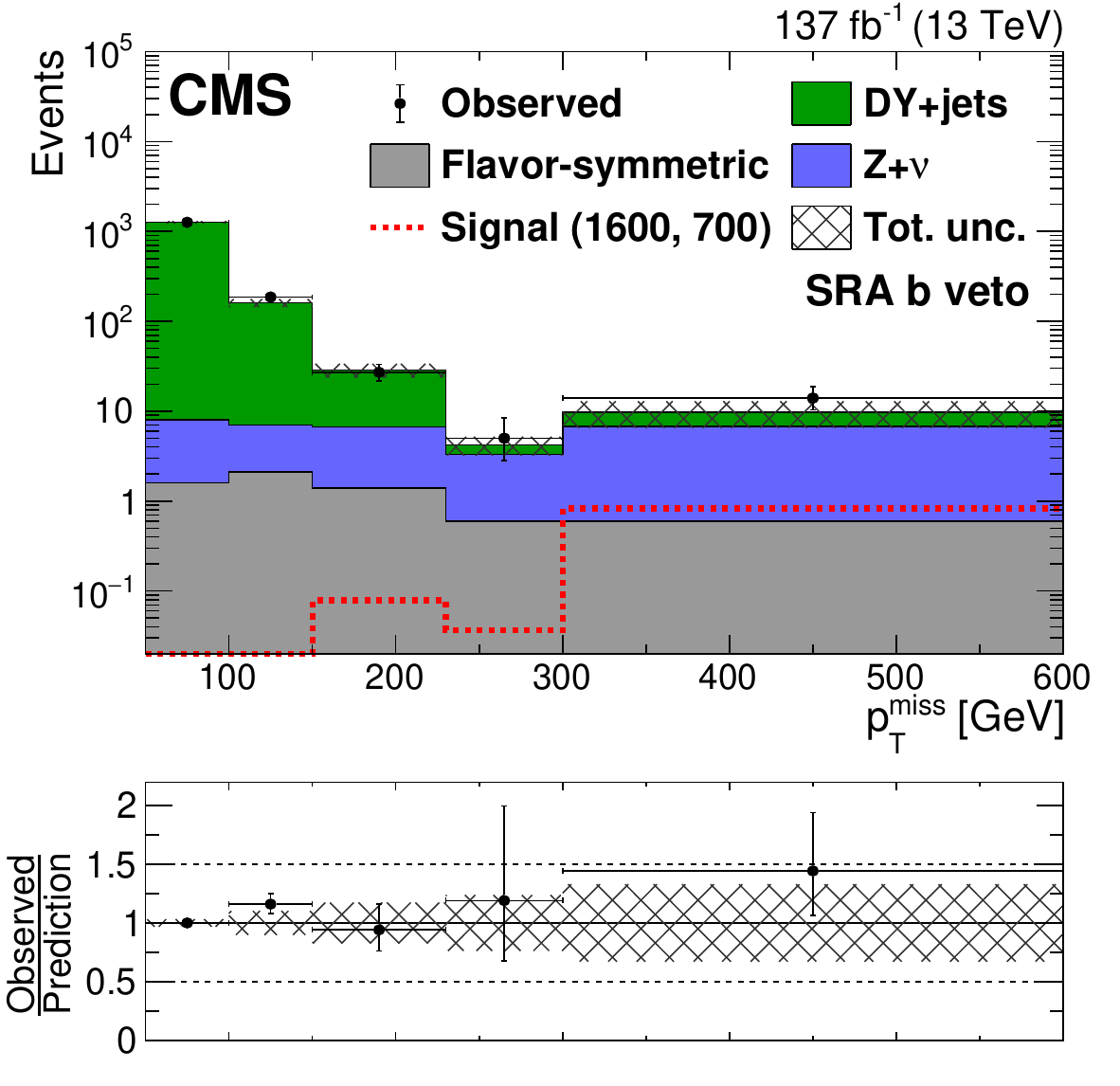}
\includegraphics[width=0.42\linewidth]{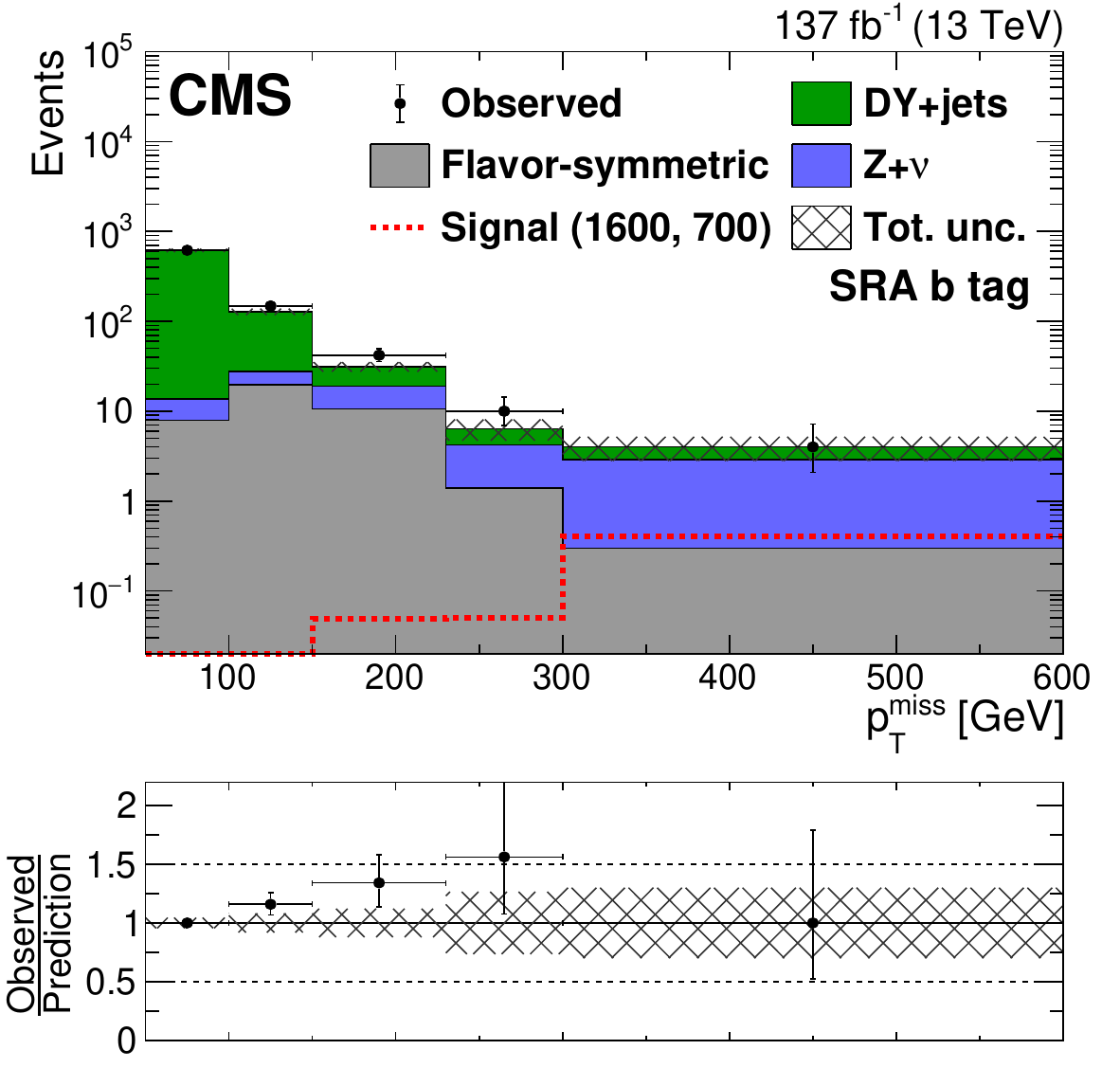}\\
\includegraphics[width=0.42\linewidth]{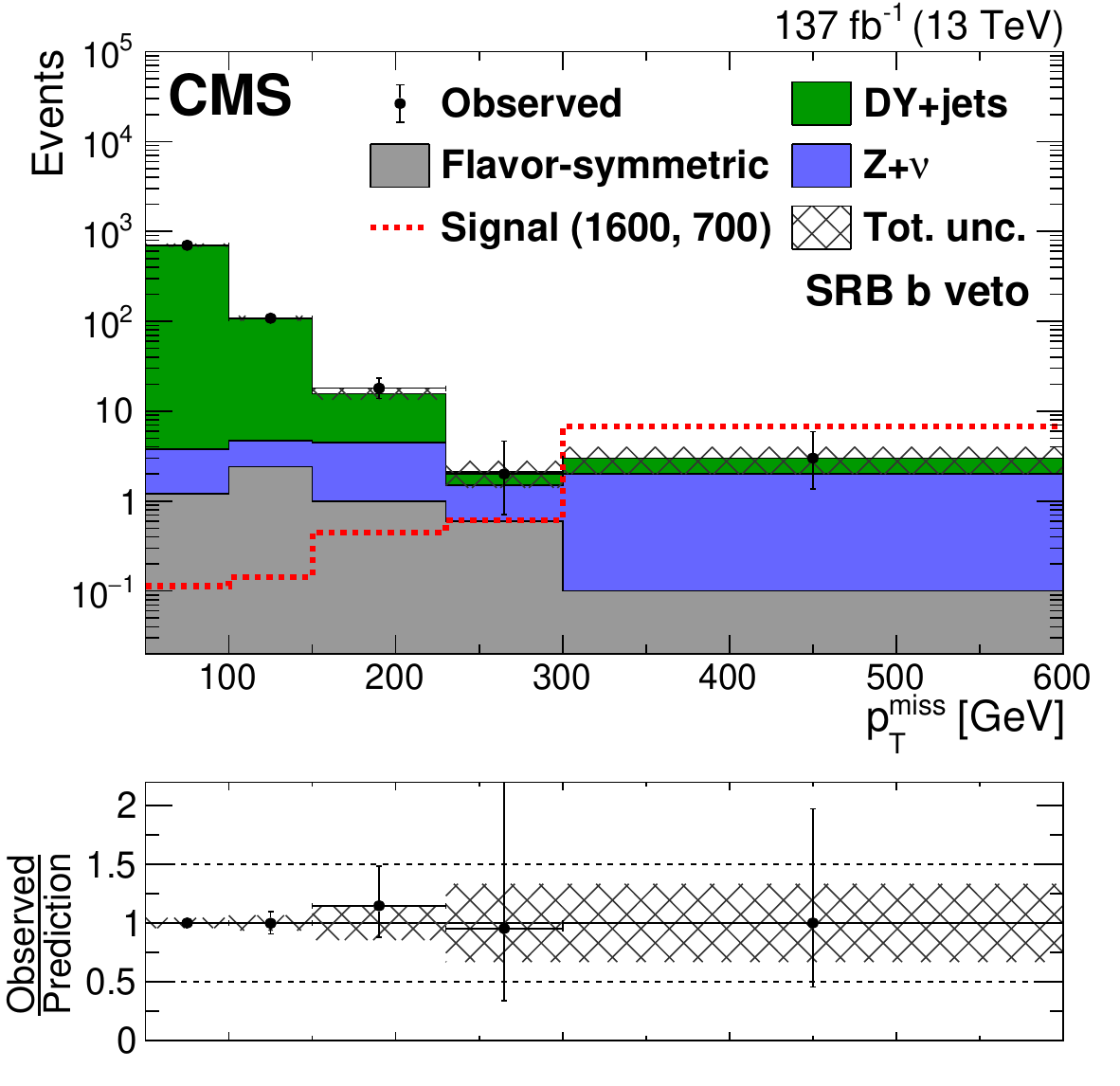}
\includegraphics[width=0.42\linewidth]{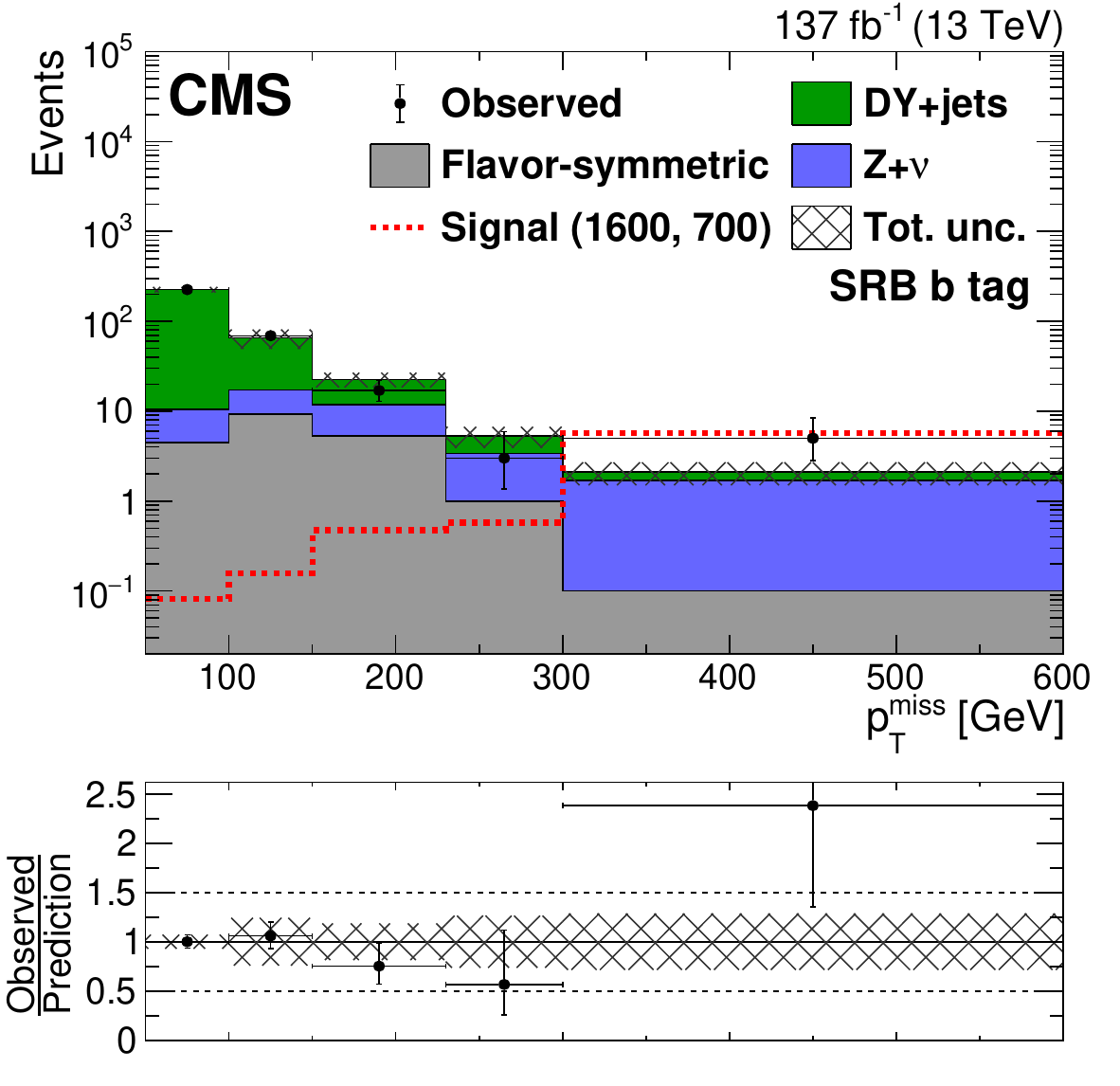}\\
\includegraphics[width=0.42\linewidth]{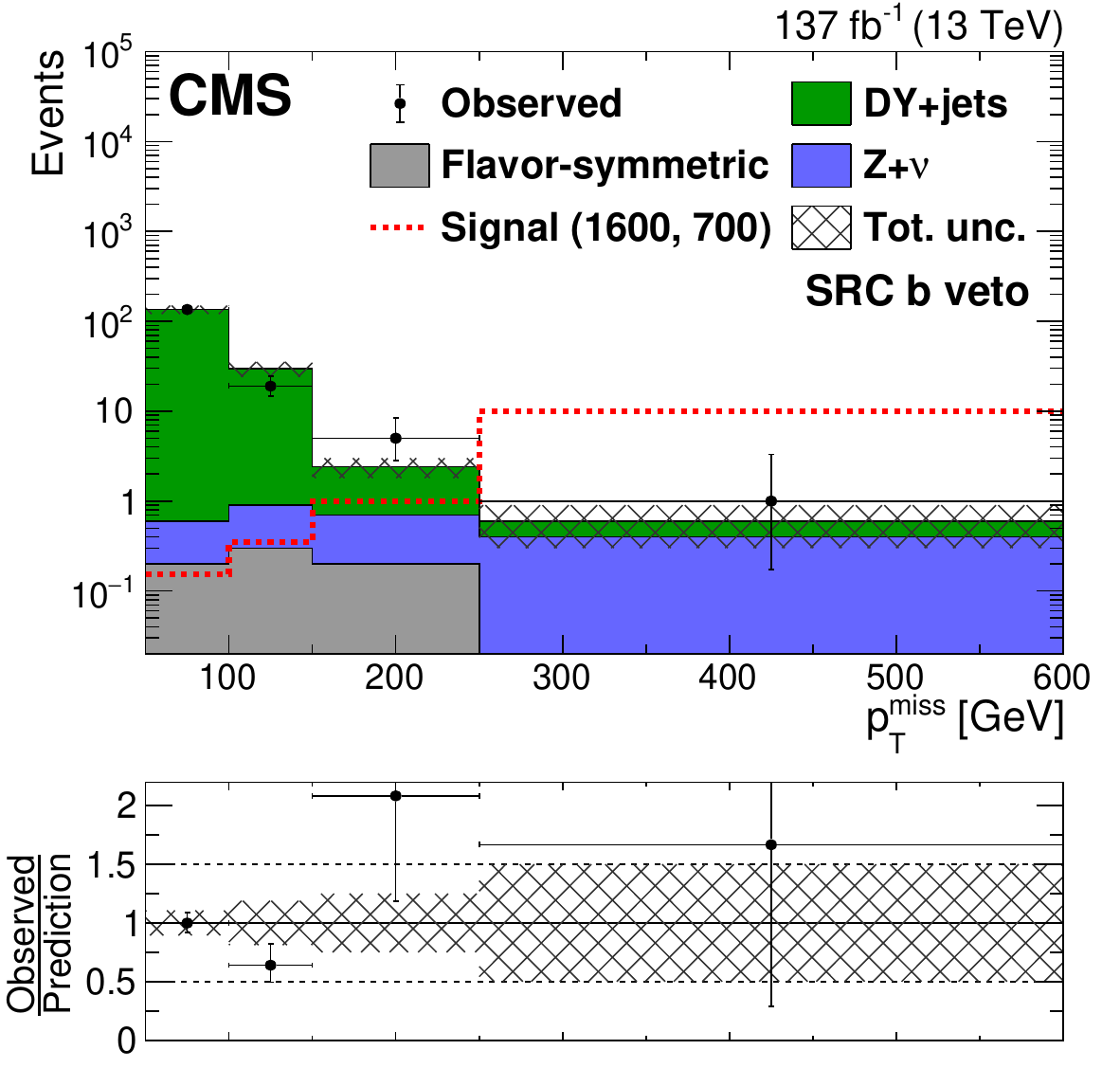}
\includegraphics[width=0.42\linewidth]{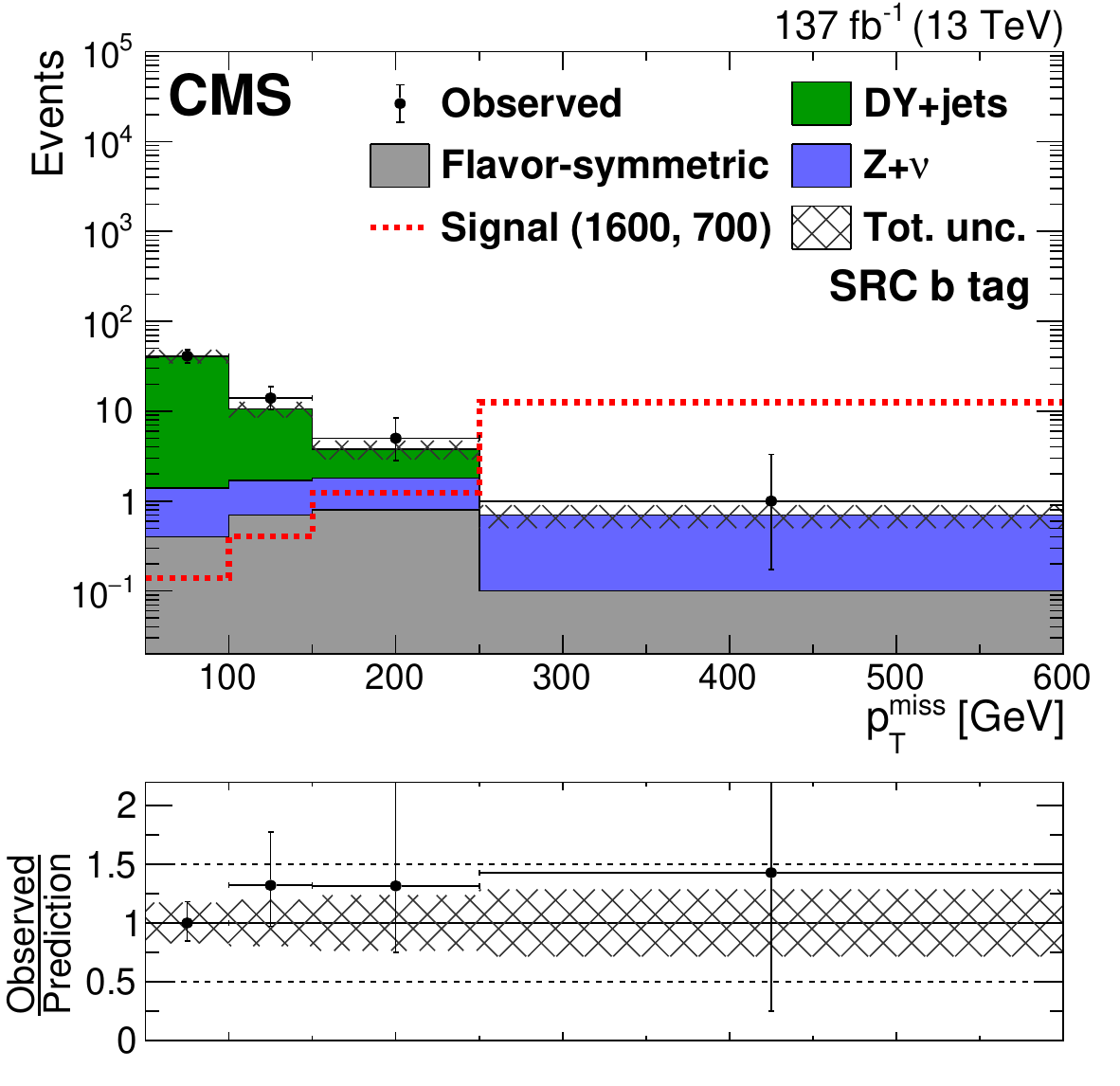}
\caption{
  The \ptmiss distribution in data is compared to the SM background prediction in the strong-production on-\PZ (upper) SRA, (middle) SRB, and (lower) SRC regions for
  (left) the \PQb veto and (right) \PQb tag categories before the fits to data discussed in Section~\ref{sec:interpretation}.
  The lower panel of each plot shows the ratio of observed data to the SM prediction in each bin of \ptmiss.
  The hashed band in the upper panels shows the total uncertainty in the background prediction
    including statistical and systematic sources. The signal \ptmiss distributions correspond to the gluino pair production model with the gluino (\firstchi) having a mass of 1600 (700)\GeV. The \ptmiss template prediction in each search region is normalized to the first \ptmiss bin of each distribution in data. The last bin includes overflow events. }
\label{fig:results_SR_str} 
\end{figure}

The results for the strong production on-\PZ SRs are summarized in Table~\ref{tab:results_SR_str}.
The corresponding \ptmiss distributions are shown in Fig.~\ref{fig:results_SR_str}.
No significant deviations are observed relative to the SM background. The largest
disagreement corresponds to one of the SRA \PQb tag categories in which 42 events are observed
and $31.4\pm3.8$ background events are expected, corresponding to a local significance of 1.4 s.d.

\begin{table}[!htbp]
\centering
\topcaption{\label{tab:results_SR_str}
  Predicted and observed event yields in the strong-production on-\PZ search regions, for each \ptmiss bin
  as defined in Table~\ref{tab:selections_signalRegions} before the fits to data discussed in Section~\ref{sec:interpretation}.
  Uncertainties include both statistical and systematic sources. The \ptmiss template prediction in each SR is
  normalized to the first \ptmiss bin of each distribution in data.
}
\cmsTableAlt{
\begin{tabular} {l l c c c c c }

    Category & SM processes & 	 \multicolumn{5}{c}{} \\ [\cmsTabSkip]
    SRA \PQb veto & \ptmiss [\GeVns{}]  & 50--100 & 100--150 & 150--230 & 230--300 & $>$300 \\ \hline 
                  & \dyjets          & $1253\pm41  $  & $153\pm16    $  & $22.0\pm4.9  $  & $0.9\pm0.8   $  & $2.9\pm3.0   $ \\
                  & Flavor-symmetric & $ 1.6\pm 0.5$  & $ 2.1 \pm 0.6$  & $ 1.4 \pm 0.5$  & $ 0.6 \pm 0.3$  & $ 0.6 \pm 0.2$ \\
                  & \znu             & $6.4 \pm 1.2$  & $4.9 \pm 0.9 $  & $5.3 \pm 1.0 $  & $2.7 \pm 0.5 $  & $6.2 \pm 1.2 $ \\
                  & Total background & $1261 \pm 41$  & $160 \pm 16  $  & $28.8 \pm 5.0$  & $4.2 \pm 1.0 $  & $9.6 \pm 3.2 $ \\
                  & Observed             & 1261           & 186             & 27              & 5               & 14             \\ [\cmsTabSkip]
    SRA \PQb tag & \ptmiss [\GeVns{}]  & 50--100 & 100--150 & 150--230 & 230--300 & $>$300 \\ \hline 
                  & \dyjets          & $602\pm28$ & $99.9\pm9.3  $ & $12.3\pm2.6  $ & $2.2\pm1.6  $ & $1.1\pm1.0  $ \\       
                  & Flavor-symmetric & $7.9\pm 1.8$ & $19.7 \pm 4.4$ & $10.6 \pm 2.4$ & $1.4 \pm 0.4$ & $0.3 \pm 0.2$ \\       
                  & \znu             & $5.8\pm0.9 $ & $8.1\pm1.2   $ & $8.4\pm1.2   $ & $2.8\pm0.5  $ & $2.6\pm0.6  $ \\             
                  & Total background & $616 \pm 28$ & $128 \pm 10  $ & $31.4 \pm 3.8$ & $6.3 \pm 1.7$ & $4.1 \pm 1.2$ \\       
                  & Observed             & 616        & 148          & 42           & 10          & 4           \\  [\cmsTabSkip]
    SRB \PQb veto & \ptmiss [\GeVns{}]  & 50--100 & 100--150 & 150--230 & 230--300 & $>$300 \\ \hline 
                  & \dyjets          & $696\pm31   $  & $103.6\pm7.1  $ & $11.2\pm2.1     $   & $0.6\pm0.6  $ & $1.0\pm0.9         $ \\
                  & Flavor-symmetric & $1.2 \pm 0.4$  & $2.4 \pm 0.7  $ & $1.0^{+0.3}_{-0.4} $   & $0.6 \pm 0.3$ & $0.1^{+0.2}_{-0.1}   $ \\
                  & \znu             & $2.6\pm0.5  $  & $2.3\pm0.4    $ & $3.5\pm0.6      $   & $0.9\pm0.2  $ & $1.9\pm0.4         $ \\
                  & Total background & $700 \pm 31 $  & $108.2 \pm 7.1$ & $15.7 \pm 2.3   $   & $2.2 \pm 0.7$ & $3.0 \pm 1.0       $ \\
                  & Observed             & 700            & 108             & 18                  & 2             & 3                   \\ [\cmsTabSkip]
    SRB \PQb tag & \ptmiss [\GeVns{}]  & 50--100 & 100--150 & 150--230 & 230--300 & $>$300 \\ \hline 
                  & \dyjets          & $215\pm16         $ & $48 \pm 16   $  & $10.7 \pm 3.8$ & $1.9 \pm 1.3$       & $0.4 \pm 0.5$       \\
                  & Flavor-symmetric & $4.5^{+1.1}_{-1.2}$     & $9.3 \pm 2.2$  & $5.3 \pm 1.3$  & $1.0^{+0.3}_{-0.4}$ & $0.1^{+0.2}_{-0.1}$ \\
                  & \znu             & $6.0 \pm 1.1      $ & $7.9 \pm 1.4 $  & $6.6 \pm 1.2$  & $2.4 \pm 0.4 $      & $1.6 \pm 0.3$       \\
                  & Total background & $225 \pm 16$        & $65 \pm 16$   & $22.7 \pm 4.2$ & $5.3 \pm 1.4$       & $2.1 \pm 0.6$       \\
                  & Observed             & 225               & 69          & 17           & 3                 & 5                 \\ [\cmsTabSkip]
    SRC \PQb veto & \ptmiss [\GeVns{}]  & 50--100 & 100--150 & 150--250 & $>$250 &\\ \hline 
                  & \dyjets          & $135 \pm 14 $ & $28.8 \pm 5.6$ & $1.7 \pm 0.5$ & $0.2\pm0.2     $    & \\
                  & Flavor-symmetric & $0.2 \pm 0.1$ & $0.3 \pm 0.2 $ & $0.2 \pm 0.1$ & $0.0^{+0.1}_{-0.0}$ & \\
                  & \znu             & $0.4\pm0.1  $ & $0.6\pm0.2   $ & $0.5\pm0.2  $ & $0.4\pm0.1      $   & \\
                  & Total background & $135 \pm 14 $ & $29.7 \pm 5.6$ & $2.4 \pm 0.6$ & $0.6 \pm 0.3    $   & \\ 
                  & Observed             & 135           & 19             & 5             & 1 & \\ [\cmsTabSkip]
    SRC \PQb tag & \ptmiss [\GeVns{}]  & 50--100 & 100--150 & 150--250 & $>$250 &\\ \hline 
                  & \dyjets          & $39.6\pm7.1  $ & $8.9\pm2.0   $ & $2.0\pm0.7  $ & $0.0\pm0.2  $ & \\
                  & Flavor-symmetric & $0.4 \pm 0.3 $ & $0.7 \pm 0.4 $ & $0.8 \pm 0.5$ & $0.1 \pm 0.1$ & \\
                  & \znu             & $1.0\pm0.2   $ & $1.0\pm0.2   $ & $1.0\pm0.2  $ & $0.6\pm0.2  $ & \\
                  & Total background & $41.0 \pm 7.1$ & $10.7 \pm 2.1$ & $3.8 \pm 0.9$ & $0.7 \pm 0.2$ & \\
                  & Observed             & 41             & 14             & 5             & 1             & \\

\end{tabular}

}
\end{table}

\begin{figure}[htb!]
\centering
  \includegraphics[width=0.45\linewidth]{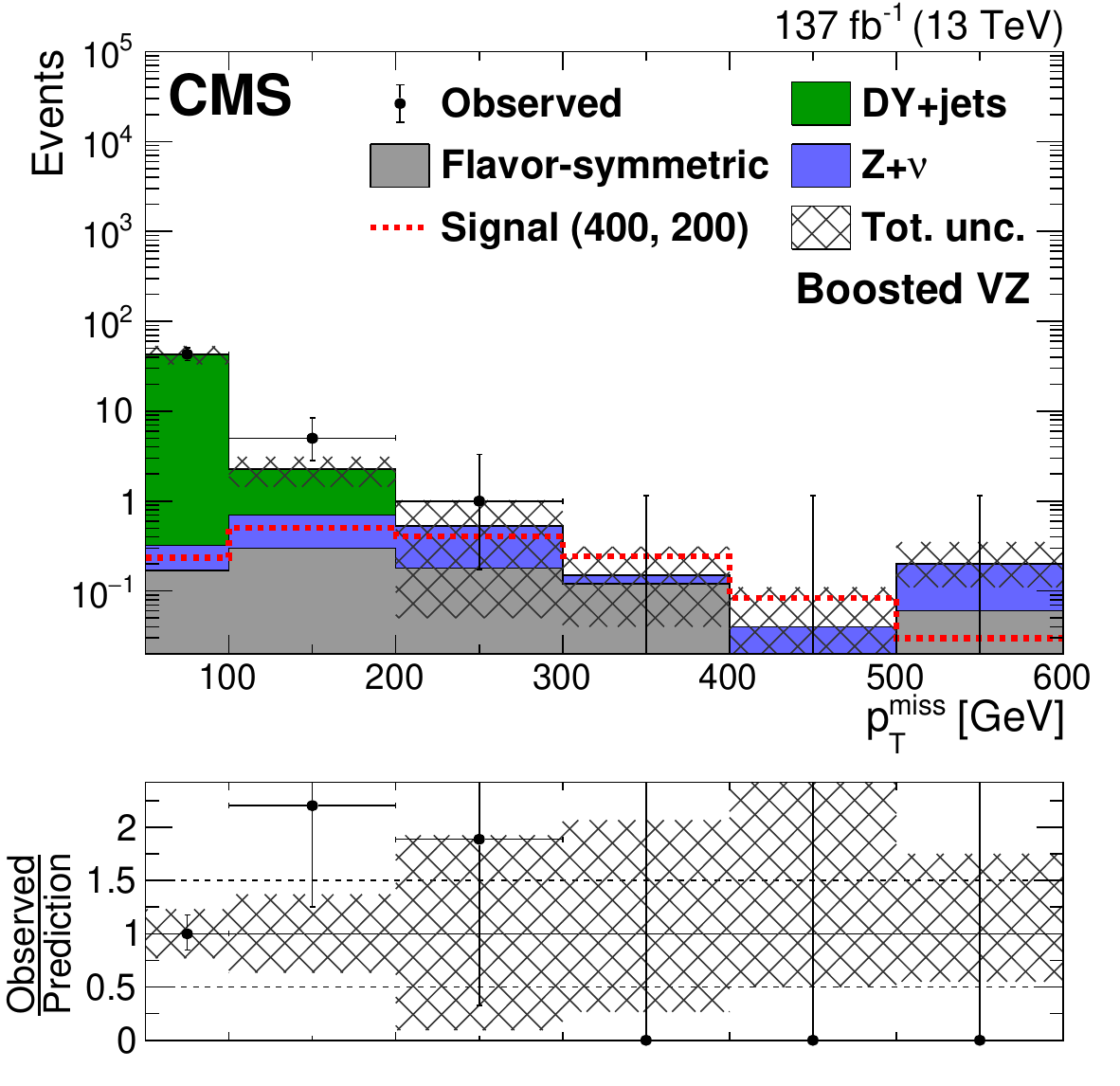} 
  \includegraphics[width=0.45\linewidth]{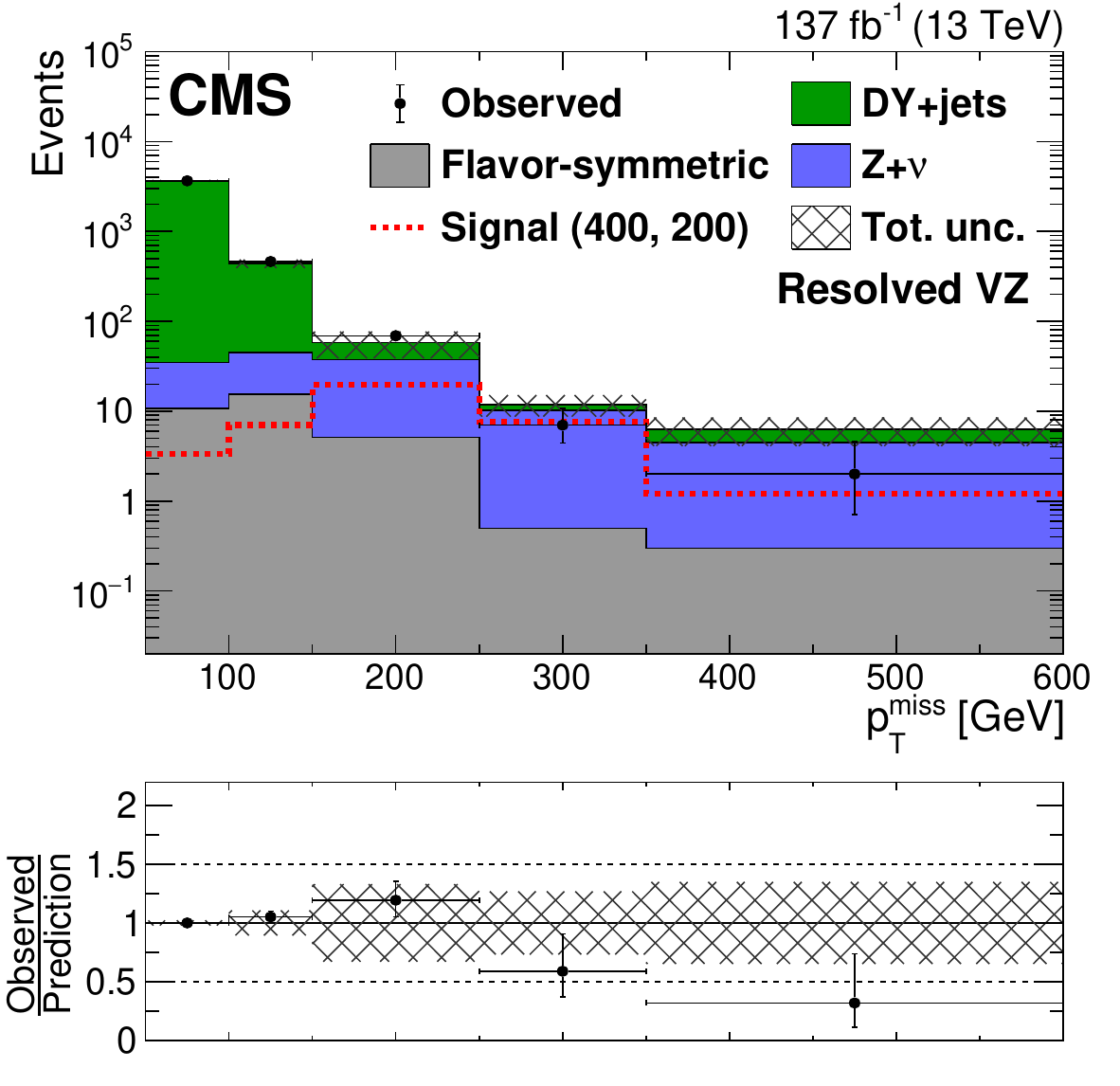}\\
  \includegraphics[width=0.45\linewidth]{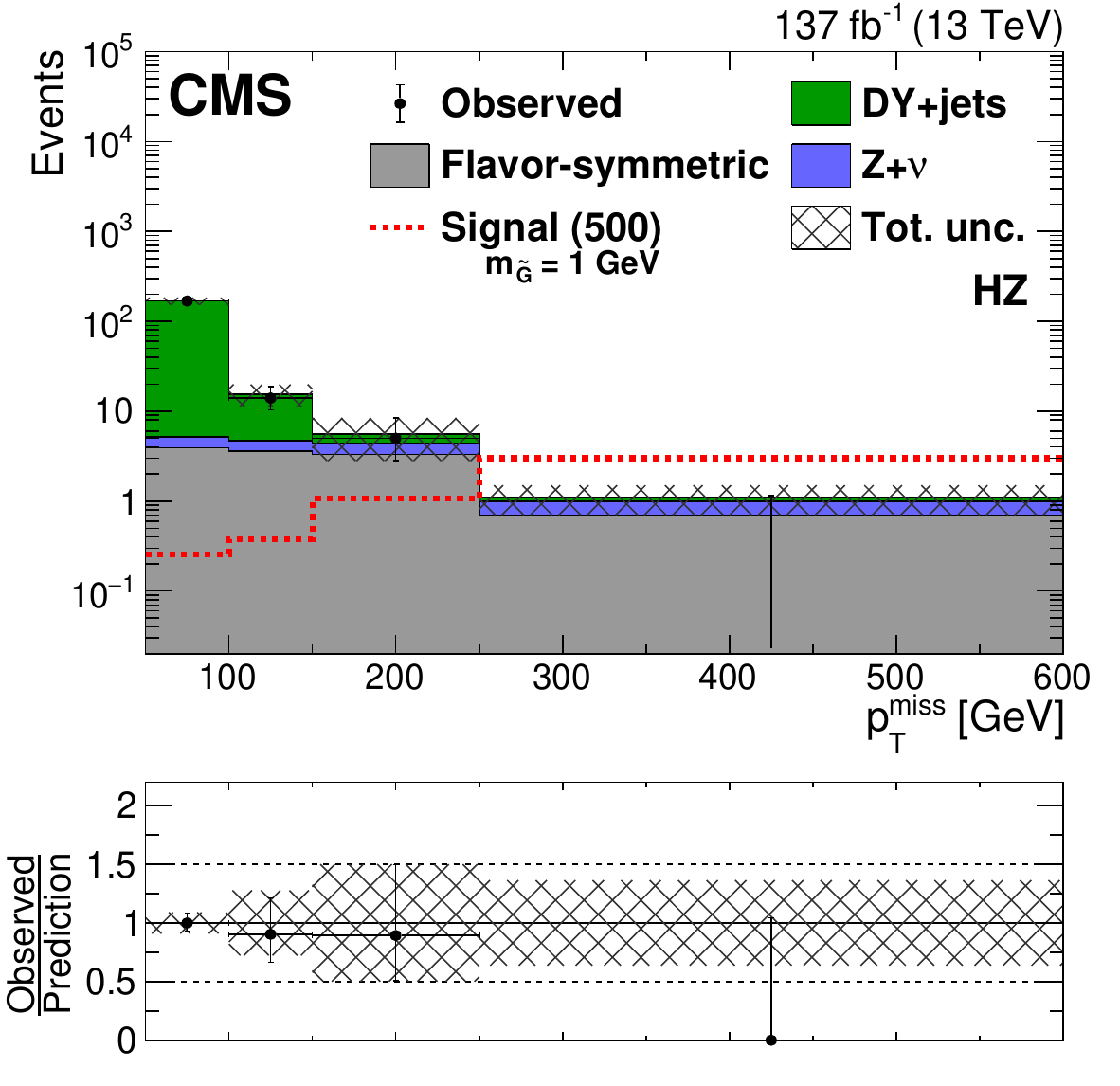}
  \caption{The \ptmiss distribution in data is compared to the SM background prediction in the EW-production on-\PZ (upper left) boosted \vz, (upper right) resolved \vz, and (lower) $\PH\PZ$ search regions before the fits to data described in Section~\ref{sec:interpretation}.   The lower panel of each figure shows the ratio of observed data to the SM prediction in each \ptmiss bin.  The hashed band shows the total uncertainty in the background prediction   including statistical and systematic sources.
    The signal \ptmiss distribution for the boosted and resolved \vz search regions correspond to the \firstcharg\secondchi production model with a \firstcharg/\secondchi (\firstchi) mass of 400 (200)\GeV,while for the $\PH\PZ$ search region the \ptmiss distribution corresponds to a \firstchi pair production model decaying into a Higgs boson, a \PZ boson and two \gravitino with the \firstchi(\gravitino) having a mass of 500 (1)\GeV. The \ptmiss template prediction in each search region is normalized to the first \ptmiss bin of each distribution in data. The last bin includes overflow events. }
\label{fig:results_SR_ewk}
\end{figure}

The results for the EW-production on-\PZ SRs are summarized in Table~\ref{tab:results_SR_ewk}.
The corresponding \ptmiss\ distributions are shown in Fig.~\ref{fig:results_SR_ewk}.
The observed data yields are consistent with the SM background predictions. 
The largest discrepancy between data and prediction occurs in the highest \ptmiss bin of the resolved \vz regions,
where 2 events are observed while $6.3\pm2.2$ are predicted, corresponding to a deficit with a local significance of $1.2$~s.d.

\begin{table}[tbh!]
\centering
\topcaption{\label{tab:results_SR_ewk}
  Predicted and observed event yields in the EW-production on-\PZ search regions, for each \ptmiss\ bin
  as defined in Table~\ref{tab:selections_signalRegions} before the fits to data described in Section~\ref{sec:interpretation}.
  Uncertainties include both statistical and systematic sources. The \ptmiss template prediction in each SR is normalized
  to the first \ptmiss bin of each distribution in data.
}
\cmsTableAlt{
\begin{tabular} {l l c c c c c c}
    Category & SM processes & 	 \multicolumn{6}{c}{} \\ [\cmsTabSkip]
    Boosted \vz  & \ptmiss [\GeVns{}] & 50--100              & 100--200       & 200--300            & 300--400            & 400--500            & $>$500        \\ \hline 
                 & \dyjets            & $42.7\pm9.9       $  & $1.6\pm0.8  $  & $0.0\pm0.5        $ & $0.0^{+0.1}_{-0.0}$ & $0.0^{+0.1}_{-0.0}$ & $0.0^{+0.1}_{-0.0}$ \\
                 & Flavor-symmetric   & $0.2^{+0.2}_{-0.1}$  & $0.3 \pm 0.2$  & $0.2^{+0.2}_{-0.1}$ & $0.1\pm 0.1       $ & $0.0^{+0.1}_{-0.0}$ & $0.1 \pm 0.1$ \\
                 & \znu               & $0.2\pm0.2        $  & $0.4\pm0.2  $  & $0.3\pm0.1        $ & $0.0^{+0.1}_{-0.0}$ & $0.0^{+0.1}_{-0.0}$ & $0.1\pm0.1  $ \\
                 & Total background   & $43.0 \pm 9.9       $  & $2.3 \pm 0.8$  & $0.5 \pm 0.5      $ & $0.2^{+0.2}_{-0.1}$ & $0.0^{+0.1}_{-0.0}$ & $0.2 \pm 0.1$ \\
                 & Observed               & 43                   & 5              & 1                   & 0                   & 0                   & 0             \\ [\cmsTabSkip]
    Resolved \vz & \ptmiss [\GeVns{}] & 50--100              & 100--150       & 150--250            & 250--350            & $>$350              &               \\ \hline 
                 & \dyjets            & $3613\pm80         $ & $394\pm46    $ & $21\pm18    $       & $1.7\pm2.4     $    & $1.8\pm1.9  $       &               \\
                 & Flavor-symmetric   & $10.7^{+3.0}_{-2.9}$ & $15.4 \pm 4.2$ & $5.1 \pm 1.5$       & $0.5 \pm 0.2   $    & $0.3 \pm 0.2$       &               \\
                 & \znu               & $24.0\pm4.1        $ & $29.5\pm5.6  $ & $32.2\pm6.5 $       & $9.7\pm2.2     $    & $4.2\pm1.1  $       &               \\
                 & Total background   & $3648 \pm 80       $ & $439 \pm 47  $ & $58 \pm 19  $       & $11.9 \pm 3.2  $    & $6.3 \pm 2.2$       &               \\
                 & Observed               & 3648                 & 461            & 69                  & 7                   & 2                   &               \\ [\cmsTabSkip]
    $\PH\PZ$       & \ptmiss [\GeVns{}] & 50--100              & 100--150       & 150--250            & $>$250              &                     &               \\ \hline 
                 & \dyjets            & $163\pm15   $        & $10.8\pm4.1  $ & $1.3\pm2.5  $       & $0.1\pm0.3   $      &                     &               \\
                 & Flavor-symmetric   & $3.9 \pm 1.4$        & $3.6 \pm 1.3 $ & $3.3 \pm 1.2$       & $0.7 \pm 0.3 $      &                     &               \\
                 & \znu               & $1.3\pm0.3  $        & $1.1\pm0.2   $ & $1.0\pm0.2  $       & $0.3\pm0.1   $      &                     &               \\
                 & Total background   & $168 \pm 15 $        & $15.6 \pm 4.3$ & $5.6 \pm 2.8$       & $1.2 \pm 0.4 $      &                     &               \\
                 & Observed               & 168                  & 14             & 5                   & 0                   &                     &               \\
\end{tabular}
}
\end{table}

\subsection{Results for the edge search samples}
\label{sub:edgeResults}

\begin{figure}[!ht]
\centering
\includegraphics[width=0.45\textwidth]{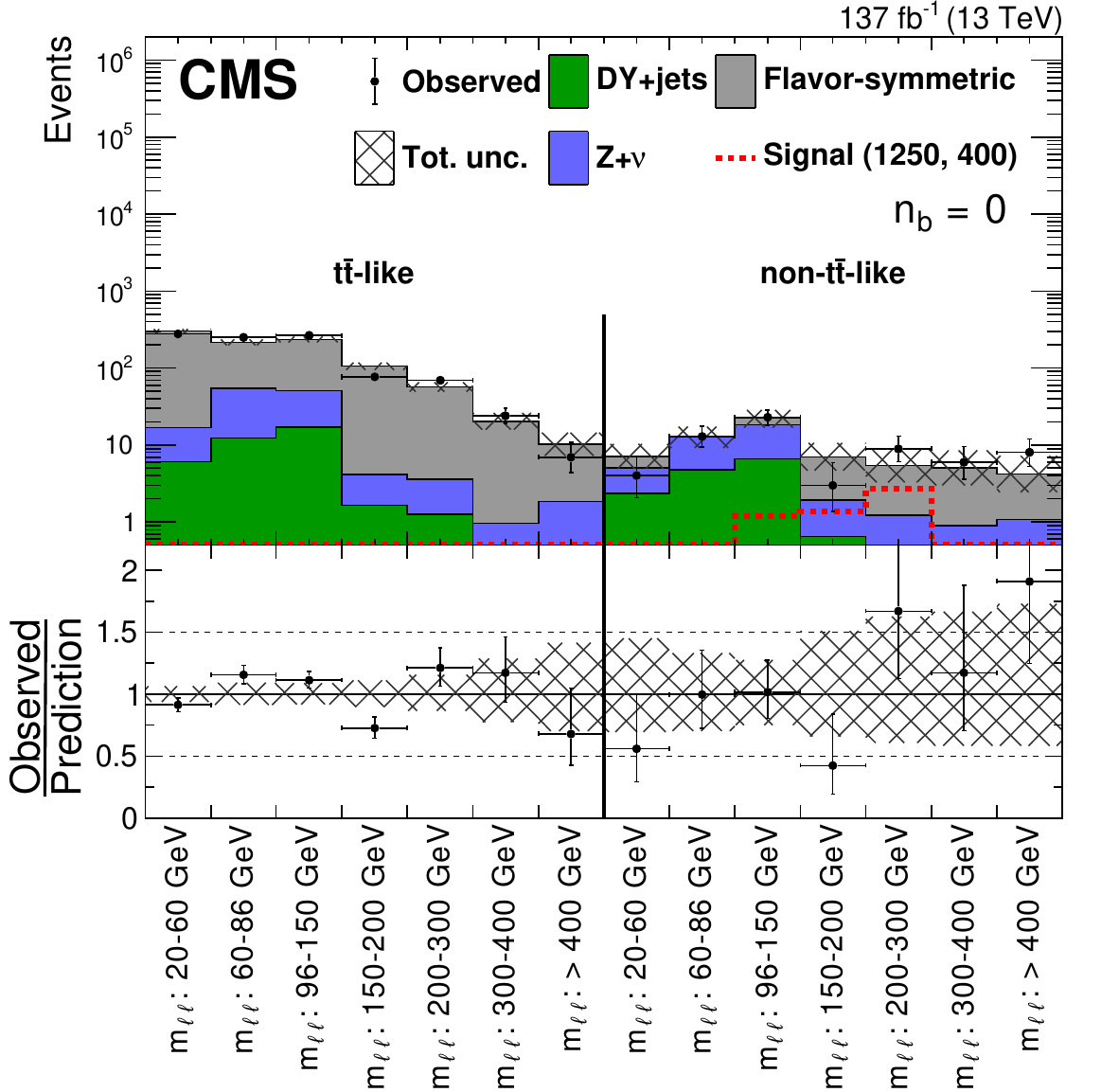}
\includegraphics[width=0.45\textwidth]{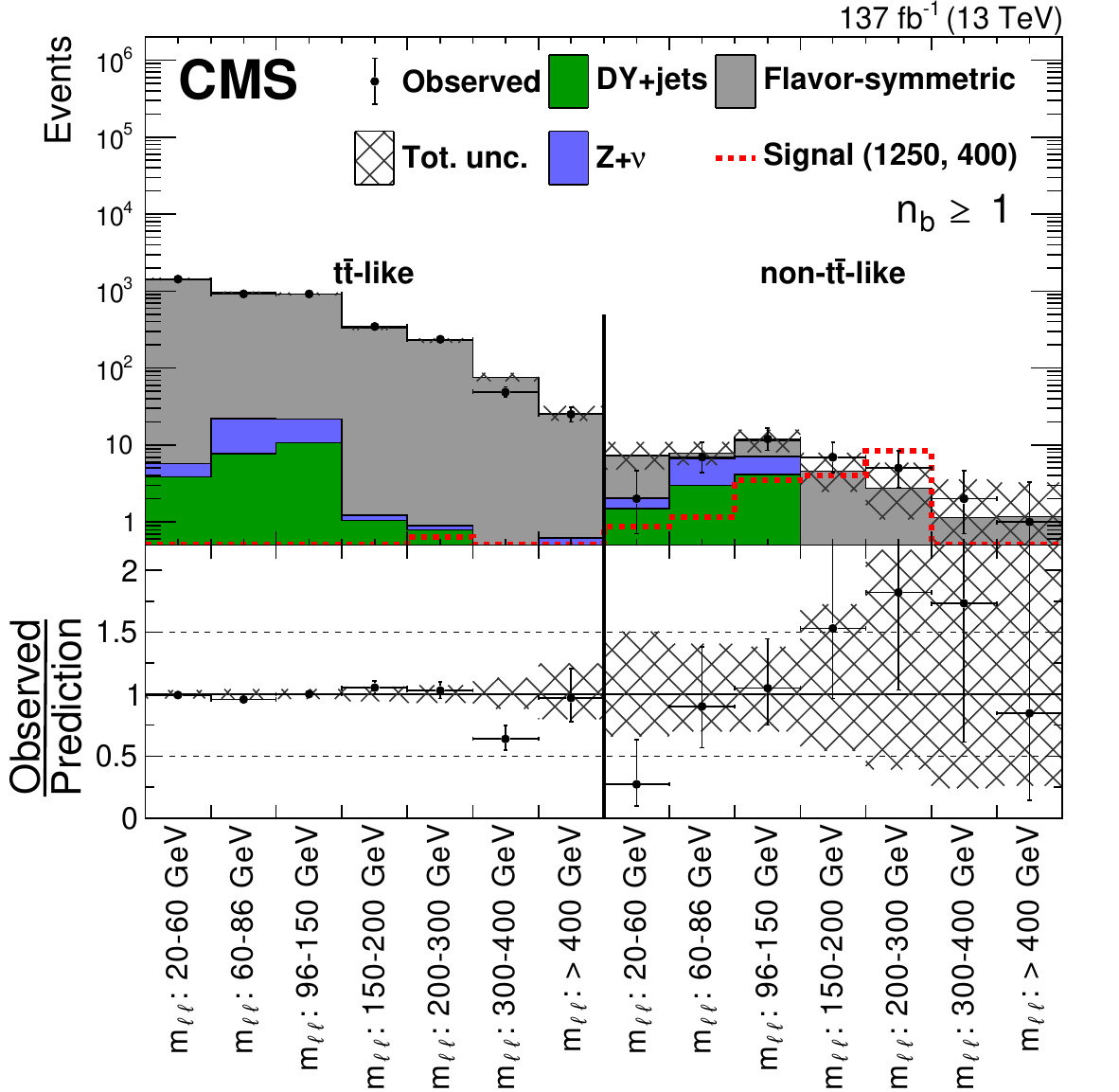}
\caption{Results of the counting experiment in the edge search regions before the fits to data described in Section~\ref{sec:interpretation}.
  In each search region, the number of observed events in data (black markers)
  is compared to the SM background prediction for the (left) \PQb veto and (right) \PQb tag categories.  
  The hashed band shows the total uncertainty in the background prediction including statistical and systematic sources. The signal distribution corresponds to the \sbottom pair production model with the \sbottom (\secondchi) having a mass of 1250 (400)\GeV.}
\label{fig:cNc_resultOverview}
\end{figure}

Comparisons between the SM predictions and the observed data in the 28 edge SRs
 are summarized in Table~\ref{tab:edgeResults}.
A graphical representation of the same results is
displayed in Fig.~\ref{fig:cNc_resultOverview}.

\begin{table}[!htbp]
\renewcommand{\arraystretch}{1.3}
\setlength{\belowcaptionskip}{6pt}
\centering
\topcaption{Predicted and observed yields in each bin of the edge search counting experiment
as defined in Table~\ref{tab:selections_signalRegions} before the fits to data described in Section~\ref{sec:interpretation}.
Uncertainties include statistical and systematic sources.
}
\label{tab:edgeResults}
\cmsTableAlt{
\begin{tabular}{ c  c  c  c  c  c c}

$n_\text{\PQb}$ & $m_{\ell\ell}$ range  [\GeVns{}] & Flavor-symmetric & \dyjets & \znu & Total background & Observed \\  \hline 
\multirow{8}{*}{} & \multicolumn{6}{c}{\ttbar-like}  \\ 
 & 20--60   & $286^{+19}_{-18}   $ & $6.1\pm3.8 $ & $10.8\pm3.1$ & $304^{+20}_{-19}   $ & 277 \\
 & 60--86   & $163^{+14}_{-13}   $ & $12.3\pm7.6$ & $42\pm12   $ & $217^{+20}_{-19}   $ & 251 \\
 & 96--150  & $187^{+15}_{-14}   $ & $17\pm11   $ & $34\pm9    $ & $238^{+21}_{-20}   $ & 265 \\
 & 150--200 & $102^{+12}_{-11}   $ & $1.7\pm1.8 $ & $2.5\pm0.8 $ & $106^{+12}_{-11}   $ & 77  \\
 & 200--300 & $53.4^{+8.7}_{-7.6}$ & $1.3\pm1.3 $ & $2.3\pm0.8 $ & $57.0^{+8.8}_{-7.8}$ & 69  \\
 & 300--400 & $19.5^{+5.8}_{-4.6}$ & $0.3\pm0.3 $ & $0.7\pm0.3 $ & $20.5^{+5.8}_{-4.7}$ & 24  \\
 & $>$400     & $8.5^{+4.2}_{-3.0} $ & $0.5\pm0.5 $ & $1.3\pm0.5 $ & $10.3^{+4.3}_{-3.1}$ & 7   \\[\cmsTabSkip]
$=0$ &  \multicolumn{6}{c}{non-\ttbar-like}   \\
& 20--60   & $ 2.1^{+2.7}_{-1.3}  $ & $ 2.4\pm1.5  $ & $ 2.7\pm0.9  $ & $7.1^{+3.2}_{-2.2}$    & 4 \\
& 60--86   & $ 0.0^{+1.8}_{-0.0}  $ & $ 4.8\pm3.0  $ & $ 8.3\pm2.5  $ & $ 13.1^{+4.3}_{-3.9} $ & 13   \\
& 96--150  & $ 4.2^{+3.3}_{-2.0}  $ & $ 6.6\pm4.1  $ & $ 11.8\pm3.3 $ & $ 22.6^{+6.2}_{-5.7} $ & 23   \\
& 150--200 & $ 5.1^{+3.5}_{-2.2}  $ & $ 0.6\pm0.7  $ & $ 1.3\pm0.5  $ & $ 7.1^{+3.6}_{-2.4}  $ & 3    \\
& 200--300 & $ 4.1^{+3.3}_{-2.0}  $ & $ 0.5\pm0.5  $ & $ 0.8\pm0.3  $ & $ 5.4^{+3.4}_{-2.1}  $ & 9    \\
& 300--400 & $ 4.2^{+3.4}_{-2.1}  $ & $ 0.1\pm0.1  $ & $ 0.8\pm0.4  $ & $ 5.1^{+3.4}_{-2.1}  $ & 6    \\
& $>$400   & $ 3.1^{+3.0}_{-1.7}  $ & $ 0.2\pm0.2  $ & $ 0.9\pm0.3  $ & $ 4.2^{+3.1}_{-1.7}  $ & 8    \\ [\cmsTabSkip] \hline
\multirow{8}{*}{} & \multicolumn{6}{c}{\ttbar-like}  \\
& 20--60   & $ 1432^{+48}_{-47}   $ & $ 3.8\pm2.4  $ & $ 1.9\pm0.6  $ & $ 1438^{+48}_{-47}   $ & 1427 \\
& 60--86   & $ 936^{+37}_{-36}    $ & $ 7.7\pm4.9  $ & $ 14.3\pm3.6 $ & $ 958^{+37}_{-37}    $ & 916  \\
& 96--150  & $ 897^{+36}_{-35}    $ & $ 10.7\pm6.8 $ & $ 10.9\pm2.8 $ & $ 918^{+37}_{-36}    $ & 918  \\
& 150--200 & $ 330^{+20}_{-19}    $ & $ 1.0\pm1.1  $ & $ 0.2\pm0.1  $ & $ 332^{+20}_{-19}    $ & 349  \\
& 200--300 & $ 227^{+17}_{-16}    $ & $ 0.8\pm0.8  $ & $ 0.1\pm0.1  $ & $ 228^{+17}_{-16}    $ & 235  \\
& 300--400 & $ 76.3^{+10}_{-9.1}  $ & $ 0.2\pm0.2  $ & $0.0^{+0.1}_{-0.1}$ & $ 76.5^{+10}_{-9.1}  $ & 49   \\
& $>$400   & $ 25.2^{+6.3}_{-5.2} $ & $ 0.3\pm0.3  $ & $ 0.3\pm0.3  $ & $ 25.8^{+6.3}_{-5.2} $ & 25   \\ [\cmsTabSkip]
$\geq 1$ & \multicolumn{6}{c}{non-\ttbar-like}   \\ 
\multirow{7}{*}{} &  20-60   &  5.2$^{+3.5}_{-2.3}$    & 1.5$\pm$0.9   & 0.6$\pm$0.3  &  7.3$^{+3.7}_{-2.5}$ & 2 \\
& 60--86   & $ 1.0^{+2.3}_{-0.8}  $ & $ 3.0\pm1.9  $ & $ 3.8\pm1.0  $ & $ 7.8^{+3.2}_{-2.3}  $ & 7    \\
& 96--150  & $ 4.3^{+3.4}_{-2.1}  $ & $ 4.2\pm2.6  $ & $ 3.0\pm0.8  $ & $ 11.5^{+4.4}_{-3.4} $ & 12   \\
& 150--200 & $ 4.1^{+3.3}_{-2.0}  $ & $ 0.4\pm0.4  $ & $ 0.1\pm0.1  $ & $ 4.6^{+3.3}_{-2.1}  $ & 7    \\
& 200--300 & $ 2.4^{+3.2}_{-1.6}  $ & $ 0.3\pm0.3  $ & $ 0.1\pm0.1  $ & $ 2.7^{+3.2}_{-1.7}  $ & 5    \\
& 300--400 & $ 1.1^{+2.4}_{-0.9}  $ & $ 0.1\pm0.1  $ & $0.0^{+0.1}_{-0.1}$ & $ 1.2^{+2.4}_{-0.9}  $ & 2    \\
& $>$400   & $ 0.9^{+2.1}_{-0.9}  $ & $ 0.1\pm0.1  $ & $ 0.2\pm0.2  $ & $ 1.2^{+2.1}_{-0.9}  $ & 1    \\

\end{tabular}
}
\end{table}

We find an agreement between the observed data  and SM predictions in all SRs.
The largest deviation is observed in the \ttbar-like region for $300<\mll<400\GeV$ and $\nb>0$,
in which 49 events are observed and $76^{+10}_{-9}$ were expected, corresponding
to a deficit in data with a local significance of $2.4$~s.d.
We also observe a slightly larger number of events than the background prediction in the high
\mll non-\ttbar-like category, but the predictions agree within one s. d.

\begin{figure}[!hbt]
\centering
\includegraphics[width=0.42\textwidth]{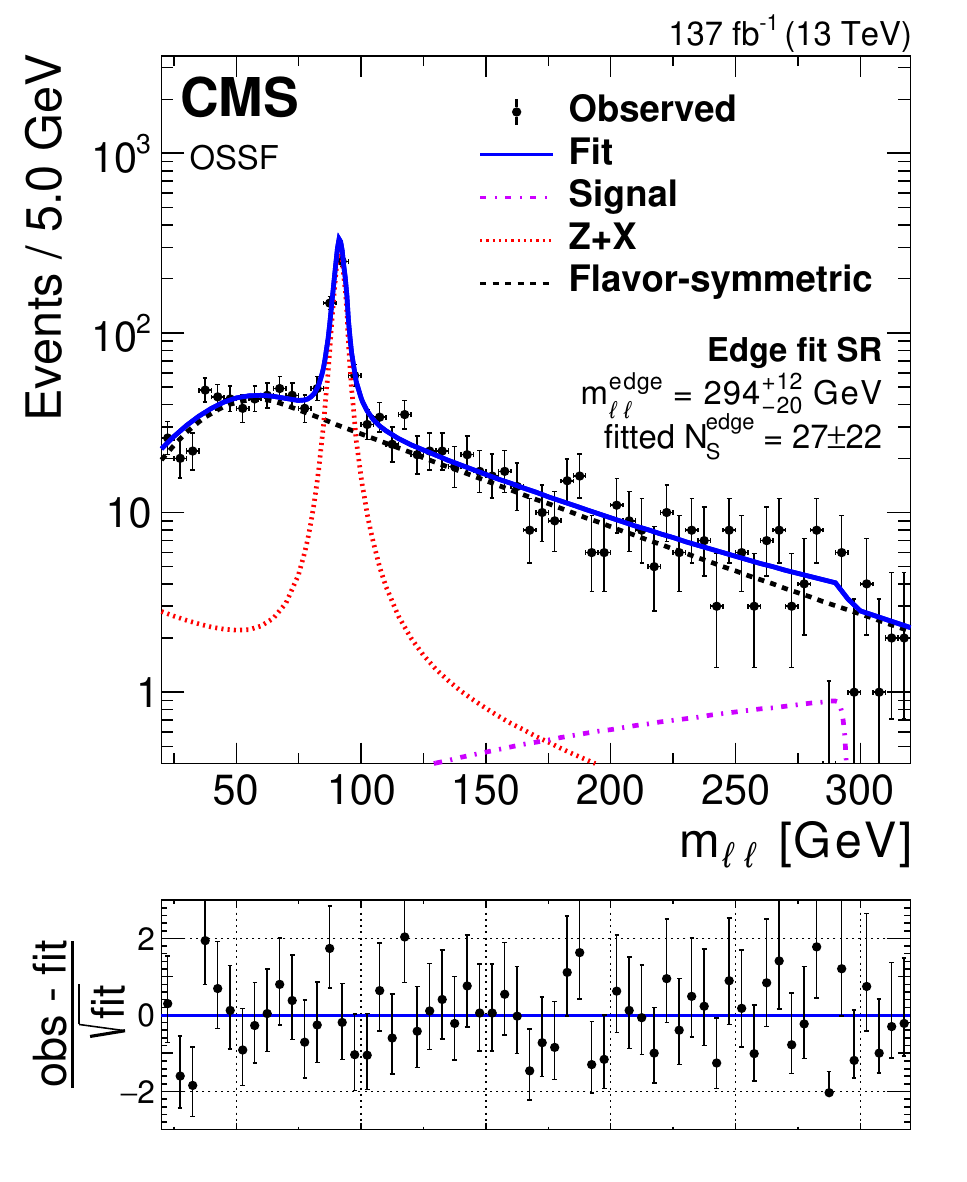}
\includegraphics[width=0.42\textwidth]{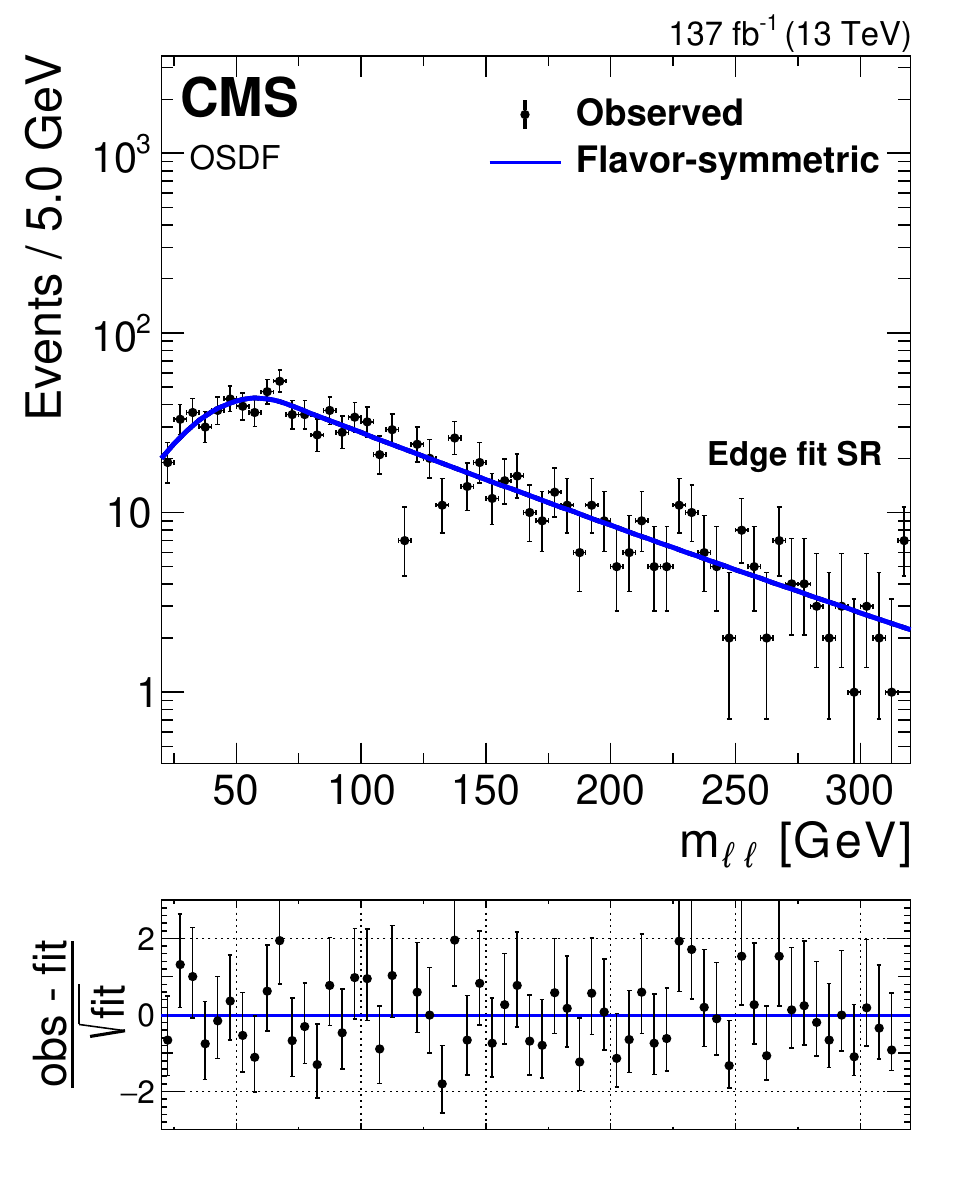}
\caption{
  Fit the \mll distributions in data in the edge fit search regions under the signal+background hypothesis
  projected onto the (left) SF and (right) DF data samples. 
  The fit shape is shown as a solid blue line.
  The individual fit components are indicated by the dashed and dotted lines.
  The flavor-symmetric background is shown as the black dashed line.
  The \ZplusX background is displayed as the red dotted line.
  The extracted signal component is displayed as the purple dash-dotted line.
  The lower panel in each plot shows the difference between the observed data yield and the fit 
  divided by the square root of the number of fitted events.}
\label{fig:Fit_data_H1}
\end{figure}

The dilepton mass distributions and the results of the kinematic edge  fit are shown in Fig.~\ref{fig:Fit_data_H1} while
Table~\ref{tab:fitResults} presents a summary of the fit results. 
A best fit signal yield of $27\pm22$ events is obtained when evaluating the signal hypothesis in the edge fit SR
with a fitted edge position of $\mll=294^{+12}_{-20}\GeV$, assuming the signal normalization to be
nonnegative.
To test the compatibility of this result with the background-only hypothesis, 
we estimate the global $p$-value~\cite{Gross:2010qma} of the result using the test statistic $-2\ln Q$,
where $Q$ denotes the ratio of the fitted likelihood value for the signal+background
hypothesis to that for the background-only hypothesis.  
The test statistic $-2\ln Q$ is evaluated in data and compared to
the corresponding quantity computed using a large sample of background-only pseudo-experiments where the edge position is not fixed to any particular value. 
The resulting $p$-value is interpreted as the one-sided tail probability
of a Gaussian distribution, and corresponds to an excess in the observed yields relative to the SM background prediction
at a global significance of 0.7 s.d. If unphysical negative signal yields are permitted, the best fit corresponds to a negative signal yield with an edge 
position of $34.4\GeV$ and a global significance of 1.8~s.d.

\begin{table}[!hbt]
\renewcommand{\arraystretch}{1.2}
\centering
\topcaption{Results of the \mll unbinned maximum likelihood fit to data in the edge fit search region
as defined in Table~\ref{tab:selections_signalRegions}.
The fitted yields of the \ZplusX and flavor-symmetric (FS) background components are tabulated
together with the fitted value of \Rsfof. The fitted signal contribution and the corresponding edge position are also shown.
The local and global signal significances are expressed in terms of s.d.
The uncertainties include both statistical and systematic sources.}
    \label{tab:fitResults}
  \begin{tabular}{l c}
  \hline
  \ZplusX yield                      & $447 \pm 28$        \\
  FS yield                &  $1019 \pm 29$        \\
  \Rsfof                          & $1.02 \pm 0.04$              \\
  Signal events                   & $27 \pm 22$       \\
  $\mll^\text{edge} $         & $294^{+12}_{-20}$\GeV  \\[\cmsTabSkip]
  Local significance                   & 1.3 s.d.          \\
  Global significance                  & 0.7 s.d.          \\ \hline
  \end{tabular}
\end{table}

\subsection{Results in the slepton search regions}
\label{sub:sleptonresults}

\begin{figure}[!htbp]
    \centering
    \includegraphics[width=0.42\textwidth]{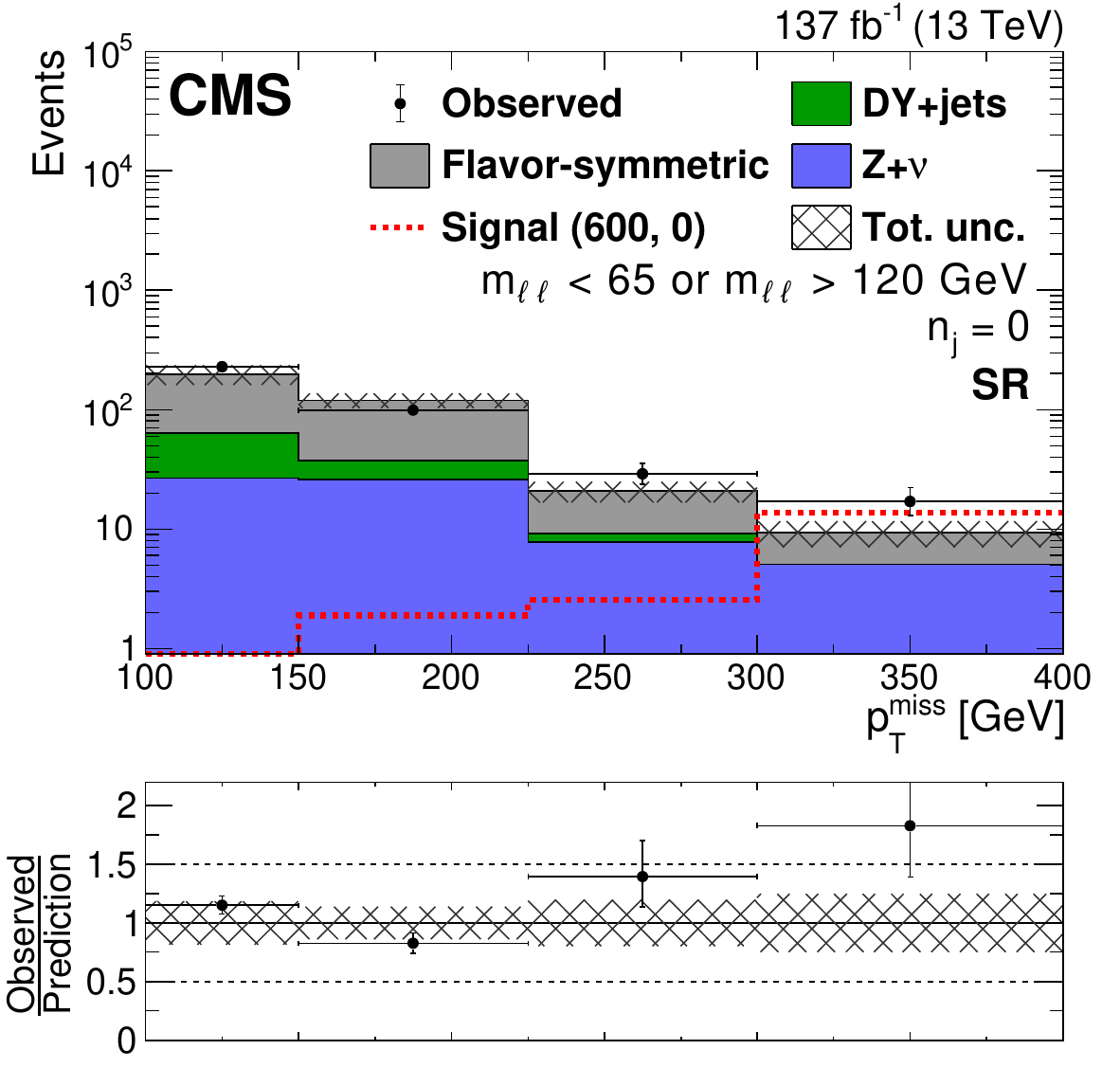}
    \includegraphics[width=0.42\textwidth]{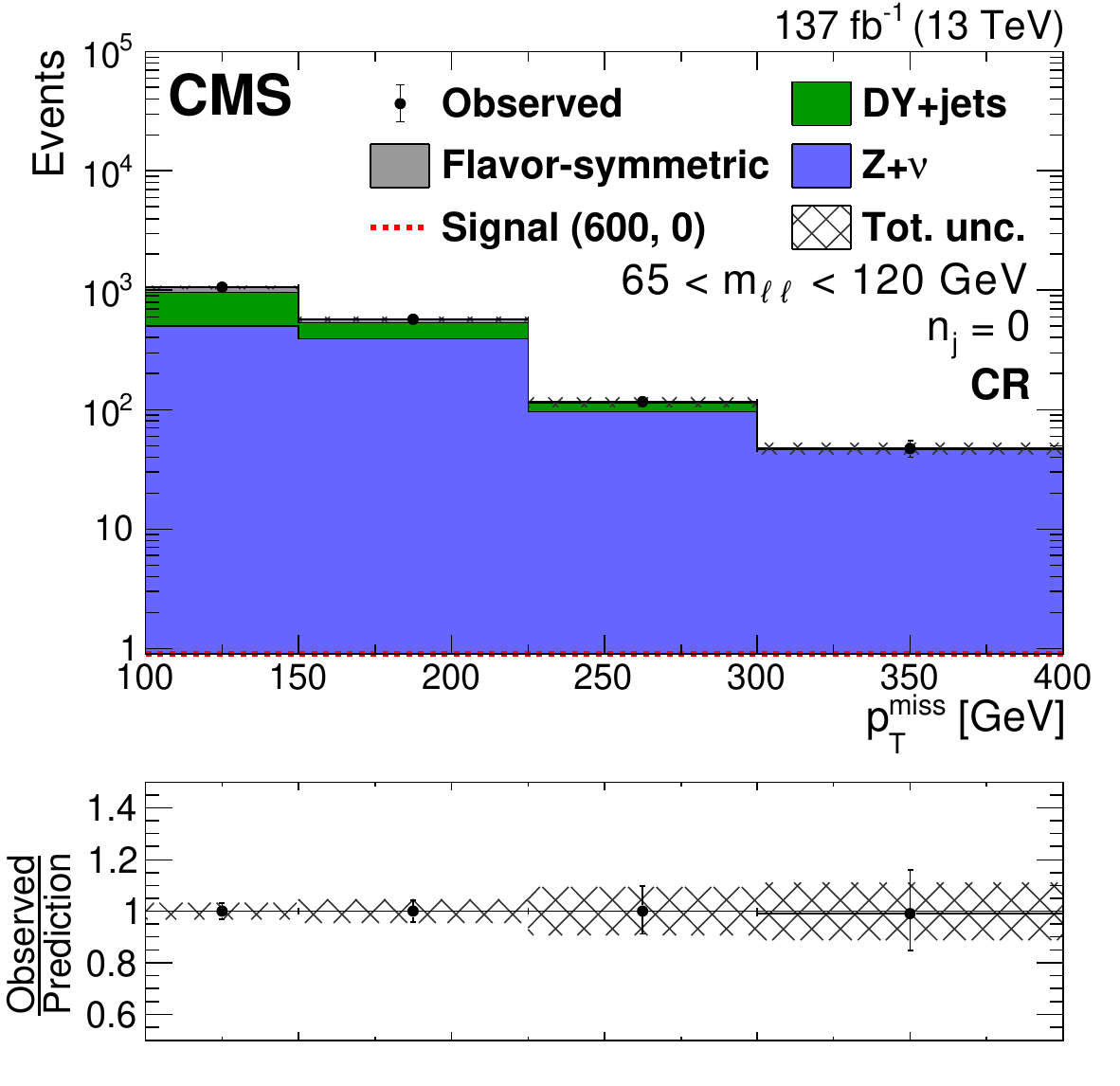}\\
    \includegraphics[width=0.42\textwidth]{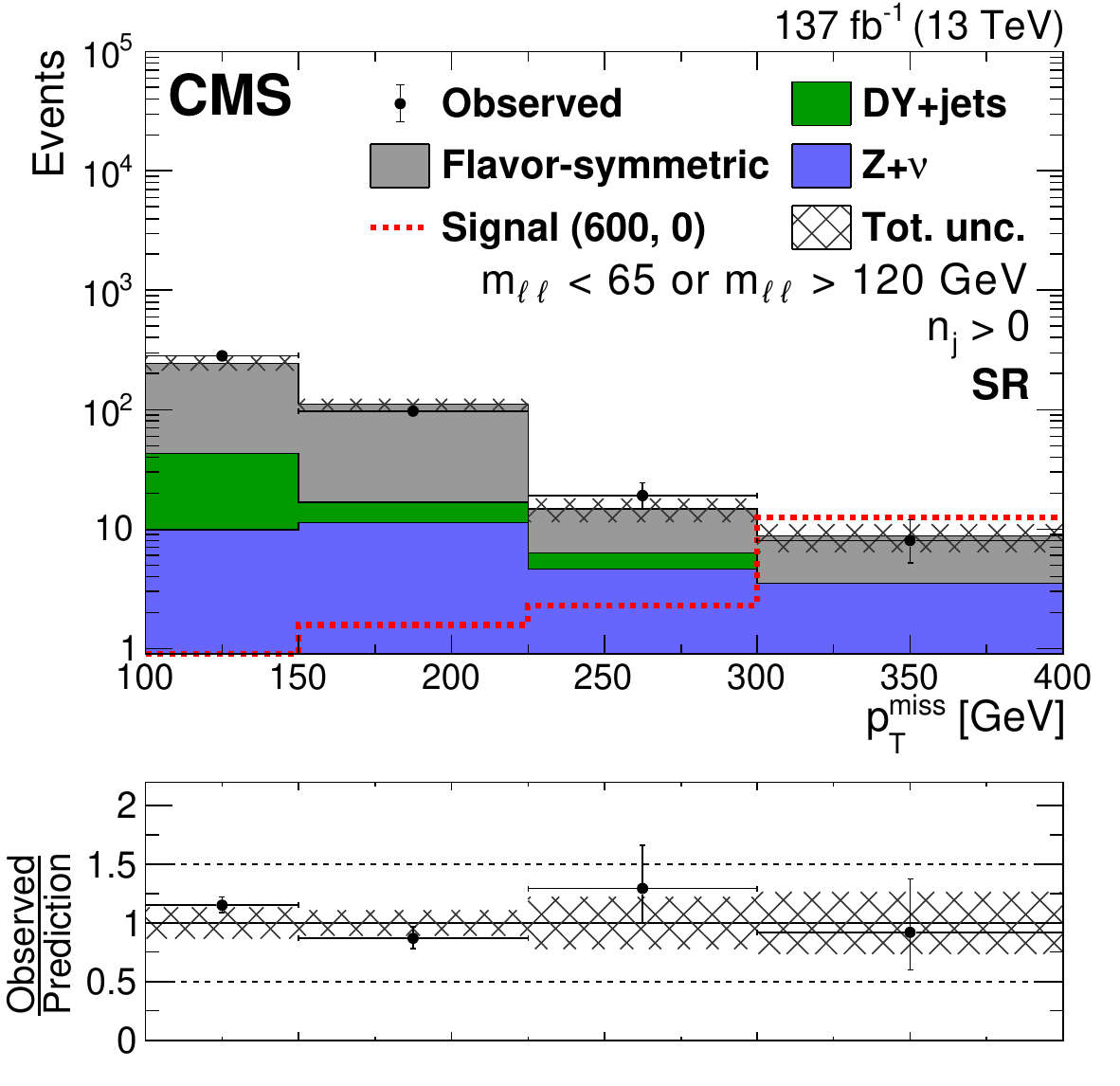}
    \includegraphics[width=0.42\textwidth]{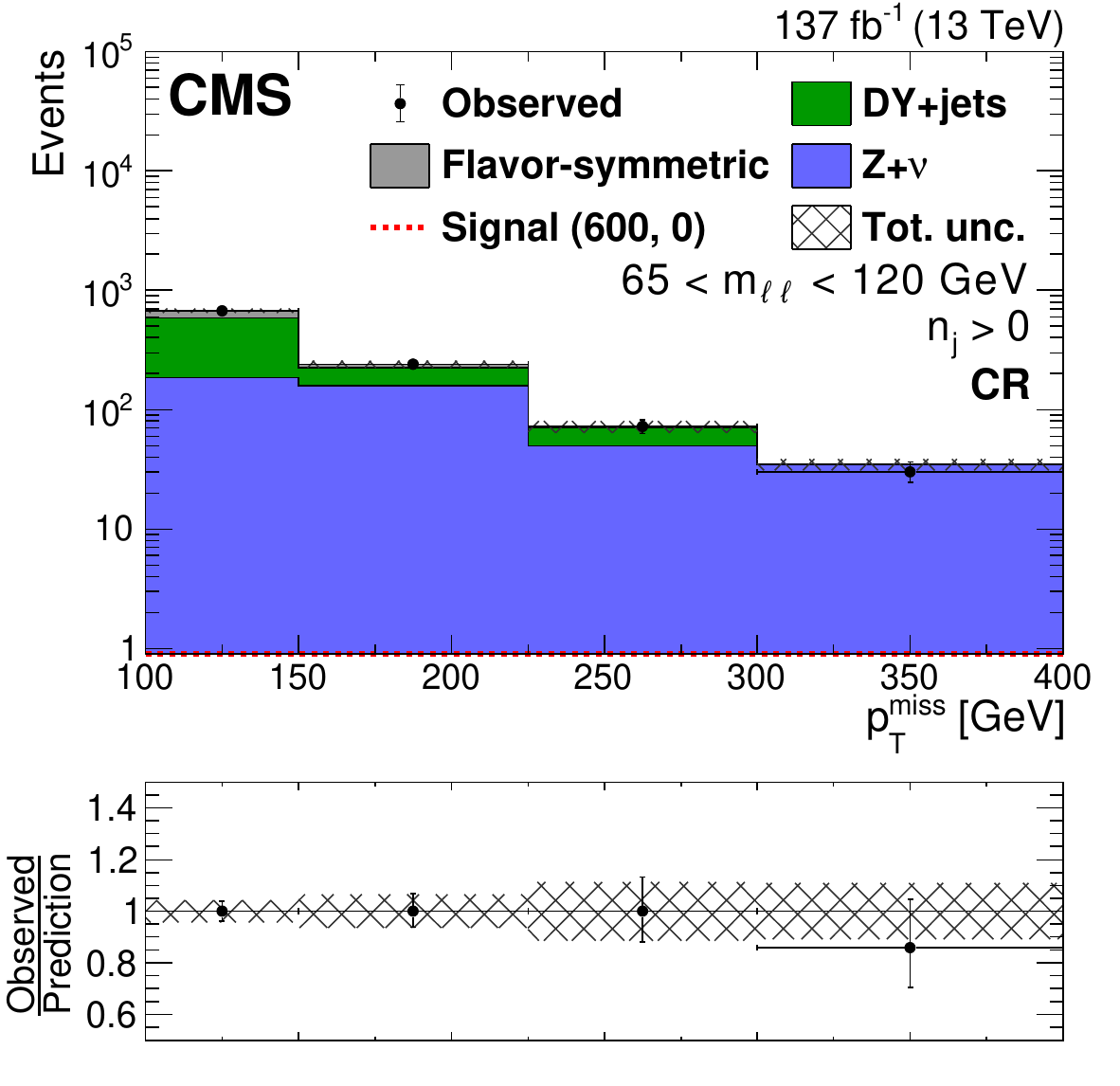}
    \caption{Distribution of \ptmiss for events in the slepton (left) search regions and (right) control regions obtained by inverting the \mll selection
      used to obtain the DY background normalization for regions (upper) without jets and (lower) with jets. A background-only fit to data in the control region has been performed to determine the \dyjets contribution as discussed in Section~\ref{sec:interpretation}.
      The lower panel of each plot shows the ratio of observed data to the SM prediction in each \ptmiss bin.
      The hashed band shows the total uncertainty in the background prediction
      including statistical and systematic sources. The signal \ptmiss distribution corresponds to the direct slepton pair production model with a slepton mass of 600\GeV and a massless \firstchi particle. The last bin includes overflow events. 
    }
    \label{fig:slepton_results}
\end{figure}

The \ptmiss distribution of data events in the slepton SRs is shown
together with the SM background predictions in Fig.~\ref{fig:slepton_results}.
Results are also summarized in Table~\ref{tab:slepton_results}.
The observed data yields are consistent with the SM predictions.
The largest discrepancy between data and SM prediction is observed in the highest \ptmiss bin
of the SR without jets where 17 events are observed and $9.3\pm2.3$ are
predicted, corresponding to a local significance of 1.6~s.d.

\begin{table}[!htb]
\renewcommand{\arraystretch}{1.3}
\setlength{\belowcaptionskip}{6pt}
\centering
\topcaption{
  Predicted and observed event yields in the slepton search and control regions. A background-only fit to observation in the CR is performed to determine the \dyjets contribution as described in Section~\ref{sec:interpretation}.
  Uncertainties include both statistical and systematic sources.}
\label{tab:slepton_results}
\begin{tabular}{lcccc}
  \ptmiss  [\GeVns{}]  &  100--150 & 150--225 & 225--300 & $>$300 \\ \hline 
 \multicolumn{5}{c}{ CR $65<\mll<120\GeV$, $\njets>0$ } \\ 
 Flavor-symmetric & $  85     \pm   11   $ & $  15.7  \pm    4.0 $ & $  1.1  \pm    0.9  $ & $  0.0^{+1.8}_{-0.0} $ \\
 \dyjets          & $  402    \pm   38   $ & $  67    \pm    21  $ & $  21.1  \pm    9.6 $ & $  0.0^{+0.1}_{-0.0} $ \\
 \znu             & $  187    \pm   20   $ & $  159   \pm    18  $ & $  49.8  \pm    6.1 $ & $  34.9  \pm    4.6  $ \\
 Total background & $  674    \pm   29   $ & $  241   \pm    16  $ & $  72.0  \pm    8.2 $ & $  34.9  \pm    3.8  $ \\
 Observed             &   674                &   241               &   72                &   30                 \\ [\cmsTabSkip]
 \multicolumn{5}{c}{ CR $65<\mll<120\GeV$, $\njets=0$ }\\
 Flavor-symmetric & $  98     \pm   11   $ & $  40.0  \pm    6.8 $ & $  2.0  \pm    1.4  $ & $  1.0  \pm    0.8   $ \\
 \dyjets          & $  458    \pm   58   $ & $  137   \pm    46  $ & $  18   \pm    13   $ & $  0.0^{+0.8}_{-0.0} $ \\
 \znu             & $  503    \pm   53   $ & $  396   \pm    46  $ & $  96   \pm    12   $ & $  46.4  \pm    6.0  $ \\
 Total background & $  1059   \pm   34   $ & $  573   \pm    26  $ & $  116  \pm    11   $ & $  47.5  \pm    5.3  $ \\
 Observed             &   1059               &   573               &   116               &   47                 \\[\cmsTabSkip]
 \multicolumn{5}{c}{ SR $\mll<65$ or $\mll>120\GeV$, $\njets>0$ }\\ 
 Flavor-symmetric & $  203    \pm     16 $ & $  95    \pm   11   $ & $  8.4  \pm    2.9  $ & $  5.2  \pm    2.3   $ \\
 \dyjets          & $  33     \pm     28 $ & $  5.4  \pm    5.6  $ & $  1.7  \pm    1.8  $ & $  0.0^{+0.1}_{-0.0} $ \\
 \znu             & $  9.9    \pm    1.4 $ & $  11.3  \pm    1.6 $ & $  4.6  \pm    0.6  $ & $  3.5  \pm    0.5   $ \\
 Total background & $  245    \pm     33 $ & $  112    \pm   12  $ & $  14.7  \pm    3.3 $ & $  8.7  \pm    2.3   $ \\
 Observed             &   283                &   97                &   19                &   8                  \\[\cmsTabSkip]
 \multicolumn{5}{c}{ SR $\mll<65$ or $\mll>120\GeV$, $\njets=0$ }\\
 Flavor-symmetric & $  134    \pm   12   $ & $  82.5  \pm    9.5 $ & $  11.6  \pm    3.3 $ & $  4.2  \pm    2.2   $ \\
 \dyjets          & $  38    \pm   34    $ & $  11    \pm   13   $ & $  1.4  \pm    2.3  $ & $  0.0^{+0.1}_{-0.0} $ \\
 \znu             & $  26.6  \pm    3.7  $ & $  26.2  \pm    3.7 $ & $  7.8  \pm    1.1  $ & $  5.1  \pm    0.7   $ \\
 Total background & $  198    \pm   37   $ & $  120   \pm     16 $ & $  20.8  \pm    4.1 $ & $  9.3  \pm    2.3   $ \\
 Observed             & 228                    & 99                    & 29                    & 17                     \\
\end{tabular}
\end{table}

\section{Interpretation of the results}
\label{sec:interpretation}

The results are interpreted in the context of models of BSM physics presented in Section~\ref{sec:introduction}.
Maximum likelihood fits are performed under either background-only or signal+background hypotheses
to the data in the SRs and some of the CRs: the event yield with $50<\ptmiss<100\GeV$ in the on-\PZ SRs and the on-\PZ category in the slepton SRs.
The uncertainties in the modeling of the backgrounds, summarized
below,
are inputs to the fitting procedure.
The likelihoods are constructed as the product of Poisson probability density functions, one for each SR, with
additional nuisance parameters constrained by log-normal terms that account for  uncertainties in the background predictions, and in signal yields when the signal is included in the hypothesis.
When a CR for a given background is included in the fit, the normalizations of both signal and that background are treated as free parameters.
This accounts for the possible presence of signal events
in the CRs (signal contamination).

The signal+background fits are used to set 95\% confidence level (\CL) upper limits on the production cross sections
of the modeled signals.
We employ a modified frequentist approach, using the \CLs criterion and 
relying on asymptotic approximations, to calculate the distribution of 
the profile likelihood test statistic~\cite{Junk:1999kv,Read:2002hq,Cowan:2010js,HiggsTool1}.
The limits are then used, together with the theoretical cross section, 
to exclude ranges of masses for the BSM particles involved in each  model.

\subsection{Systematic uncertainties in the signal }
\label{sec:systs}

We include uncertainties in the expected signal  for all of the SMS models,
as summarized in Table~\ref{tab:systs}.
The integrated luminosities of the 2016, 2017, and 2018 data-taking periods are individually
known with uncertainties in the 2.3--2.5\% range~\cite{LUM-17-001,LUM-17-004,LUM-18-002},
while the total Run~2 (2016--2018) integrated luminosity has an uncertainty of 1.8\%, the
improvement in precision reflecting the (uncorrelated) time evolution of some systematic effects.
We also include uncertainties in
the lepton trigger, identification, and isolation efficiencies,
in the \PQb tagging efficiencies and the mistag probability.
To check the modeling of ISR in the EW (strong) signal simulations we obtain distributions
in the number of ISR jets in data samples enriched in \dyjets (\ttbar). We derive weights as ratios
of these distributions to simulation. The systematic uncertainty in the ISR modeling is then given
by the difference between weighted and unweighted simulations of the signals.

Additional uncertainties arise from the potential mismodeling of pileup, from JES,
from the choice of the $\mu_{\mathrm{R}}$ and $\mu_{\mathrm{F}}$
scales used in the event generator~\cite{Cacciari:2003fi,Catani:2003zt,Frederix:2011ss}, and from the limited number of simulated events.
Finally,
any further possible mismodeling of lepton efficiencies, jet energy response, \PQb tag efficiency, and the \ptmiss distributions
associated with the CMS fast simulation framework is accounted for with an additional uncertainty.
The assumed correlations in the signal uncertainties across SRs are the same as those
described in Section~\ref{sec:backgrounds}
for the background processes. Uncertainties in ISR modeling and fast simulation are treated
as correlated across the SRs and across the data samples.

\begin{table}[htb]
\renewcommand*{\arraystretch}{1.1}
\centering
\topcaption{\label{tab:systs}
Summary of the systematic uncertainties in the signal yields 
together with their typical sizes across the search regions and the SMS models under consideration.}
\begin{tabular}{l c}
\hline
Source of uncertainty                  & Uncertainty (\%)   \\
\hline 
Integrated luminosity                  & 1.8                \\
Limited size of simulated samples      & 1--15              \\
Trigger efficiency                     & 3                  \\
Lepton efficiency                      & 5                  \\
Fast simulation lepton efficiency      & 4                  \\
Fast simulation \PQb tag efficiency    & 0--5               \\
Jet energy scale                       & 0--5               \\
Pileup modeling                        & 1--2               \\
ISR modeling                           & 0--2.5             \\
$\mu_{\mathrm{R}}$ and $\mu_{\mathrm{F}}$ variation & 1--3  \\
Fast simulation \ptmiss modeling        & 0--4              \\ \hline
\end{tabular}
\end{table}

\subsection{Interpretations of the results using simplified SUSY models}

\begin{figure}[!hbt]
 \centering
 \includegraphics[width=0.60\textwidth]{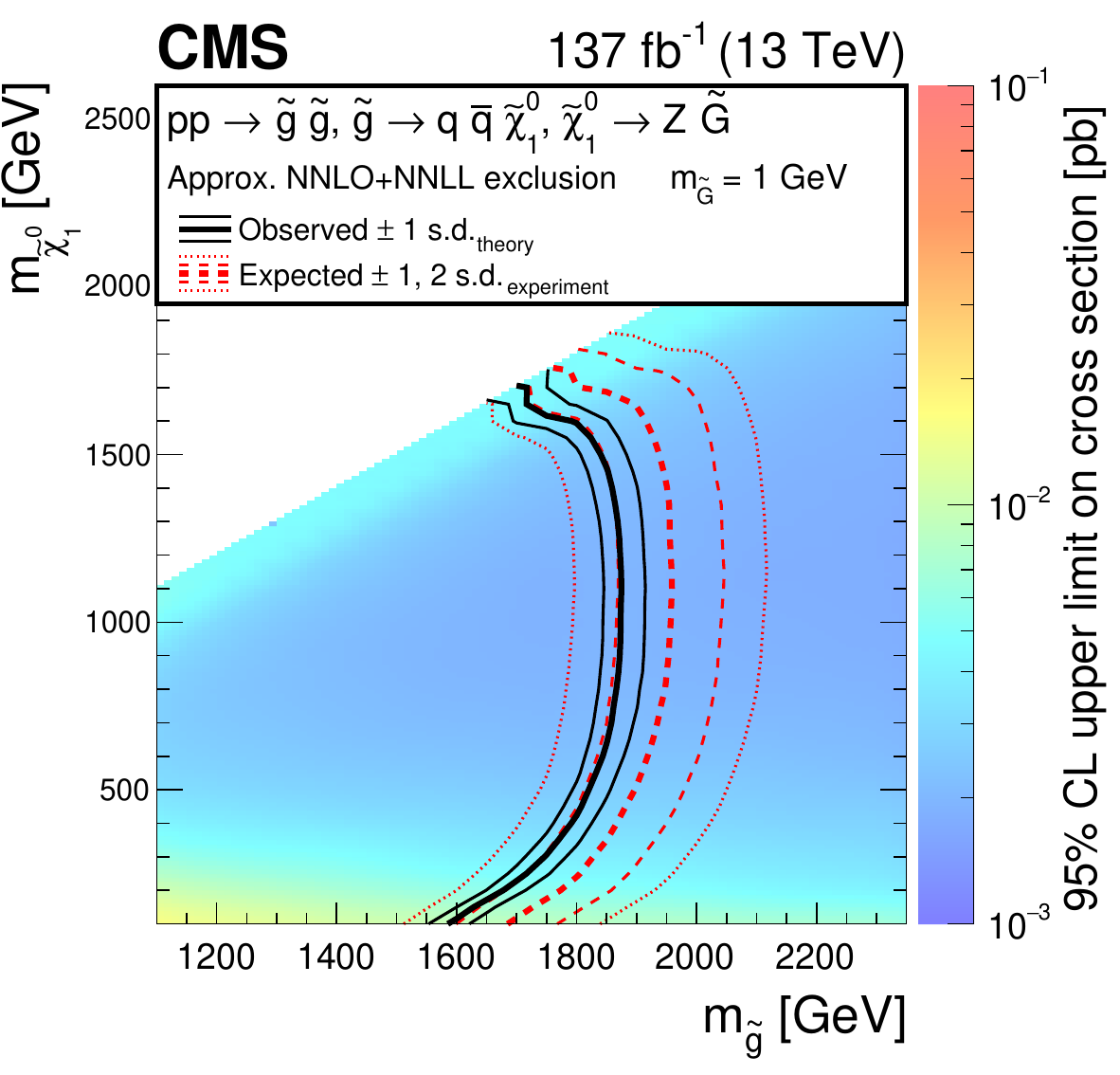}
   \caption{\label{fig:Limits1}
     Cross section upper limits and exclusion contours at 95\% \CL for an SMS model of GMSB gluino pair production, 
     as a function of the \gluino and \firstchi masses,
     obtained from the results in the strong-production on-\PZ search regions.
     The area enclosed by the thick black curve represents the observed exclusion region, 
     while the dashed red lines indicate the expected limits and their $\pm1$ and $\pm2$ s.d. ranges. 
     The thin black lines show the effect of the theoretical uncertainties on the signal cross section.
   }
\end{figure}

The results in the strong-production on-\PZ search regions are interpreted using 
an SMS model of gluino pair production, discussed in Section~\ref{sec:introduction}.
This signal is characterized by final states with substantial activity (energy in jets). 
All strong-production on-\PZ SRs are included in the maximum likelihood fit
in order to maximize the acceptance in
models in which the  \gluino and \firstchi masses are close, where less jet activity is expected.
The upper limit at 95\% \CL on the signal production cross section is shown in Fig.~\ref{fig:Limits1},
as a function of the \gluino and \firstchi masses, 
together with the expected and observed exclusion contours. 
We exclude gluino masses up to 1600--1870\GeV, depending
on the mass of \firstchi, 
extending the reach of previous CMS results~\cite{SUS-16-034} by approximately 100\GeV.

\begin{figure}[hbt!]
 \centering
 \includegraphics[width=0.6\textwidth]{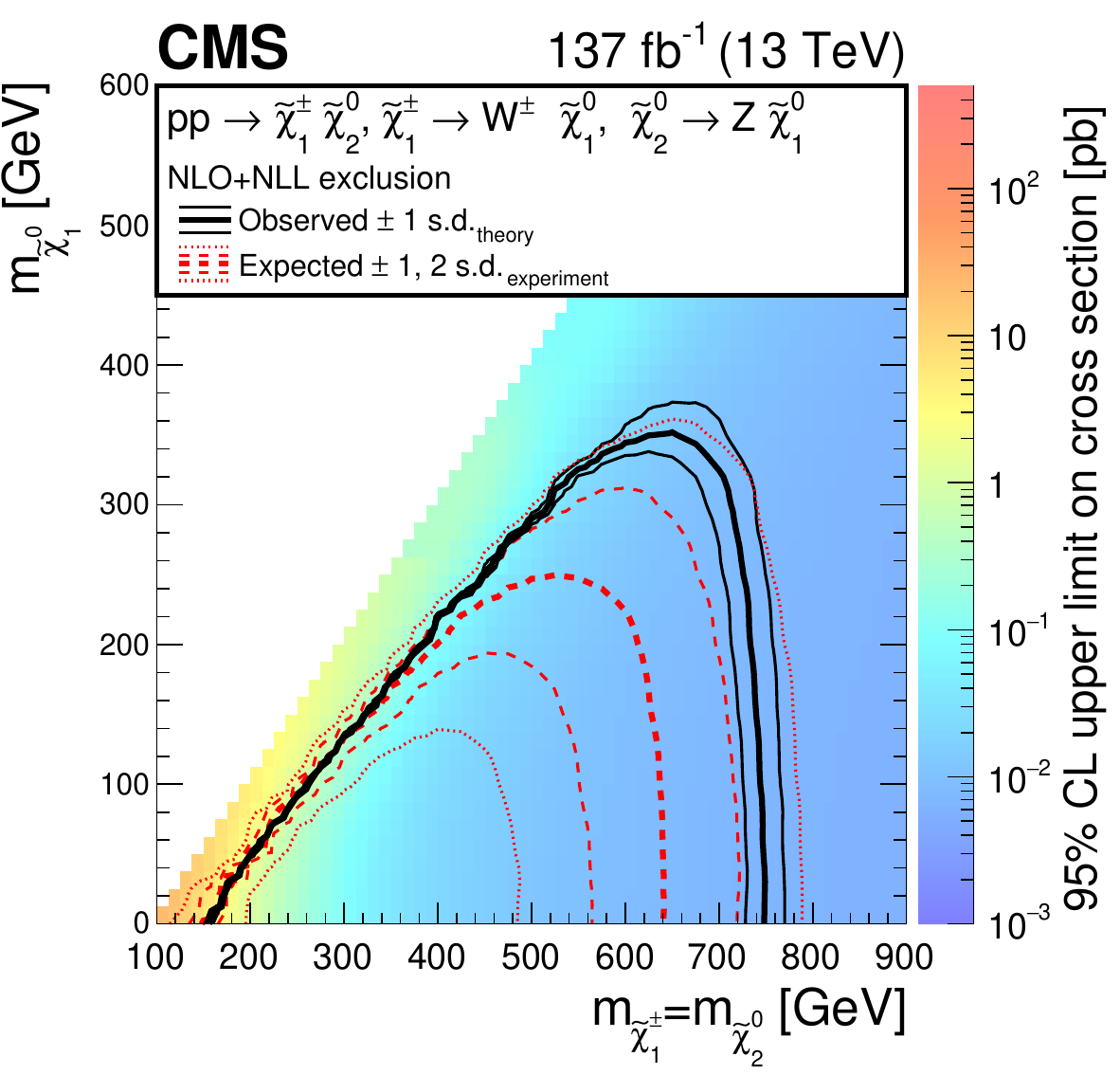}
   \caption{\label{fig:LimitTChiWZ}
     Cross section upper limits and exclusion contours at 95\% \CL for an SMS model of \firstcharg\secondchi production,
     with signatures containing a \PW and a \PZ bosons,
     as a function of the \firstcharg/\secondchi and \firstchi masses,
     obtained from the results in the EW-production on-\PZ search regions.
     The area enclosed by the thick black curve represents the observed exclusion region, 
     while the dashed red lines indicate the expected limits and their $\pm1$ and $\pm2$ s.d. ranges. 
     The thin black lines show the effect of the theoretical uncertainties on the signal cross section.
   }
\end{figure}

Upper limits at 95\% \CL on the  cross section of models for \firstcharg\secondchi and for \firstchi pair production
 are set using the results of the EW-production on-\PZ SRs. 
For \firstcharg\secondchi production with decays to $\PW\PZ$, the \vz search regions provide most of the sensitivity. 
While the resolved \vz SR is sensitive to a wide range of \firstcharg/\secondchi and \firstchi mass hypotheses,
the use of the boosted \vz region improves  the sensitivity for
masses of the \firstcharg/\secondchi  much larger than
the mass of \firstchi, where the bosons produced in the decay chain receive a large Lorentz boost. 
Figure~\ref{fig:LimitTChiWZ} shows the cross section upper limits and the exclusion contours at 95\% \CL obtained for this model
as a function of the \firstcharg/\secondchi and \firstchi masses.
We exclude \firstcharg/\secondchi masses up to 750\GeV, extending the reach of Ref.~\cite{SUS-16-034}
by approximately 100\GeV.

\begin{figure}[htb!]
\centering
\includegraphics[width=0.49\textwidth]{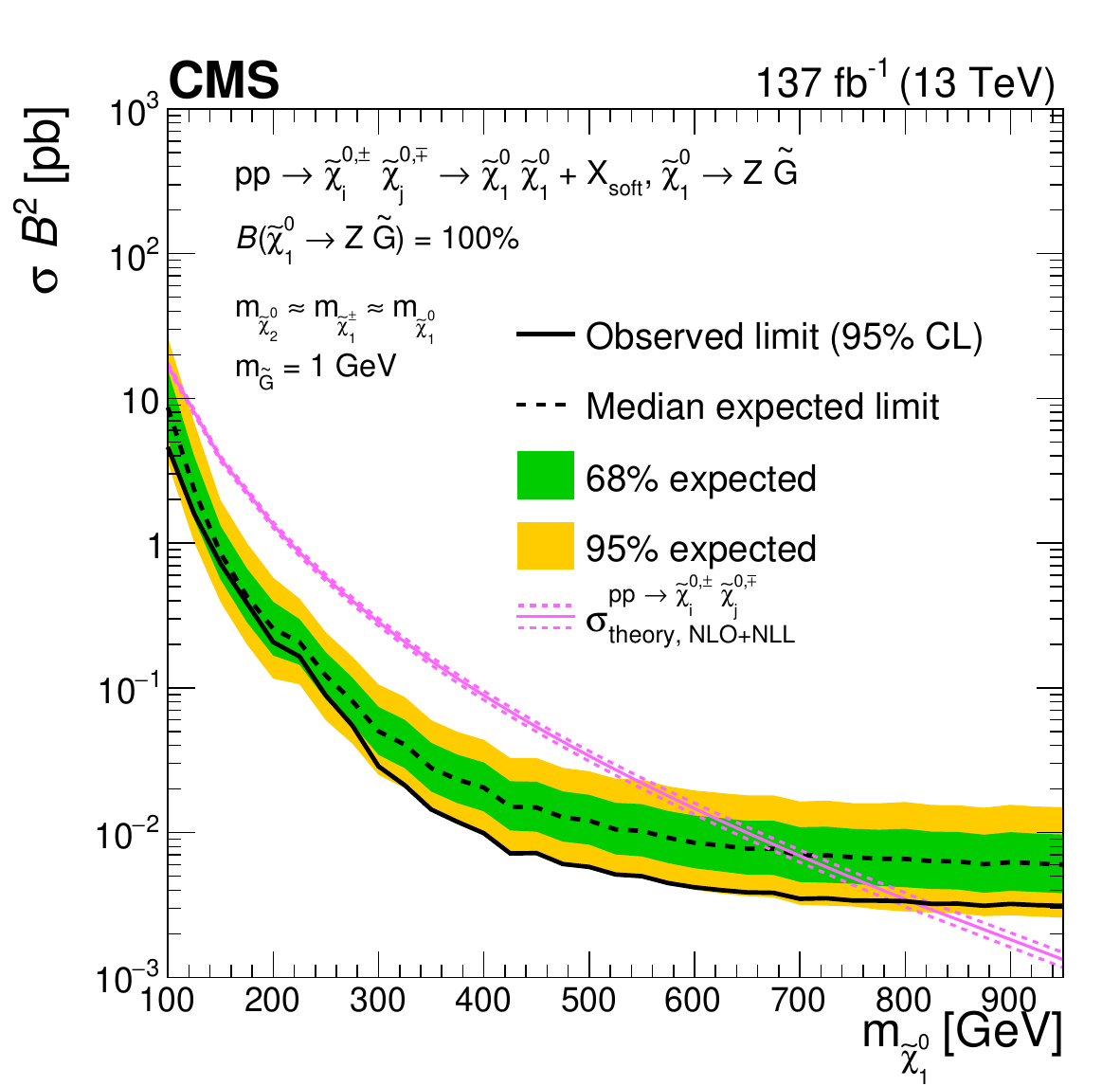}
\includegraphics[width=0.49\textwidth]{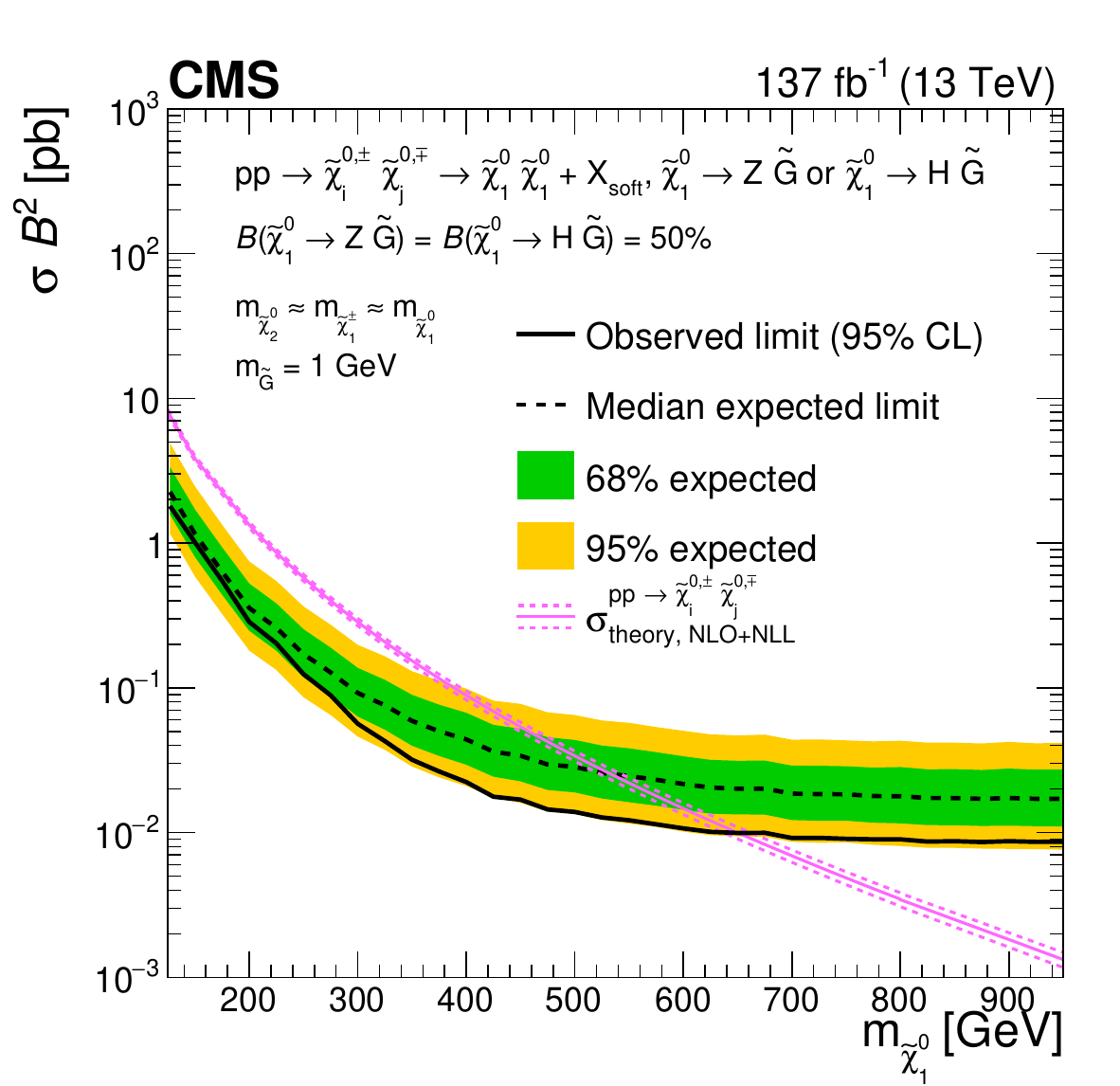}
\caption{
  Production cross section upper limits at 95\% \CL as a function of the \firstchi mass,
  for a model of EW \firstchi pair production, 
  where either (left) both \firstchi decay into a \PZ boson with a 100\% branching fraction ($\mathcal{B}$),
  or (right) each \firstchi can decay to a \PZ or an \PH with equal probability.
  The model assumes the production of mass-degenerate neutralinos and charginos that decay into \firstchi possibly
  emitting soft particles, labeled as $\PX_{\text{soft}}$.
  The magenta curve shows the theoretical production cross section with its uncertainty.
  The solid (dashed) black line represents the observed (median expected) exclusion.
  The inner green (outer yellow) band indicates the region containing 68\,(95)\% of the distribution of limits 
  expected under the background-only hypothesis.
}
\label{fig:LimitTChiZZHZ}
\end{figure}

Two models are considered for \firstchi pair production. 
One assumes that both \firstchi decay into a \PZ boson with a 100\% branching fraction.
The other assumes that each \firstchi can decay to either \PZ or \PH with equal probability.
The first model  leads to signatures with a pair of \PZ bosons,
with most of the signal events expected to populate the \vz SRs.
On the other hand, signal events where an \PH  decays to \bbbar are expected to populate the $\PH\PZ$ region.
The observed and expected upper limits at 95\% \CL on the production cross section times branching fraction product for both  models
are shown in Fig.~\ref{fig:LimitTChiZZHZ}.
In these two scenarios, we are able to exclude \firstchi masses up to 800 and 650\GeV respectively, 
extending the reach of Ref.~\cite{SUS-16-034} by approximately 150\GeV.
Figures~\ref{fig:LimitTChiWZ} and~\ref{fig:LimitTChiZZHZ} show observed exclusion limits that are more stringent than the expected ones. In both cases this arises from the downward fluctuation in the  observed data yields in the two highest \ptmiss SRs of the resolved \vz search region, as discussed in  Section~\ref{sub:onZResults}.

\begin{figure}[!htbp]
  \centering
  \includegraphics[width=0.49\textwidth]{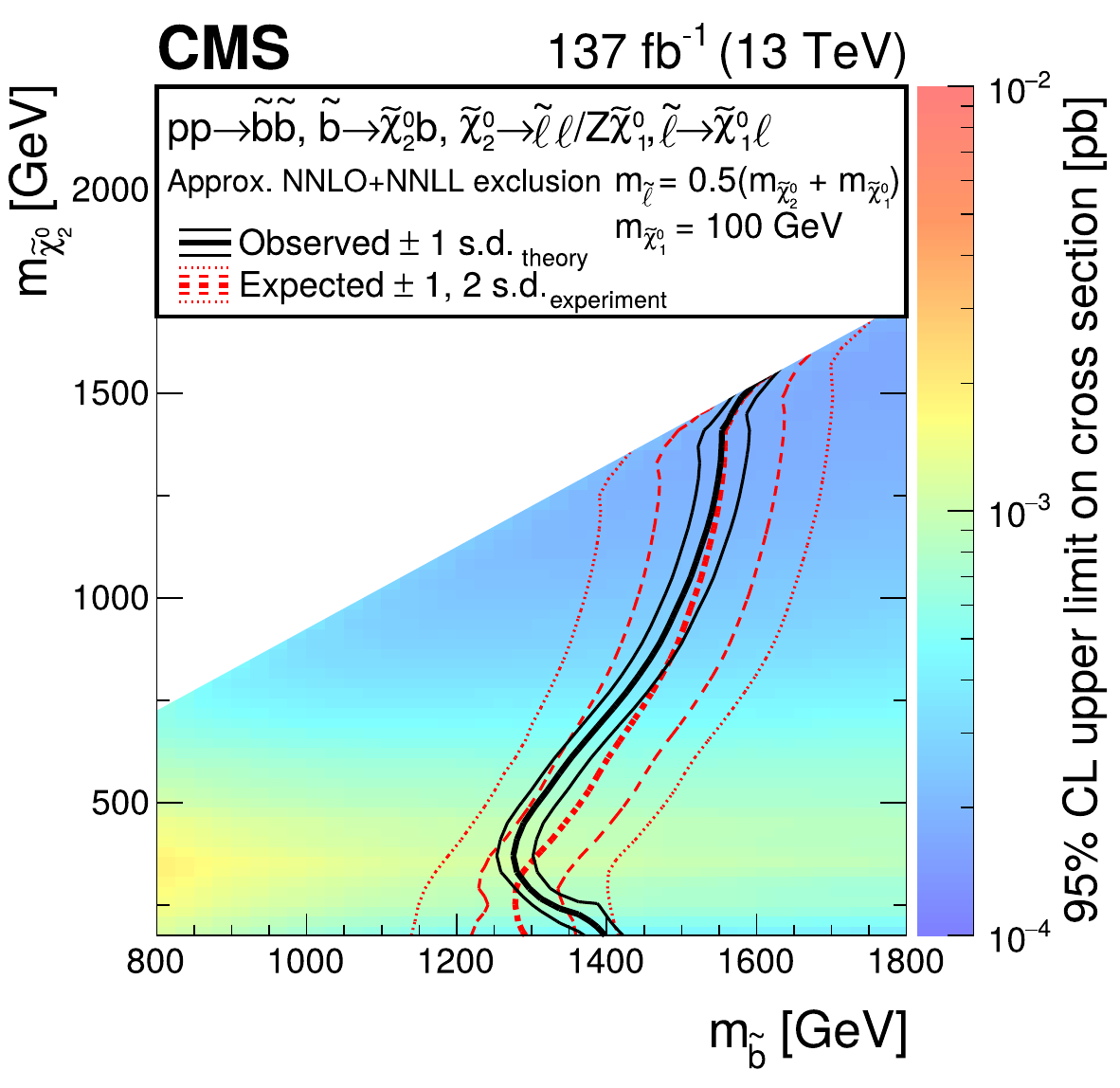}
  \includegraphics[width=0.49\textwidth]{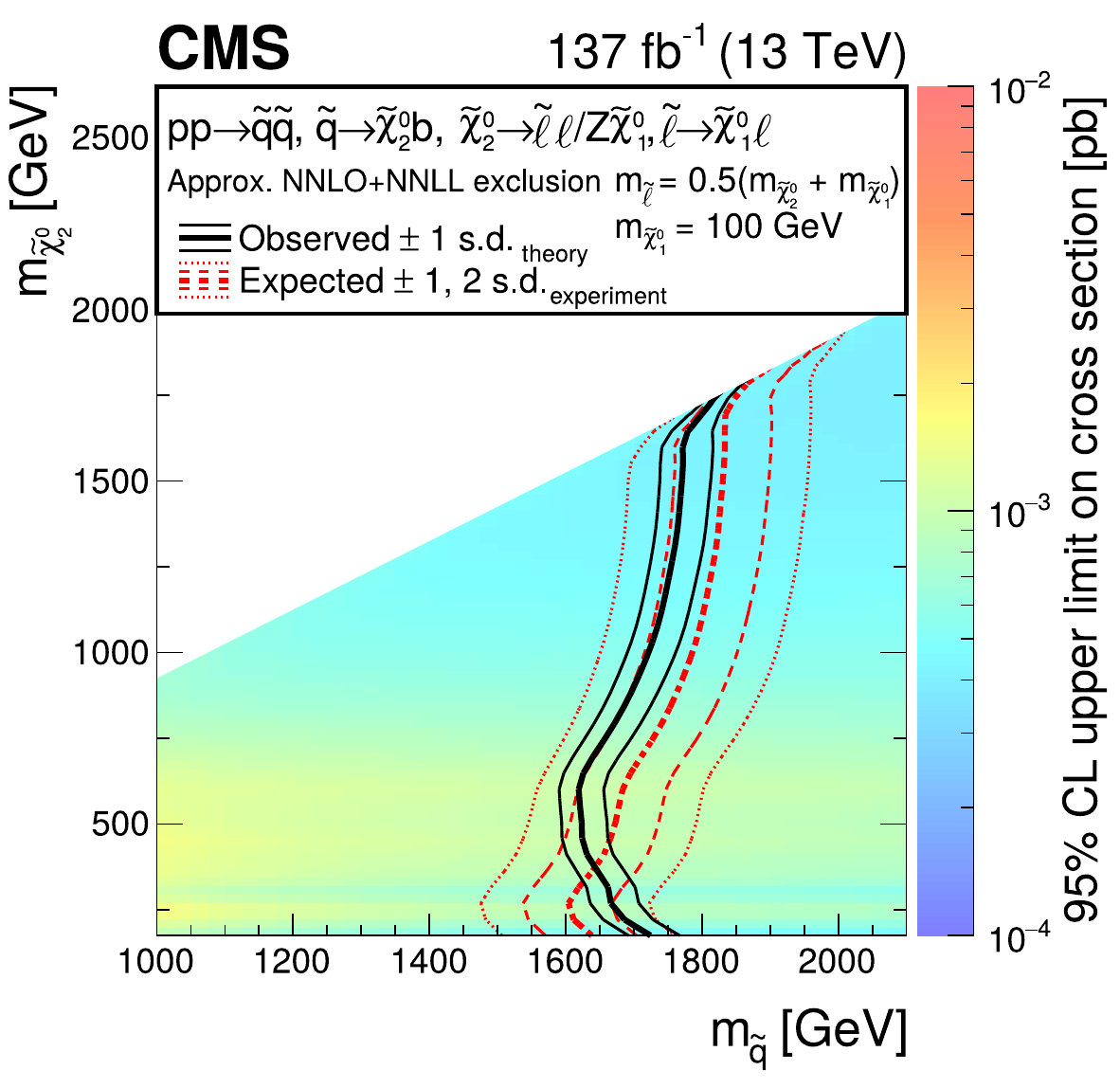}
    \caption{
     Cross section upper limits and exclusion contours at 95\% \CL for SMS models of 
     (left) bottom and (right) light-flavor squark pair production. In these models,
     each squark decays into a quark and a \secondchi, and the \secondchi then decays via an intermediate slepton, 
     forming a kinematic edge in the \mll distribution. 
     The limits are obtained from the results in the edge search regions, 
     and are shown as a function of the (left) \sbottom or (right) \squark and \secondchi masses. 
     The thick black curve represents the observed upper limit on the squark mass,
     while the dashed red lines indicate the expected limits and their $\pm1$ and $\pm2$ s.d. ranges. 
     The thin black lines show the effect of the theoretical uncertainties on the signal cross section.
    }
    \label{fig:Limits2}
\end{figure}

The edge search regions used in the counting experiment serve to constrain the two slepton edge models presented in Section~\ref{sec:introduction} 
(Fig.~\ref{sig:feynman_strong}~left and middle diagrams).
Figure~\ref{fig:Limits2} shows the upper limits at 95\% \CL on the production cross section 
for both of these models. 
We exclude bottom (light-flavor) squark (\squark) masses up to 1300--1600\,(1600--1800)\GeV,
depending on the assumed \secondchi mass.
For the case of the bottom squark pair production, 
we improve the results from Ref.~\cite{SUS-16-034} by up to 300\GeV. 
The observed exclusion limits are more stringent than expected for models with small \secondchi mass. 
This is caused by a mild deficit of observed events 
in the non-\ttbar-like edge regions for $20 < \mll < 60\GeV$.

\begin{figure}[!htbp]
\centering
  \includegraphics[width=0.49\textwidth]{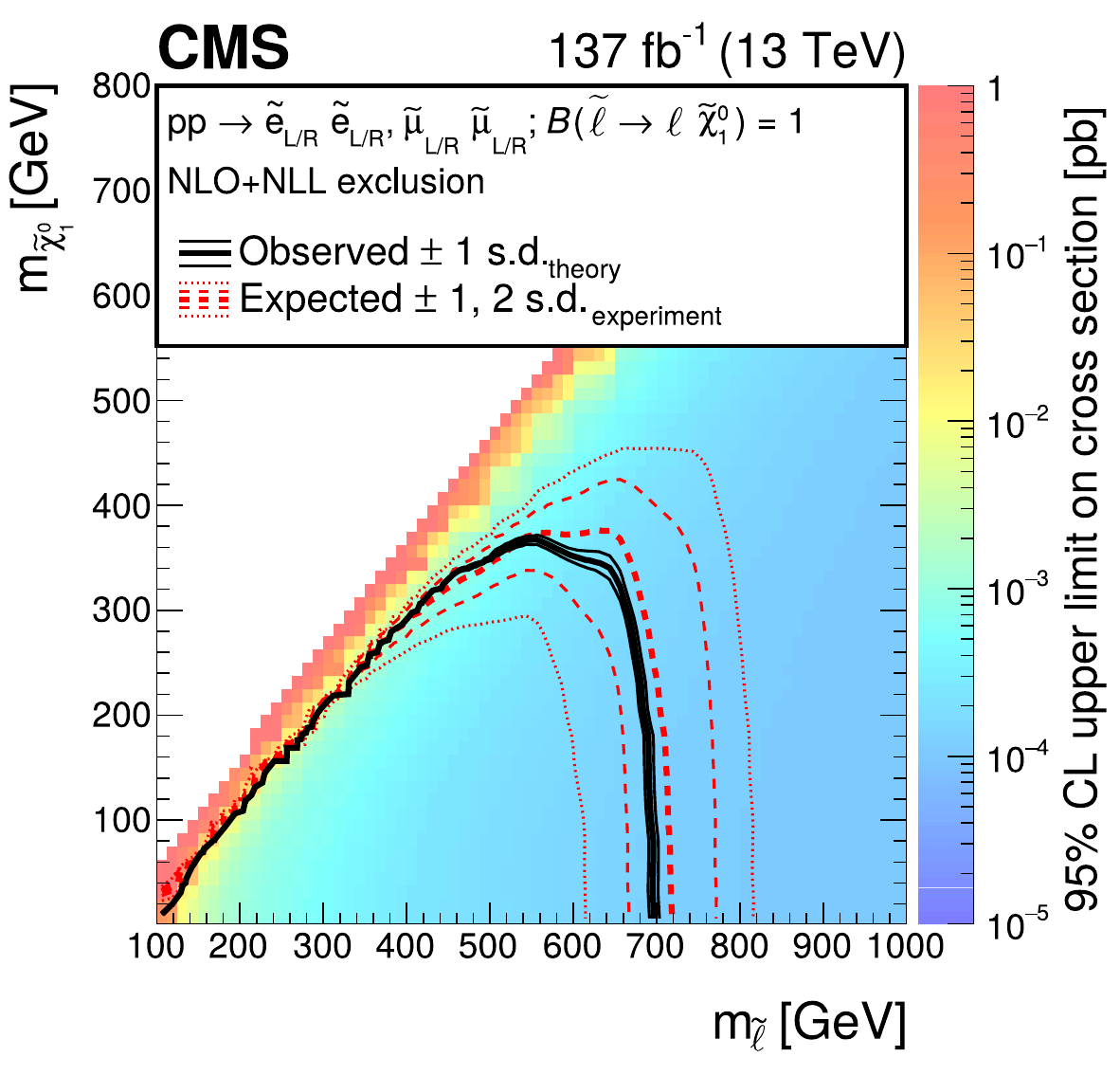}
  \caption{
     Cross section upper limits and exclusion contours at 95\% \CL for an SMS model of slepton pair production,
     as a function of the slepton and \firstchi masses,
     obtained from the results in the slepton search regions.
     The area enclosed by the thick black curve represents the observed exclusion region, 
     while the dashed red lines indicate the expected limits and their $\pm1$ and $\pm2$ s.d. ranges. 
     The thin black lines show the effect of the theoretical uncertainties in the signal cross section.
  }
  \label{fig:slepton_interpr}
\end{figure}

The results in the slepton SRs are 
interpreted in the context of a slepton pair
production model, introduced in Section~\ref{sec:introduction}.
Upper limits at 95\% \CL on the signal production cross section are shown in
Fig.~\ref{fig:slepton_interpr}. 
Slepton masses up to 700\GeV are excluded for small \firstchi masses, improving the previous CMS results~\cite{Sirunyan:2018nwe} by approximately 200\GeV.

\clearpage

\section{Summary}
\label{sec:summary}

A search is presented for phenomena beyond the standard model in events with two oppositely charged same-flavor leptons and
missing transverse momentum in the final state. 
The search is performed in a sample of proton-proton collisions at $\sqrt{s}=13\TeV$
collected with the CMS detector  corresponding to an integrated
luminosity of \lint. 
Search regions are defined to be sensitive to a wide range of signatures.
The observed event yields and distributions are found to be consistent with the expectations from the standard model. 
The results are used to set upper limits on the production cross sections of simplified models of supersymmetry.
Gluino masses are excluded up to 1870\GeV,
light-flavor (bottom) squark masses up to 1800\,(1600)\GeV,
chargino (neutralino) masses up to 750\,(800)\GeV, and
slepton masses up to 700\GeV,
typically extending the reach over previous CMS results by a few hundred~GeV.

\begin{acknowledgments}
  We congratulate our colleagues in the CERN accelerator departments for the excellent performance of the LHC and thank the technical and administrative staffs at CERN and at other CMS institutes for their contributions to the success of the CMS effort. In addition, we gratefully acknowledge the computing centers and personnel of the Worldwide LHC Computing Grid for delivering so effectively the computing infrastructure essential to our analyses. Finally, we acknowledge the enduring support for the construction and operation of the LHC and the CMS detector provided by the following funding agencies: BMBWF and FWF (Austria); FNRS and FWO (Belgium); CNPq, CAPES, FAPERJ, FAPERGS, and FAPESP (Brazil); MES (Bulgaria); CERN; CAS, MoST, and NSFC (China); COLCIENCIAS (Colombia); MSES and CSF (Croatia); RIF (Cyprus); SENESCYT (Ecuador); MoER, ERC PUT and ERDF (Estonia); Academy of Finland, MEC, and HIP (Finland); CEA and CNRS/IN2P3 (France); BMBF, DFG, and HGF (Germany); GSRT (Greece); NKFIA (Hungary); DAE and DST (India); IPM (Iran); SFI (Ireland); INFN (Italy); MSIP and NRF (Republic of Korea); MES (Latvia); LAS (Lithuania); MOE and UM (Malaysia); BUAP, CINVESTAV, CONACYT, LNS, SEP, and UASLP-FAI (Mexico); MOS (Montenegro); MBIE (New Zealand); PAEC (Pakistan); MSHE and NSC (Poland); FCT (Portugal); JINR (Dubna); MON, RosAtom, RAS, RFBR, and NRC KI (Russia); MESTD (Serbia); SEIDI, CPAN, PCTI, and FEDER (Spain); MOSTR (Sri Lanka); Swiss Funding Agencies (Switzerland); MST (Taipei); ThEPCenter, IPST, STAR, and NSTDA (Thailand); TUBITAK and TAEK (Turkey); NASU (Ukraine); STFC (United Kingdom); DOE and NSF (USA).
   
  \hyphenation{Rachada-pisek} Individuals have received support from the Marie-Curie program and the European Research Council and Horizon 2020 Grant, contract Nos.\ 675440, 724704, 752730, and 765710 (European Union); the Leventis Foundation; the A.P.\ Sloan Foundation; the Alexander von Humboldt Foundation; the Belgian Federal Science Policy Office; the Fonds pour la Formation \`a la Recherche dans l'Industrie et dans l'Agriculture (FRIA-Belgium); the Agentschap voor Innovatie door Wetenschap en Technologie (IWT-Belgium); the F.R.S.-FNRS and FWO (Belgium) under the ``Excellence of Science -- EOS" -- be.h project n.\ 30820817; the Beijing Municipal Science \& Technology Commission, No. Z191100007219010; the Ministry of Education, Youth and Sports (MEYS) of the Czech Republic; the Deutsche Forschungsgemeinschaft (DFG) under Germany's Excellence Strategy -- EXC 2121 ``Quantum Universe" -- 390833306; the Lend\"ulet (``Momentum") Program and the J\'anos Bolyai Research Scholarship of the Hungarian Academy of Sciences, the New National Excellence Program \'UNKP, the NKFIA research grants 123842, 123959, 124845, 124850, 125105, 128713, 128786, and 129058 (Hungary); the Council of Science and Industrial Research, India; the HOMING PLUS program of the Foundation for Polish Science, cofinanced from European Union, Regional Development Fund, the Mobility Plus program of the Ministry of Science and Higher Education, the National Science Center (Poland), contracts Harmonia 2014/14/M/ST2/00428, Opus 2014/13/B/ST2/02543, 2014/15/B/ST2/03998, and 2015/19/B/ST2/02861, Sonata-bis 2012/07/E/ST2/01406; the National Priorities Research Program by Qatar National Research Fund; the Ministry of Science and Higher Education, project no. 0723-2020-0041 (Russia); the Tomsk Polytechnic University Competitiveness Enhancement Program; the Programa Estatal de Fomento de la Investigaci{\'o}n Cient{\'i}fica y T{\'e}cnica de Excelencia Mar\'{\i}a de Maeztu, grant MDM-2015-0509 and the Programa Severo Ochoa del Principado de Asturias; the Thalis and Aristeia programs cofinanced by EU-ESF and the Greek NSRF; the Rachadapisek Sompot Fund for Postdoctoral Fellowship, Chulalongkorn University and the Chulalongkorn Academic into Its 2nd Century Project Advancement Project (Thailand); the Kavli Foundation; the Nvidia Corporation; the SuperMicro Corporation; the Welch Foundation, contract C-1845; and the Weston Havens Foundation (USA).\end{acknowledgments}

\bibliography{auto_generated}

\cleardoublepage \appendix\section{The CMS Collaboration \label{app:collab}}\begin{sloppypar}\hyphenpenalty=5000\widowpenalty=500\clubpenalty=5000\input{SUS-20-001-authorlist.tex}\end{sloppypar}
\end{document}

%% file: SUS-20-001-authorlist.tex
\vskip\cmsinstskip
\textbf{Yerevan Physics Institute, Yerevan, Armenia}\\*[0pt]
A.M.~Sirunyan$^{\textrm{\dag}}$, A.~Tumasyan
\vskip\cmsinstskip
\textbf{Institut f\"{u}r Hochenergiephysik, Wien, Austria}\\*[0pt]
W.~Adam, T.~Bergauer, M.~Dragicevic, A.~Escalante~Del~Valle, R.~Fr\"{u}hwirth\cmsAuthorMark{1}, M.~Jeitler\cmsAuthorMark{1}, N.~Krammer, L.~Lechner, D.~Liko, I.~Mikulec, F.M.~Pitters, N.~Rad, J.~Schieck\cmsAuthorMark{1}, R.~Sch\"{o}fbeck, M.~Spanring, S.~Templ, W.~Waltenberger, C.-E.~Wulz\cmsAuthorMark{1}, M.~Zarucki
% ...
\vskip\cmsinstskip
\dag: Deceased\\
1:  Also at Vienna University of Technology, Vienna, Austria\\
% ...
